\documentclass[3p,times,twocolumn]{elsarticle}
\usepackage{ecrc}

\volume{00}
\firstpage{1}
\journalname{Nuclear Physics B Proceedings Supplement}
\runauth{Uwe C. T\"auber}
\jid{nuphbp}
\jnltitlelogo{Nuclear Physics B Proceedings Supplement}
\usepackage{amssymb}

\def\underbrac#1{\mathop{\vbox{\ialign{##\crcr\noalign{\kern-13.4pt}
                \underbracfill\crcr\noalign{\kern-13.4pt\nointerlineskip}
                 $\hfil\displaystyle{#1}\hfil$\crcr}}}\limits}
\def\underbracfill{${\llulin }\leaders\hrule\hfill{\rrulin }$}
\newcommand{\rrulin}{\unitlength=1.00mm \linethickness{0.4pt}
            \begin{picture}(1,0)(0,-0.02) \line(0,1){1} \end{picture} }
\newcommand{\llulin}{\unitlength=1.00mm \linethickness{0.4pt}
            \begin{picture}(1,0)(-1,-0.02) \line(0,1){1} \end{picture} }

\begin{document}
\begin{frontmatter}




\title{Renormalization Group: Applications in Statistical Physics}

\author{Uwe C. T\"auber}

\address{Department of Physics, Virginia Tech, Blacksburg, VA 24061-0435, USA}

\ead{tauber@vt.edu}  

\begin{abstract}
These notes aim to provide a concise pedagogical introduction to some important
applications of the renormalization group in statistical physics.
After briefly reviewing the scaling approach and Ginzburg--Landau theory for 
critical phenomena near continuous phase transitions in thermal equilibrium,
Wilson's momentum shell renormalization group method is presented, and the
critical exponents for the scalar $\Phi^4$ model are determined to first order
in a dimensional $\epsilon$ expansion about the upper critical dimension 
$d_c = 4$.
Subsequently, the physically equivalent but technically more versatile 
field-theoretic formulation of the perturbational renormalization group for 
static critical phenomena is described.
It is explained how the emergence of scale invariance connects ultraviolet 
divergences to infrared singularities, and the renormalization group equation 
is employed to compute the critical exponents for the $O(n)$-symmetric 
Landau--Ginzburg--Wilson theory to lowest non-trivial order in the $\epsilon$ 
expansion.
The second part of this overview is devoted to field theory representations of
non-linear stochastic dynamical systems, and the application of renormalization
group tools to critical dynamics.
Dynamic critical phenomena in systems near equilibrium are efficiently captured
through Langevin stochastic equations of motion, and their mapping onto the
Janssen--De~Dominicis response functional, as exemplified by the 
field-theoretic treatment of purely relaxational models with non-conserved 
(model A) and conserved order parameter (model B).
As examples for other universality classes, the Langevin description and 
scaling exponents for isotropic ferromagnets at the critical point (model J) 
and for driven diffusive non-equilibrium systems are discussed. 
Finally, an outlook is presented to scale-invariant phenomena and 
non-equilibrium phase transitions in interacting particle systems.
It is shown how the stochastic master equation associated with chemical 
reactions or population dynamics models can be mapped onto imaginary-time, 
non-Hermitian ``quantum'' mechanics.
In the continuum limit, this Doi--Peliti Hamiltonian is in turn represented 
through a coherent-state path integral action, which allows an efficient and
powerful renormalization group analysis of, e.g., diffusion-limited 
annihilation processes, and of phase transitions from active to inactive, 
absorbing states.
\end{abstract}


\end{frontmatter}

\section{Introduction}
\label{intro}

Since Ken Wilson's seminal work in the early 1970s \cite{Wilson74}, based also 
on the groundbreaking foundations laid by Leo Kadanoff, Ben Widom, Michael 
Fisher \cite{Fisher74}, and others in the preceding decade, the renormalization
group (RG) has had a profound impact on modern statistical physics.
Not only do renormalization group methods provide a powerful tool to 
analytically describe and quantitatively capture both static and dynamic 
critical phenomena near continuous phase transitions that are governed by 
strong interactions, fluctuations, and correlations, they also allow us to 
address physical properties associated with the emerging generic scale 
invariance in certain entire thermodynamic phases, many non-equilibrium steady 
states, and in relaxation phenomena towards either equilibrium or 
non-equilibrium stationary states.
In fact, the renormalization group presents us with a conceptual framework and
mathematical language that has become ubiquitous in the theoretical description
of many complex interacting many-particle systems encountered in nature.
One may even argue that the fundamental RG notions of universality and 
relevance or irrelevance of interactions and perturbations, and the 
accompanying systematic coarse-graining procedures are of crucial importance 
for any attempt at capturing natural phenomena in terms of only a few meso- or
macroscopic degrees of freedom, and thus also form the essential philosophical 
basis for any computational modeling, including Monte Carlo simulations.

In these lecture notes, I aim to give a pedagogical introduction and concise 
overview of first the classic applications of renormalization group methods to
equilibrium critical phenomena, and subsequently to the study of critical
dynamics, both near and far away from thermal equilibrium.
The second half of this article will specifically explain how the stochastic
dynamics of interacting many-particle systems, mathematically described either
through (coupled) non-linear Langevin or more ``microscopic'' master equations,
can be mapped onto dynamical field theory representations, and then analyzed by
means of RG-improved perturbative expansions. 
In addition, it will be demonstrated how exploiting the general structure of 
the RG flow equations, fixed point conditions, and prevalent symmetries yields 
certain exact statements.
Other authors contributing to this volume will discuss additional applications 
of renormalization group tools to a broad variety of physical systems and 
problems, and also cover more recently developed efficient non-perturbative 
approaches.

\section{Critical Phenomena}

We begin with a quick review of Landau's generic mean-field treatment of
continuous phase transitions in thermal equilibrium, define the critical 
exponents that characterize thermodynamic singularities, and then venture to an
even more general description of critical phenomena by means of scaling theory.
Next we generalize to spatially inhomogeneous configurations, investigate
critical infrared singularities in the two-point correlation function, and 
analyze the Gaussian fluctuations for the ensuing Landau--Ginzburg--Wilson 
Hamiltonian (scalar Euclidean $\Phi^4$ field theory).
This allows us to identify $d_c = 4$ as the upper critical dimension below 
which fluctuations crucially impact the critical power laws.
Finally, we introduce Wilson's momentum shell renormalization group approach,
reconsider the Gaussian model, discuss the general emerging structure, and at
last perturbatively compute the fluctuation corrections to the critical 
exponents to first order in the dimensional expansion parameter 
$\epsilon = d_c - d$.
Far more detailed expositions of the contents of this chapter can be found in
the excellent textbooks \cite{Ma76}--\cite{Mazenko03}
and in chap.~1 of Ref.~\cite{Tauberxx}.

\subsection{Continuous phase transitions}

Different thermodynamic phases are characterized by certain macroscopic, 
usually extensive state variables called order parameters; examples are the 
magnetization in ferromagnetic systems, polarization in ferroelectrics, and the
macroscopically occupied ground-state wave function for superfluids and
superconductors.
We shall henceforth set our order parameter to vanish in the high-temperature
disordered phase, and to assume a finite value in the low-temperature ordered
phase.
Landau's basic construction of a general mean-field description for phase
transitions relies on an expansion of the free energy (density) in terms of the
order parameter, naturally constrained by the symmetries of the physical system
under consideration.
For example, consider a scalar order parameter $\phi$ with discrete inversion 
or $Z_2$ symmetry that in the ordered phase may take either of two degenerate 
values $\phi_\pm = \pm |\phi_0|$. 
We shall see that the following generic expansion (with real coefficients) 
indeed describes a {\em continuous} or second-order phase transition:
\begin{equation}
  f(\phi) = \frac{r}{2} \, \phi^2 + \frac{u}{4 !} \, \phi^4 + \ldots 
  - h \, \phi \ ,
\label{landex}
\end{equation}
if the temperature-dependent parameter $r$ changes sign at $T_c$.
For simplicity, and again in the spirit of a regular Taylor expansion, we let
$r = a (T - T_c^0)$, where $T_c^0$ denotes the {\em mean-field critical 
temperature}.
Stability requires that $u > 0$ (otherwise more expansion terms need to be
added); near the critical point we can simply take $u$ to be a constant. 
Note that the external field $h$, thermodynamically conjugate to the order
parameter, {\em explicitly} breaks the assumed $Z_2$ symmetry $\phi \to -\phi$.

Minimizing the free energy with respect to $\phi$ then yields the thermodynamic
ground state. 
Thus, from $f'(\phi) = 0$ we immediately infer the {\em equation of state}
\begin{equation}
  h(T,\phi) = r(T) \, \phi + \frac{u}{6} \, \phi^3 \ ,
\label{stateq}  
\end{equation}
and the minimization or stability condition reads 
$0 < f''(\phi) = r + \frac{u}{2} \, \phi^2$.
At $T = T_c^0$, (\ref{stateq}) reduces to the {\em critical isotherm} 
$h(T_c^0,\phi) = \frac{u}{6} \, \phi^3$.
For $r > 0$, the {\em spontaneous order parameter} at zero external field 
$h = 0$ vanishes; for $r < 0$, one obtains $\phi_\pm = \pm \phi_0$, where
\begin{equation}
  \phi_0 = (6 |r| / u)^{1/2} \ .
\label{spordp}
\end{equation}
Note the emergence of characteristic power laws in the thermodynamic functions
that describe the properties near the critical point located at $T = T_c^0$, 
$h = 0$.

The continuous, but non-analytic onset of spontaneous ordering is the hallmark
of a second-order phase transition, and induces additional thermodynamic 
singularities at the critical point:
The isothermal order parameter {\em susceptibility} becomes 
$V \chi_T^{-1} = (\partial h / \partial \phi)_T = r + \frac{u}{2} \, \phi_0^2$,
whence
\begin{equation}
  \frac{\chi_T}{V} = \left\{ \begin{array}{cc} 1 / r & \ r > 0 \\ 1 / 2 |r| & 
  \ r < 0 \end{array} \right. \ ,
\label{opsusc}
\end{equation}
diverging as $|T - T_c^0|^{-1}$ on both sides of the phase transition, with
{\em amplitude ratio} 
$\chi_T(T \downarrow T_c^0) / \chi_T(T \uparrow T_c^0) = 2$.
Inserting (\ref{spordp}) into the Landau {\em free energy} (\ref{landex}) one
finds for $T < T_c^0$ and $h = 0$
\begin{equation}
  f(\phi_\pm) = \frac{r}{4} \, \phi_0^2 = - \frac{3 r^2}{2 u} \, ,
\label{frenor}  
\end{equation}
and consequently for the {\em specific heat}
\begin{equation}
  C_{h=0} = - V T \left( \frac{\partial^2 f}{\partial T^2} \right)_{h=0}
  = V T \, \frac{3 a^2}{u} \ ,
\label{spheat}
\end{equation}
whereas per construction $f(0) = 0$ and $C_{h=0} = 0$ in the disordered phase.
Thus, Landau's mean-field theory predicts a critical point {\em discontinuity} 
$\Delta C_{h=0} = V T_c^0 \, \frac{3 a^2}{u}$ for the specific heat.
Experimentally, one indeed observes singularities in thermodynamic observables
and power laws at continuous phase transitions, but often with critical 
exponents that differ from the above mean-field predictions.
Indeed, the divergence of the order parameter susceptibility (\ref{opsusc})
indicates violent fluctuations, inconsistent with any mean-field description 
that entirely neglects such fluctuations and correlations.

\subsection{Scaling theory}

The emergence of scale-free power laws suggests the following general scaling 
hypothesis for the free energy, namely that its singular contributions near a 
critical point ($T = T_c$, $h = 0$) can be written as a generalized 
{\em homogeneous} function
\begin{equation}
  f_{\rm sing}(\tau,h) = |\tau|^{2 - \alpha} \, {\hat f}_\pm
  \left( h / |\tau|^\Delta \right) \ , 
\label{fschyp}
\end{equation}
where $\tau = \frac{T - T_c}{T_c}$ measures the deviation from the (true)
critical temperature $T_c$.
Thus, the free energy near criticality is not an independent function of the 
two intensive control parameters $T$ or $\tau$ and $h$, but satisfies a 
remarkable two-parameter scaling law, with analytic {\em scaling functions} 
${\hat f}_\pm(x)$ respectively for $T > T_c$ and $T < T_c$ that only depend on
the ratio $x = h / |\tau|^\Delta$, and satisfy ${\hat f}_\pm(0) = {\rm const.}$
In Landau theory, the corresponding {\em critical exponents} are 
$\alpha = 0$, compare (\ref{frenor}), and $\Delta = 3/2$, as can be inferred by
combining Eqs.~(\ref{stateq}) and (\ref{spordp}).
The associated {\em specific heat} singularity follows again via
\begin{equation}
  C_{h=0} = - \frac{V T}{T_c^2} 
  \left( \frac{\partial^2 f_{\rm sing}}{\partial \tau^2} \right)_{h=0} 
  = C_\pm \, |\tau|^{-\alpha} \ ,
\label{spsing}  
\end{equation}
indicating a divergence if $\alpha > 0$, and a cusp singularity for 
$\alpha < 0$.
Similarly, one obtains the {\em equation of state}
\begin{equation}
  \phi(\tau,h) = - \left( \frac{\partial f_{\rm sing}}{\partial h}
  \right)_\tau = - |\tau|^{2 - \alpha - \Delta} \ 
  {\hat f}_\pm'\left( h / |\tau|^\Delta \right) \ ,
\label{sceqst}  
\end{equation}
and therefrom the {\em coexistence line} at $h = 0$, $\tau < 0$
\begin{equation}
  \phi(\tau,0) = - |\tau|^{2 - \alpha - \Delta} \ {\hat f}_-'(0) \propto
  |\tau|^\beta \ , 
\label{screl1}
\end{equation}
where we have identified $\beta = 2 - \alpha - \Delta$.

Additional {\em scaling relations}, namely identities that relate different
critical exponents, can be easily derived; for example, on the {\em critical 
isotherm} at $\tau = 0$, the $\tau$-dependence in ${\hat f}_\pm'$ on the r.h.s.
of (\ref{sceqst}) must cancel the singular prefactor, i.e., 
${\hat f}_\pm'(x \to \infty) \sim x^{(2 - \alpha - \Delta) / \Delta}$, and
\begin{equation}
  \phi(0,h) \propto h^{(2 - \alpha - \Delta) / \Delta} = h^{1 / \delta} \ , \
  {\rm with} \ \delta = \Delta / \beta \ .
\label{screl2}
\end{equation}
Finally, the isothermal {\em susceptibility} becomes
\begin{equation}
  \frac{\chi_\tau}{V} = \left( \frac{\partial \phi}{\partial h} 
  \right)_{\tau, \, h=0} = \chi_\pm \, |\tau|^{- \gamma} \, , \ 
  \gamma = \alpha + 2 (\Delta - 1) \ , \
\label{screl3}  
\end{equation}
and upon eliminating $\Delta = \beta \delta$, one arrives at the following set 
of scaling relations
\begin{equation}
  \alpha + \beta (1 + \delta) = 2 = \alpha + 2 \beta + \gamma \, , \ 
  \gamma = \beta (\delta - 1) \ . \
\label{screls}  
\end{equation}
Clearly, as consequence of the two-parameter scaling hypothesis (\ref{fschyp}),
there can only be {\em two independent} thermodynamic critical exponents.
In the framework of Landau's mean-field approximation, the set of critical
exponents reads $\alpha = 0$, $\beta = \frac12$, $\gamma = 1$, $\delta = 3$, 
and $\Delta = \frac32$; note that these integer or rational numbers really just
follow from straightforward dimensional analysis.
In both computer and real experiments, one typically measures different 
critical exponent values, yet these still turn out to be {\em universal} in the
sense that at least for short-range interaction forces they depend only on 
basic symmetry properties of the order parameter and the spatial dimensionality
$d$, but {\em not} on microscopic details such as lattice structure, nature and
strength of interaction potentials, etc. 
Indeed, the Ising ferromagnetic and the liquid-gas critical points, both 
characterized by a scalar real order parameter, are governed by identical power
laws, as is the critical behavior for planar magnets with a two-component 
vector order parameter and the normal- to superfluid transition in helium 4, 
with a complex scalar or equivalently, a real two-component order parameter. 
The striking emergence of thermodynamic self-similarity in the vicinity of 
$T_c$ has been spectacularly demonstrated in the latter system, with the 
$\Lambda$-like shape of the specific heat curve appearing identical on milli- 
and micro-Kelvin temperature scales.

\subsection{Landau--Ginzburg--Wilson Hamiltonian}

In order to properly include the effects of fluctuations, we need to generalize
the Landau expansion (\ref{landex}) to spatially varying order parameter
configurations $S(x)$, which leads us to the {\em coarse-grained} effective
{\em Landau--Ginzburg--Wilson (LGW) Hamiltonian} 
\begin{eqnarray}
  &&{\cal H}[S] = \int \! d^dx \, \biggl[ \frac{r}{2} \, S(x)^2 
  + \frac{1}{2} \, [\nabla S(x)]^2 \nonumber \\ 
  &&\qquad\qquad\qquad + \frac{u}{4 !} \, S(x)^4 - h(x) \, S(x) \biggr] \ ,
\label{lgwham}
\end{eqnarray}
where $r = a (T - T_c^0)$ and $u > 0$ as before, and $h(x)$ now represents a 
local external field.
Under the natural assumption that spatial inhomogeneities are energetically
unfavorable, the gradient term $\sim [\nabla S(x)]^2$ comes with a positive
coefficient that has been absorbed into the scalar order parameter field.
Within the canonical framework of statistical mechanics, the {\em probability 
density} for a configuration $S(x)$ is given by the {\em Boltzmann factor}
${\cal P}_s[S] = \exp (- {\cal H}[S] / k_{\rm B} T) / {\cal Z}[h]$. 
Here, the {\em partition function} ${\cal Z}[h]$ and expectation values of  
observables $A[S]$ are represented through functional integrals:
\begin{eqnarray}
  &&{\cal Z}[h] = \int \! {\cal D}[S] \ e^{- {\cal H}[S] / k_{\rm B} T} \ , 
\label{canpar} \\
  &&\langle A[S] \rangle = \int\! {\cal D}[S] \ A[S(x)] \, {\cal P}_s[S] \ .
\label{canobs}  
\end{eqnarray}
At $h = 0$, for example, the $k$th order parameter moments follow via 
functional derivatives 
\begin{equation}
  \Big\langle \prod_{j=1}^k S(x_j) \Big\rangle = (k_{\rm B} T)^k \prod_{j=1}^k 
  \frac{\delta}{\delta h(x_j)} \, {\cal Z}[h] \Big\vert_{h = 0} , \
\label{opmoms}
\end{equation}
and similarly the associated cumulants can be obtained from functional 
derivatives of $\ln {\cal Z}[h]$; the partition function thus also serves as a 
{\em generating function}.
For explicit calculations, one requires the integral measure in (\ref{canobs}),
e.g., through discretizing $x \to x_i$ on, say, a $d$-dimensional cubic
hyperlattice, whence simply ${\cal D}[S] = \prod_i dS(x_i)$.
Alternatively, one may employ the Fourier transform 
$S(x) = \int \frac{d^dq}{(2 \pi)^d} \, S(q) \, e^{i q \cdot x}$; noting that 
$S(-q) = S(q)^*$ since $S(x)$ is real, and consequently the real and imaginary
parts of $S(q)$ are not independent, one only needs to integrate over wave 
vector half-space,
\begin{equation}
  {\cal D}[S] = \prod_{q , q_1 > 0} 
  \frac{d \, {\rm Re} \, S(q) \ d \, {\rm Im} \, S(q)}{V} \ .
\label{intmft}
\end{equation}

In the {\em Ginzburg--Landau approximation}, one considers only the most
likely configuration $S(x)$, which is readily found by the method of
steepest descent for the path integrals in (\ref{canobs}), leading to the
classical field or Ginzburg--Landau equation 
\begin{equation}
  0 = \frac{\delta {\cal H}[S]}{\delta S(x)} = \left[ r - \nabla^2 +
  \frac{u}{6} \, S(x)^2 \right] S(x) - h(x) . \
\label{glaneq}
\end{equation}
In the spatially homogeneous case, (\ref{glaneq}) reduces to the mean-field 
equation of state (\ref{stateq}).
Let us next expand in the fluctuations $\delta S(x) = S(x) - \phi$ about the 
mean order parameter $\phi = \langle S \rangle$ and linearize, which yields 
$\delta h(x) \approx \left( r - \nabla^2 + \frac{u}{2}\, \phi^2 \right) 
\delta S(x)$.
Through Fourier transform one then immediately obtains the order parameter
response function in the mean-field approximation, also known as 
{\em Ornstein--Zernicke susceptibility}
\begin{equation}
  \chi_0(q) = \frac{\partial S(q)}{\partial h(q)} \bigg\vert_{h = 0} = 
  \frac{1}{\xi^{-2}+ q^2} \, , \  
\label{ornzer}
\end{equation}
where we have introduced the characteristic {\em correlation length}
$\xi = (r + \frac{u}{2} \, \phi_0^2)^{-1/2}$, i.e.,
\begin{equation}  
  \xi = \left\{ \begin{array}{cc} 1 / r^{1/2} & \ r > 0 \\ 1 / |2 r|^{1/2} & 
  \ r < 0 \end{array} \right. \ .
\label{corlen}
\end{equation}
On the other hand, consider the connected zero-field two-point 
{\em correlation function} (cumulant)
\begin{eqnarray}
  &&C(x-x') = \langle S(x) \, S(x') \rangle - \langle S \rangle^2 \nonumber \\
  &&\qquad\qquad = (k_{\rm B} T)^2 \, \frac{\delta^2 \, \ln {\cal Z}[h]} 
  {\delta h(x) \, \delta h(x')} \bigg\vert_{h=0} ;
\label{corfun}
\end{eqnarray} 
in a spatially translation-invariant system, we may define its Fourier
transform as $C(x) = \int \frac{d^dq}{(2 \pi)^d} \, C(q) \, e^{i q \cdot x}$,
and through comparison with the definition of the susceptibility in 
(\ref{ornzer}) arrive at the {\em fluctuation-response theorem} 
$C(q) = k_{\rm B} T \, \chi(q)$, valid in thermal equilibrium.

Generalizing the Ginzburg--Landau mean-field result (\ref{ornzer}), we may
formulate the {\em scaling hypothesis} for the two-point correlation function
in terms of the following scaling ansatz, which defines both the {\em Fisher 
exponent} $\eta$ and the critical exponent $\nu$ that describes the 
divergence of the correlation length $\xi$ at $T_c$:
\begin{equation} 
  C(\tau,q) = |q|^{- 2 + \eta} \, {\hat C}_\pm(q \xi) \ , \quad 
  \xi = \xi_\pm \, |\tau|^{- \nu} \ .
\label{corsch}
\end{equation}
The thermodynamic susceptibility then becomes
\begin{equation}
  \chi(\tau,q=0) \propto \xi^{2 - \eta} \propto |\tau|^{- \gamma} \, , \ 
  {\rm with} \ \gamma = \nu (2 - \eta) \ , \
\label{screl4}
\end{equation}
providing us with yet another scaling relation that connects the thermodynamic
critical exponent $\gamma$ with $\eta$ and $\nu$.
Consequently, we see that the thermodynamic critical point singularities are 
induced by the diverging spatial correlations. 
Fourier back-transform gives
\begin{equation}
   C(\tau,x) = |x|^{- (d - 2 + \eta)} \, {\widetilde C}_\pm(x / \xi) \propto
   \xi^{- (d - 2 + \eta)} 
\label{spcorr}
\end{equation}
at large distances $|x| \to \infty$.
In this limit, one expects 
$\langle S(x) \, S(0) \rangle \to \phi^2 \propto (-\tau)^{2 \beta}$, and 
comparison with (\ref{spcorr}) therefore implies the {\em hyperscaling} 
relations
\begin{equation}
  \beta = \frac{\nu}{2} \, (d - 2 + \eta) \ \ {\rm and} \ \ 
  2 - \alpha = d \nu \ .
\label{hypscr}
\end{equation}
The Ornstein--Zernicke function (\ref{ornzer}) satisfies the scaling law 
(\ref{corsch}) with the mean-field values $\nu = \frac12$ and $\eta = 0$.
Notice that the set of mean-field critical exponents obeys (\ref{hypscr}) only 
in $d = 4$ dimensions.

\subsection{Gaussian approximation}

We now proceed to analyze the LGW Hamiltonian (\ref{lgwham}) in the Gaussian 
approximation, where non-linear fluctuation contributions are neglected.
In the high-temperature phase, we have $\phi = 0$, and thus simply omit the
terms $\sim u \, S(x)^4$, leaving the Gaussian Hamiltonian
\begin{equation}
  {\cal H}_0[S] = \int_q \left[ \frac{r + q^2}{2} \, |S(q)|^2 - h(q) S(-q) 
  \right] \ ,
\label{gauham}
\end{equation}
with the abbreviation $\int_q = \int \! \frac{d^dq}{(2 \pi)^d}$.
The associated Gaussian partition function is readily computed by completing
the square in (\ref{gauham}), or the linear field transformation 
${\widetilde S}(q) = S(q) - h(q) / (r + q^2)$,
\begin{eqnarray}
  &&{\cal Z}_0[h] = \int \! {\cal D}[S] \ e^{- {\cal H}_0[S] / k_{\rm B}T}  
  \nonumber \\
  &&\quad = \exp \left( \frac{1}{2 k_{\rm B}T} \int_q \frac{|h(q)|^2}{r + q^2}
  \right) \, {\cal Z}_0[h = 0] \ , \
\label{gaupar}
\end{eqnarray} 
which yields the Gaussian two-point correlator
\begin{eqnarray}
  &&\langle S(q) \, S(q') \rangle_0 = \frac{(k_{\rm B} T)^2}{{\cal Z}_0[h]} \, 
  \frac{(2 \pi)^{2d} \, \delta^2 {\cal Z}_0[h]}{\delta h(-q)\, \delta h(-q')}
  \bigg\vert_{h=0} \nonumber \\ 
  &&= C_0(q) \, (2 \pi)^d \delta(q + q') \, , \ 
  C_0(q) = \frac{k_{\rm B} T}{r + q^2} \, . \
\label{gaucor}
\end{eqnarray}

Gaussian integrations give the free energy 
$F_0[h] = - k_{\rm B} T \ln {\cal Z}_0[h]$ of the model (\ref{gauham}),
\begin{equation}
  F_0[h] = - \frac{1}{2} \int_q \left( \frac{|h(q)|^2}{r + q^2} 
  + k_{\rm B} T V \, \ln \frac{2 \pi \, k_{\rm B} T}{r + q^2} \right) \, .
\label{gaufre}
\end{equation}
Let us explore the leading singularity near $T_c^0$ in the {\em specific heat}
$C_{h=0} = -T (\partial^2 F_0 / \partial T^2)_{h=0}$ that originates from 
derivatives with respect to the control parameter $r$,
\begin{equation}
  \frac{C_{h=0}}{V} \approx \frac{k_{\rm B}}{2} \, (a T_c^0)^2 
  \int_q \frac{1}{(r + q^2)^2} \ .
\label{gausph}
\end{equation}
\begin{itemize}
  \item In high dimensions $d > 4$, the r.h.s. integral is UV-divergent, but 
        can be regularized by a Brillouin zone boundary cutoff 
        $\Lambda \sim 2 \pi / a_0$ stemming from the original underlying 
        lattice. 
        Consequently the fluctuation contribution (\ref{gausph}) is finite as
        $r \to 0$ and $\alpha = 0$ as in mean-field theory. 
  \item In low dimensions $d < 4$, we set $k = q / \sqrt{r} = q \xi$ to render
        the fluctuation integral, which is UV-finite, dimensionless.
        With the $d$-dimensional unit sphere surface area 
        $K_d = 2 \pi^{d/2} / \Gamma(d/2)$, one finds:
  \begin{equation}
    \frac{C_{h = 0}}{V} \approx 
    \frac{k_{\rm B} (a T_c^0)^2 \, \xi^{4-d}}{2^d \pi^{d/2} \, \Gamma(d/2)}
    \int_0^\infty \! \frac{k^{d-1}}{(1 + k^2)^2} \ dk \, . \
  \label{sphd<4}
  \end{equation}
        As the critical temperature is approached, the correlation length 
        prefactor diverges $\propto |T - T_c^0|^{- \frac{4-d}{2}}$; already the
        lowest-order fluctuation contribution contains a strong infrared (IR)
        singularity that dominates over the mean-field power law.
  \item At $d = d_c = 4$, the integral diverges logarithmically as 
        either $\Lambda$ or $\xi \to \infty$:
  \begin{equation}
    \int_0^{\Lambda \xi} \frac{k^3}{(1 + k^2)^2}\ dk \sim \ln (\Lambda \xi) \ ;
  \label{sphd=4}
  \end{equation}
        note that at this {\em upper critical dimension}, ultraviolet and 
        infrared divergences are intimately coupled.
\end{itemize}
Above the critical dimension, we thus expect the mean-field scaling exponents 
to correctly describe the critical power laws.
In dimensions $d \leq d_c$, however, fluctuation contributions become prevalent
and modify the mean-field scaling laws.
Given the strongly fluctuating and correlated nature of critical systems, 
standard theoretical approaches such as perturbation, cluster, or high- and 
low-temperature expansions typically fail to yield reliable approximations.
Fortunately, the renormalization group provides a powerful method to tackle 
interacting many-particle systems dominated by fluctuations and correlations, 
especially in a scale-invariant regime.

\subsection{Wilson's momentum shell renormalization group}

The {\em renormalization group program in statistical physics} can be 
summarized as follows: 
The goal is to establish a mathematical framework that can properly capture the
infrared (IR) singularities appearing in thermodynamic properties as well as
correlation functions near a continuous phase transitions and in related 
situations that are {\em not} perturbatively accessible.
To this end, one exploits a fundamental new symmetry that emerges at a critical
point, namely {\em scale invariance}, induced by the divergence of the dominant
characteristic correlation length scale $\xi$.  
Approximation schemes need to carefully avoid the region where the physical IR
singularities become manifest; instead, one analyzes the theory in the 
ultraviolet (UV) regime, by means of either of various equivalent methods:
In Wilson's momentum shell RG approach, one integrates out short-wavelength 
modes; in the field-theoretic version of the RG, one explicitly renormalizes 
the UV divergences.
Either method quantifies the weight of fluctuation contributions to certain
coarse-grained or ``renormalized'' physical parameters and couplings.
One then maps the resulting system back to the original theory given in terms
of some ``effective'' Hamiltonian, which in Wilson's scheme entails a rescaling
of both control parameters and field degrees of freedom.
Thus one obtains recursion relations for {\em effective}, now scale-dependent 
{\em running couplings}.  
Subject to a recursive sequence of such renormalization group transformations, 
these effective couplings will
\begin{itemize}
  \item either grow, and ultimately tend to infinity: to access a
        scale-invariant regime, one therefore has to set these {\em relevant} 
        parameters to zero at the outset, which defines the {\em critical 
        surface} of the problem; 
  \item or diminish, and eventually approach zero: these {\em irrelevant} 
        couplings consequently do not affect the asymptotic critical scaling 
        properties; 
  \item certain {\em marginal} parameters may also approach an 
        {\em infrared-stable fixed point}, provided their initial value is 
        located in the fixed point's basin of attraction: clearly, 
        scale-invariant behavior thus emerges near an IR-stable fixed point, 
        and the independence from a wide range of initial conditions along with
        the automatic disappearance of the irrelevant couplings constitute the 
        origin of {\em universality}.
\end{itemize}

The central idea now is to take advantage of the emerging scale invariance at a
critical fixed point as a means to infer the proper infrared scaling behavior 
from an at least approximative analysis of the ultraviolet regime, where, e.g.,
perturbation theory is feasible.
Thus one may establish a solid theoretical foundation for scaling laws such as
(\ref{fschyp}) and (\ref{corsch}), thereby derive scaling relations, and also
construct a systematic approximation scheme to compute critical exponents and
even scaling functions.
We shall soon see that an appropriate small parameter for a perturbational
expansion is given through a {\em dimensional expansion} in terms of the 
deviation from the upper critical dimension $\epsilon = d_c - d$.

Wilson's momentum shell renormalization group approach consists of two RG 
transformation steps:
\begin{enumerate}
  \item[(1)] Carry out the partition integral over all Fourier components 
             $S(q)$ with wave vectors residing in the spherical momentum shell 
             $\Lambda / b \leq |q| \leq \Lambda$, where $b > 1$: this 
             effectively {\em eliminates} the short-wavelength modes.
  \item[(2)] Perform a {\em scale transformation} with the same scale parameter
             $b > 1$: $x \to x' = x / b$, $q \to q' = b \, q$.
             Accordingly, one also needs to rescale the fields:
  \begin{eqnarray}
    &&S(x) \to S'(x') = b^\zeta S(x) \ , \nonumber \\ 
    &&S(q) \to S'(q') = b^{\zeta - d} S(q) \ ,
  \label{fldrsc}
  \end{eqnarray}
             with a proper choice of $\zeta$ ensuring that the rescaled 
             residual Hamiltonian assumes the original form.
\end{enumerate}
Subsequent iterations of this procedure yield {\em scale-dependent effective 
couplings}, and the task will be to analyze their dependence on the scale
parameter $b$.
Notice the {\em semi-group} character of the above RG transformations: there
obviously exists no unique inverse, since the elimination step (1) discards
detailed information about fluctuations in the UV regime.

The mechanism and efficacy of the momentum shell RG is best illuminated by 
first considering the exactly tractable Gaussian model.
Introducing the short-hand notations 
$\int_q^< = \int_{|q| < \Lambda/b} \frac{d^dq}{(2 \pi)^d}$ and
$\int_q^> = \int_{\Lambda / b \leq |q| \leq \Lambda} \frac{d^dq}{(2 \pi)^d}$, 
one may readily decompose the Hamiltonian (\ref{gauham}) into distinct additive
Fourier mode contributions
$$ {\cal H}_0[S] = \left( \int_q^<\! + \int_q^>\! \right) 
   \left[ \frac{r + q^2}{2} \, |S(q)|^2 - h(q) \, S(-q) \right] . $$
Integrating out the momentum shell fluctuations then just gives a constant
contribution to the free energy.
We now wish to achieve that ${\cal H}_0[S^<] \to {\cal H}_0[S']$ under the 
scale transformations in step (2).
For the term $\sim q^2 \, |S(q)|^2$, this is accomplished through the choice 
$\zeta = \frac{d - 2}{2}$ in (\ref{fldrsc}); the other contributions then 
immediately result in the following recursion relations for the control
parameters $r$ and $h$: $r \to r' = b^2 r$, and 
$h(q) \to h'(q') = b^{- \zeta} h(q)$, whence 
$h(x) \to h'(x') = b^{d - \zeta} h(x)$.
Both the temperature variable $r$ and the external field thus constitute 
{\em relevant} parameters, and the {\em critical surface} in parameter space is
given by $r = 0 = h$.
As any other length scale, the correlation length scales according to 
$\xi \to \xi' = \xi / b$; eliminating the scale parameter $b$ one arrives at 
the relation $\xi \propto r^{-1/2}$, or $\nu = \frac12$.
Likewise, for the rescaled correlation function one finds 
$C'(x') = b^{2 \zeta} \, C(x)$, whence $\eta = 0$: for the Gaussian theory, we 
recover the mean-field scaling exponents.   

We can gain additional non-trivial information by considering further 
couplings; e.g., imagine adding contributions of the form 
$c_s \int \! d^dx \, (\nabla^s S)^2$ to the Hamiltonian (\ref{gauham}) that 
represent higher-order terms in a gradient expansion for spatial order 
parameter fluctuations, subject to preserving the $Z_2$ and spatial inversion 
symmetries.
One readily confirms that 
$c_s \to c_s' = b^{d - 2 s - 2 \zeta} c_s = b^{-2 (s-1)} c_s$, 
which implies that all these additional couplings $c_s$ are {\em irrelevant} 
for $s > 1$ and scale to zero under repeated scale transformations.
The inversion symmetry $S(x) \to -S(x)$ permits general local non-linearities 
of the from $u_p \int \! d^dx\, S(x)^{2p}$.
Under Gaussian model RG transformation, these scale as
$u_p \to u_p' = b^{d - 2 p \zeta} u_p = b^{2 p - (p - 1) d} u_p$; these 
couplings are consequently {\em marginal} at $d_c(p) = 2 p / (p - 1)$, 
{\em relevant} for $d < d_c(p)$, and {\em irrelevant} for $d > d_c(p)$. 
The upper critical dimension decreases monotonously for $p \geq 2$, with the
asymptote $d_c(\infty) = 2$.
For the quartic coupling in the LGW Hamiltonian (\ref{lgwham}), this confirms
$d_c(2) = 4$, while $d_c(3) = 3$ for a sixth-order term 
$v \int \! d^dx\, S(x)^6$: $v \to v' = b^{6 - 2 d} v$, which becomes 
{\em irrelevant} near the upper critical dimension of the quartic term: 
$v' = b^{-2} v$ at $d_c(2) = 4$.
In general, the coupling ratio 
$\frac{u_{p+1}'}{u_p'} = b^{2 - d} \frac{u_{p+1}}{u_p}$ renormalizes to zero in
dimensions $d > 2$.
At two dimensions, the fields $S(x)$ become dimensionless, $\zeta = 0$, and
consequently all these non-linearities scale identically.
The LGW Hamiltonian thus does not represent the correct asymptotic field 
theory, and one must resort to other effective descriptions (e.g., the 
non-linear sigma model).

The above considerations already allow a discussion of the general structure of
the momentum shell RG procedure.
According to (\ref{spcorr}), the general field rescaling (\ref{fldrsc}) should 
contain Fisher's exponent $\eta$, $\zeta = \frac{d - 2 + \eta}{2}$, whence
$h'(x) = b^{(d + 2 - \eta)/2} h(x)$, and (\ref{corsch}) implies 
$\tau' = b^{1 / \nu} \tau$ with the correlation length critical exponent $\nu$.
In the simplest scenario, there are thus only {\em two} relevant parameters 
$\tau$ and $h$.
Let us further assume the presence of (a few) {\em marginal} perturbations 
$u_i \to u_i' = u_i^* + b^{- x_i} u_i$, while other couplings are 
{\em irrelevant}: $v_i \to v_i' = b^{- y_i} v_i$, with both $x_i > 0$ and 
$y_i > 0$.
After a single RG transformation step, the free energy density becomes
\begin{eqnarray}
  &&f_{\rm sing}(\tau,h,\{ u_i \},\{ v_i \}) = 
\label{fenrg1} \\
  &&\quad b^{-d} f_{\rm sing}\Bigl( b^{1/\nu} \tau, b^{d - \zeta} h, \Big\{ 
  u_i^* + \frac{u_i}{b^{x_i}} \Big\}, \Big\{ \frac{v_i}{b^{y_i}} \Big\} \Bigr) 
  \ . \nonumber
\end{eqnarray}
After sufficiently many ($\ell \gg 1$) RG transformations, the marginal
couplings have reached their fixed point values $u_i^*$, whereas the irrelevant
perturbations have scaled to zero,
\begin{eqnarray}
  &&f_{\rm sing}(\tau,h,\{ u_i \},\{ v_i \}) = 
\label{fenrga} \\
  &&\quad b^{-\ell d} f_{\rm sing}\Bigl( b^{\ell/\nu} \tau, 
  b^{\ell (d + 2 - \eta) / 2} h, \{ u_i^* \}, \{ 0 \} \Bigr) \ . \nonumber
\end{eqnarray}
Upon choosing $b^\ell |\tau|^\nu = 1$ for the scale parameter $b^\ell$, one
arrives at the {\em scaling form}
\begin{equation}
    f_{\rm sing}(\tau,h) = |\tau|^{d \nu} \, {\hat f}_\pm \left(
    h / |\tau|^{\nu (d + 2 - \eta) / 2} \right) \ ,
\label{fenrgs}
\end{equation}
with ${\hat f}_\pm(x) = f_{\rm sing}( \pm 1, x, \{ u_i^* \}, \{ 0 \})$.
With the exponent identities (\ref{screls}) and (\ref{hypscr}), this is 
equivalent to the scaling hypothesis (\ref{fschyp}).
In a similar manner, one readily derives the correlation function scaling law 
(\ref{spcorr}), employing the {\em matching condition} $b^\ell = \xi / \xi_\pm$
for
\begin{equation}
  C(\tau,x,\{ u_i \},\{ v_i \}) = b^{- 2 \ell \zeta} \, C\Bigl( b^{\ell / \nu} 
  \tau, \frac{x}{b^\ell}, \{ u_i^* \}, \{ 0 \} \Bigr) \, . \
\label{corrga}
\end{equation}

\subsection{Dimensional expansion and critical exponents}

We are now ready to treat the non-linear fluctuation corrections by means
of a systematic {\em perturbation expansion}.
The quartic contribution to the Hamiltonian reads in Fourier space
\begin{equation}
  {\cal H}_{\rm int} = \frac{u}{4 !} \! \int_{|q_i| < \Lambda} \!\!\!\!\!\!\!\!
  S(q_1) S(q_2) S(q_3) S(-q_1-q_2-q_3) \ . \
\label{nonlin}
\end{equation}
Both the full partition function for the Hamiltonian (\ref{lgwham}) and any
associated $N$-point correlation functions can then be rewritten in terms of
expectation values in the {\em Gaussian ensemble} (we shall henceforth set 
$k_{\rm B} T = 1$)
\begin{eqnarray}
  &&{\cal Z}[h] = {\cal Z}_0[h] \, \Big\langle e^{- {\cal H}_{\rm int}[S]} 
  \Big\rangle_0 \, , \nonumber \\
  &&\Big\langle \prod_i S(q_i) \Big\rangle = 
    \frac{\big\langle \prod_i S(q_i) \, e^{- {\cal H}_{\rm int}[S]} 
    \big\rangle_0}{\big\langle e^{- {\cal H}_{\rm int}[S]} \big\rangle_0} \ .
\label{expgau}
\end{eqnarray}
Note that all expectation values of an odd number of fields $S(q_i)$ obviously
vanish in the symmetric high-temperature phase, when the external field 
$h = 0$.
Defining the {\em contraction} of two fields as the Gaussian two-point function
or {\em propagator} in Fourier space, 
$$ \underbrac{S(q) S}(q') = \langle S(q) S(q') \rangle_0 = 
   C_0(q) \, (2 \pi)^d \delta(q + q') \ , \nonumber $$ 
we may write down {\em Wick's theorem} for Gaussian correlators containing an 
even number of fields, here a straightforward property of Gaussian 
integrations:
\begin{eqnarray}
  &&\langle S(q_1) S(q_2) \ldots S(q_{N-1}) S(q_N) \rangle_0 = 
\label{wickth} \\
  &&\!\!\!\!\!\!\!\!\!\sum_{{\rm permutations} \atop i_1(1) \ldots i_N(N)} \!\!
  \underbrac{S(q_{i_1(1)}) S}(q_{i_2(2)}) \, \ldots \,
  \underbrac{S(q_{i_{N-1}(N-1)}) S}(q_{i_N(N)}) \ . \nonumber
\end{eqnarray}
Consequently, any arbitrary expectation value (\ref{expgau}) can now be 
perturbatively evaluated via a series expansion with respect to the non-linear 
coupling $u$, and by means of (\ref{wickth}) expressed through sums and 
integrals of products of Gaussian propagators $C_0(q_i)$.

As an example, we consider the first-order fluctuation correction to the 
zero-field two-point function 
$\langle S(q) S(q') \rangle = C(q) \, (2 \pi)^d \delta(q + q')$:
$\big\langle S(q) S(q') \, \bigl[ 1 - \frac{u}{4 !}$ $\int_{|q_i| < \Lambda} 
S(q_1) S(q_2) S(q_3) S(-q_1-q_2-q_3) \bigr] \big\rangle_0$.
According to Wick's theorem (\ref{wickth}), there are two types of 
contributions: 
(i) Contractions of external legs $\underbrac{S(q) S}(q')$ yield terms that 
precisely cancel with the denominator in (\ref{expgau}), leaving just the 
Gaussian propagator $\langle S(q) S(q') \rangle_0$.
(ii) The twelve remaining contributions are all of the form
$\int_{|q_i| < \Lambda} \underbrac{S(q) \, S}(q_1) \, 
\underbrac{S(q_2) \, S}(q_3) \, \underbrac{S(-q_1-q_2-q_3) \, S}(q') = 
(2 \pi)^d \delta(q + q') \, C_0(q)^2 \int_{|p| < \Lambda} C_0(p)$.
Collecting all terms, one obtains
\begin{equation}
  C(q) = C_0(q) \, \biggl[ 1 - \frac{u}{2} \, C_0(q) \! \int_{|p| < \Lambda} 
  \!\!\! C_0(p) + O(u^2) \biggr] \, ; \
\label{1lpcor}
\end{equation}
interpreting the bracket as the lowest-order contribution in Dyson's equation 
(see Chap.3.1 below), the integral turns out to be the associated self-energy 
to $O(u)$, and (\ref{1lpcor}) can be recast in the form 
\begin{equation}
  C(q)^{-1} = r + q^2 + \frac{u}{2} \int_{|p| < \Lambda} \frac{1}{r + p^2} 
  + O(u^2) \ . \
\label{1lpsen}  
\end{equation}
Notice that to order $u$, fluctuations here merely renormalize the ``mass'' 
$r$, but there is no modification of the momentum dependence in the two-point 
correlation function $C(q)$, implying that $\eta$ will remain zero in this
approximation.
In a similar manner, one readily finds the first-order fluctuation correction 
to the four-point function at vanishing external wave vectors, i.e., the 
non-linear coupling $u$ to be $- \frac32 u^2  \int_{|p| < \Lambda} C_0(p)^2$.

It is now a straightforward task to translate these perturbation theory results
into first-order recursion relations for the couplings $r$ and $u$ by means of 
Wilson's RG procedure.
To this end, we split the field variables in outer ($S_>$: $S(q)$ with 
$\Lambda / b \leq |q| \leq \Lambda$) and inner ($S_<$: $S(q)$ with 
$|q| < \Lambda / b$) momentum shell contributions; we then realize that there
are four types of contributions:
\begin{itemize}
  \item terms involving merely inner shell fields that are not integrated, e.g.
        $\sim u \int^> S_<^4 e^{-{\cal H}_0[S]}$ just need to be 
        re-exponentiated;
  \item integrals such as $u \int^> S_<^3 \, S_> e^{-{\cal H}_0[S]}$ vanish;
  \item contributions $\sim u \int^> S_>^4 e^{-{\cal H}_0[S]}$ that contain 
        only outer shell fields become constants that directly contribute to 
        the free energy;
  \item for terms of the form $\sim u \int^> S_<^2 \, S_>^2 e^{-{\cal H}_0}$, 
        one has to perform Gaussian integrations over the outer shell fields 
        $S_>$, yielding corrections to the propagator for the inner shell 
        modes.
\end{itemize}
Employing (\ref{1lpsen}), using $\eta = 0$, and introducing 
$S_d = K_d / (2 \pi)^d = 1 / 2^{d-1} \pi^{d/2} \Gamma(d/2)$, one thus finds to 
$O(u)$:
\begin{eqnarray}
    &&r' = b^2 \, \Bigl[ r + \frac{u}{2} \, A(r) \Bigr] \nonumber \\
    &&\quad = b^2 \, \biggl[ r + \frac{u}{2} \, S_d \int_{\Lambda/b}^\Lambda 
    \frac{p^{d-1}}{r + p^2} \, dp \biggr] \ , 
\label{rectem} \\
    &&u' = b^{4-d} u \, \Bigl[ 1 - \frac{3 u}{2} \, B(r) \Bigr] \nonumber \\ 
    &&\quad = b^{4-d} u \, \biggl[ 1 - \frac{3 u}{2} S_d 
    \int_{\Lambda/b}^\Lambda \frac{p^{d-1} \, dp}{(r + p^2)^2} \biggr] \ .
\label{reccpl}
\end{eqnarray}
For $T \gg T_c$, or $r \to \infty$, the fluctuation corrections become 
suppressed, and one recovers the recursion relations $r' = b^2 r$ and 
$u' =  b^{4-d} u$ of the Gaussian theory.
Near the critical point, i.e. for $r \ll 1$, one may expand   
\begin{eqnarray}
   &&A(r) = S_d \Lambda^{d-2} \frac{1 - b^{2-d}}{d-2} \nonumber \\
   &&\qquad\quad - r \, S_d \Lambda^{d-4} \, \frac{1 - b^{4-d}}{d-4} + O(r^2) 
   \ , \\
   &&B(r) = S_d \Lambda^{d-4} \, \frac{1 - b^{4-d}}{d-4} + O(r) \ .
\label{reccrt}
\end{eqnarray}

It is useful to consider instead differential RG flow equations that result
from infinitesimal RG transformations.
Setting $b = e^{\delta \ell}$ with $\delta \ell \to 0$, 
(\ref{rectem})-(\ref{reccrt}) turn into
\begin{eqnarray}
  &&\frac{d {\tilde r}(\ell)}{d \ell} = 2 {\tilde r}(\ell) 
  + \frac{{\tilde u}(\ell)}{2} \, S_d \Lambda^{d-2} \nonumber \\
  &&\qquad\quad\ - \frac{{\tilde r}(\ell) {\tilde u}(\ell)}{2} \, S_d 
  \Lambda^{d-4} + O({\tilde u} {\tilde r}^2,{\tilde u}^2) \ , 
\label{difrcr} \\
  &&\frac{d {\tilde u}(\ell)}{d \ell} = (4-d) \, {\tilde u}(\ell) 
  - \frac{3}{2} \, {\tilde u}(\ell)^2 S_d \Lambda^{d-4} \nonumber \\
  &&\qquad\quad\ + O({\tilde u} {\tilde r},{\tilde u}^2) \ .
\label{difrcu}
\end{eqnarray}
We specifically seek renormalization group {\em fixed points} $(r^*,u^*)$ that 
describe scale-invariant behavior, to be determined by the conditions
$d {\tilde r}(\ell) / d \ell = 0 = d {\tilde u}(\ell) / d \ell$.
There is obviously always the {\em Gaussian} fixed point $u_0^* = 0$;
linearizing (\ref{difrcu}) in terms of the deviation 
$\delta {\tilde u}_0(\ell) = {\tilde u}(\ell) - u_0^*$, one finds 
$d \delta {\tilde u}_0(\ell) / d \ell \approx (d-4) \delta {\tilde u}_0(\ell)$;
$u_0^*$ is hence stable for $d > d_c = 4$, but unstable for $d < 4$.
Below the upper critical dimension, there exists also a positive {\em Ising}
fixed point $u_{\rm I}^* S_d = \frac{2}{3} \, (4-d) \Lambda^{4-d}$, which is
then also stable since $d \delta {\tilde u}_I(\ell) / d \ell \approx (4 - d) 
\delta {\tilde u}_{\rm I}(\ell)$ for 
$\delta {\tilde u}_{\rm I}(\ell) = {\tilde u}(\ell) - u_{\rm I}^*$.
Correspondingly, the critical behavior is governed by the Gaussian fixed point
and associated scaling exponents in dimensions $d > 4$, but by the non-trivial
Ising fixed point in low dimensions $d < d_c = 4$.
Notice also that the numerical value of the Ising fixed point becomes small 
near the upper critical dimension, and indeed $u_{\rm I}^*$ emerges 
continuously from $u_0^* = 0$ as $\epsilon = 4 - d$ is increased from zero.
At the non-trivial RG fixed point, $\epsilon$ may serve to provide a small 
effective expansion parameter for the perturbation expansion.

To lowest order in the non-linear coupling, (\ref{difrcr}) yields at the Ising
fixed point $r_{\rm I}^* = - \frac{1}{4} \, u_{\rm I}^* S_d \Lambda^{d-2} 
= - \frac{1}{6} \, \epsilon \Lambda^2$, which describes a non-universal, 
fluctuation-induced downward shift of the critical temperature.
Introducing the deviation $\tau = r - r_{\rm I}^* = a (T - T_c)$ from the true 
critical temperature $T_c$, one may rewrite the flow equation (\ref{difrcr}) to
obtain the recursion relation for this modified relevant {\em running 
coupling}:
\begin{equation}
  \frac{d {\tilde \tau}(\ell)}{d \ell} = {\tilde \tau}(\ell) 
  \left[ 2 - \frac{{\tilde u}(\ell)}{2} S_d \Lambda^{d-4} \right] \ .
\label{difrct}
\end{equation}
Its solution in the vicinity of the Ising fixed point reads 
${\tilde \tau}(\ell) = {\tilde \tau}(0) \exp\bigl[ \bigl( 2 - 
\frac{\epsilon}{3} \bigr) \ell \bigr]$.
Combining this result with ${\widetilde \xi}(\ell) = \xi(0) \, e^{-\ell}$, one 
identifies the correlation length exponent $\nu^{-1} = 2 - \frac{\epsilon}{3}$,
or, in a consistent expansion to first order in $\epsilon = 4 - d$:
\begin{equation}
  \nu = \frac{1}{2} + \frac{\epsilon}{12} + O(\epsilon^2) \ , \quad
  \eta = 0 + O(\epsilon^2) \ .
\label{icrtex}
\end{equation}
As anticipated, the critical exponents depend only on the spatial dimension,
not on the strength of the non-linear coupling or other non-universal 
parameters.
Note that at $d_c = 4$, (\ref{difrcu}) is solved by 
${\tilde u}(\ell) = {\tilde u}(0) / [1 + 3 {\tilde u}(0) \ell / 16 \pi^2]$, a
very slow approach to the Gaussian fixed point that induces {\em logarithmic 
corrections} to the mean-field critical exponents, see (\ref{xilgcr}) below.
We finally remark that the  RG procedure generates novel coupling terms
$\sim S^6$, $\nabla^2 S^4$, etc.
To order $\epsilon^3$, their feedback into the recursion relations can however
be safely neglected.

In summary, the renormalization group procedure as outlined above in Wilson's
momentum shell formulation allows us to {\em derive} hitherto phenomenological
scaling laws, and thereby gain deeper insights into scale-invariant features.
We have also seen that the number of relevant couplings (two at standard 
critical points, namely $\tau$ and $h$) equals the number of {\em independent} 
critical exponents.
Below the {\em upper critical dimension}, fluctuation corrections modify the
critical scaling drastically as compared to the mean-field predictions.
Finally, perturbative calculations that are safely carried out in the UV regime
may be employed to systematically compute scaling exponents through a power 
series in the dimensional parameter $\epsilon = d_c - d$.

\section{Field Theory Approach to Critical Phenomena}

While Wilson's momentum shell scheme renders the basic philosophy of the
renormalization group transparent, it becomes computationally quite cumbersome 
once nested momentum integrals appear beyond the first order in perturbation 
theory.
Unnecessary technical complications in evaluating fluctuation loops can be 
avoided by extending the UV cutoff to $\Lambda \to \infty$, at the price of 
divergences in dimensions $d \geq d_c$. 
However, we already know that the Gaussian theory governs the infrared 
properties in that dimensional regime; thus in statistical physics theses UV 
singularities do not really pose a troublesome issue.
We may however employ powerful various tools from quantum field theory, proceed
to formally renormalize the UV divergences, and thereby gain crucial 
information about the desired IR scaling limit, provided an IR-stable RG fixed 
point can be identified that allows us to connect the UV and IR regimes. 
This chapter provides a succinct overview of how to construct the perturbation
expansion in terms of {\em Feynman diagrams} for one-particle irreducible 
vertex functions, proceeds to analyze the resulting UV singularities, and
finally utilizes the renormalization group equation to identify fixed points 
and determine the accompanying critical exponents.
For more extensive treatments of the field-theoretic RG approach to critical
phenomena, see Refs.~\cite{Ramond81}--\cite{Zinn93} and other excellent texts.

\subsection{Perturbation expansion and Feynman diagrams}

We now generalize our analysis to a LGW Hamiltonian with {\em continuous} 
$O(n)$ order parameter symmetry
\begin{eqnarray}
  &&{\cal H}[{\vec S}] = \int \! d^dx \sum_{\alpha = 1}^n \biggl[ \frac{r}{2} 
  \, S^\alpha(x)^2 + \frac{1}{2} \, [\nabla S^\alpha(x)]^2 \nonumber \\
  &&\qquad\qquad\qquad + \frac{u}{4!} \sum_{\beta = 1}^n S^\alpha(x)^2 \,
  S^{\beta}(x)^2 \biggr] \ ,
\label{lgwhmn}
\end{eqnarray}
which encapsulates the critical behavior for the Heisenberg model for a 
three-component vector order parameter ($n = 3$), the planar XY model (and 
equivalently, superfluids with complex scalar order parameter) for $n = 2$,
reduces to Ising $Z_2$ symmetry for $n = 1$, and in fact describes the scaling
properties of self-avoiding polymers in the limit $n \to 0$.
As in (\ref{expgau}), one constructs the perturbation expansion for arbitrary
$N$-point functions $\langle \prod_i S^{\alpha_i} \rangle$ in terms of averages
within the Gaussian ensemble with $u = 0$ (keeping $k_{\rm B} T = 1$).
Diagrammatically, the Gaussian two-point functions or {\em propagators} 
$C_0(q) \, \delta^{\alpha \beta} = \delta^{\alpha \beta} / (r + q^2)$, 
diagonal in the field component indices, are represented through lines, to be
connected through the non-linear {\em vertices} $- \frac{u}{6}$:

\smallskip
\centerline{\includegraphics[width=0.6\columnwidth]{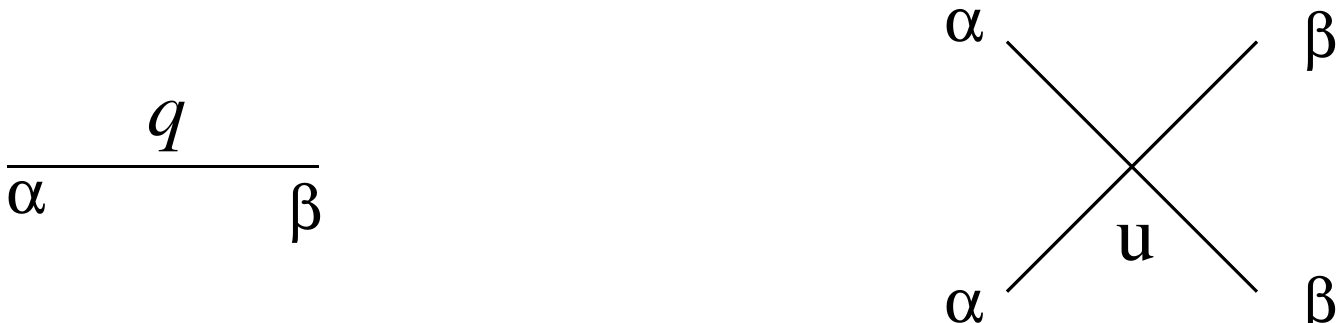}}
\smallskip

In the presence of external fields $h^\alpha$, the partition function serves as
{\em generating functional} for correlation functions (cumulants):
\begin{eqnarray} 
  &&{\cal Z}[h] = \Big\langle \exp \! \int \!\! d^dx \sum_\alpha h^\alpha 
  S^\alpha \Big\rangle \ , \nonumber \\
  &&\Big\langle \prod_i S^{\alpha_i} \Big\rangle_{(c)} = \prod_i 
  \frac{\delta (\ln) {\cal Z}[h]}{\delta h^{\alpha_i}} \Big\vert_{h = 0} \ .
\label{genfcc}
\end{eqnarray} 
The {\em cumulants} are graphically represented through {\em connected 
Feynman diagrams}; e.g., for the propagator:

\smallskip
\centerline{\includegraphics[width=\columnwidth]{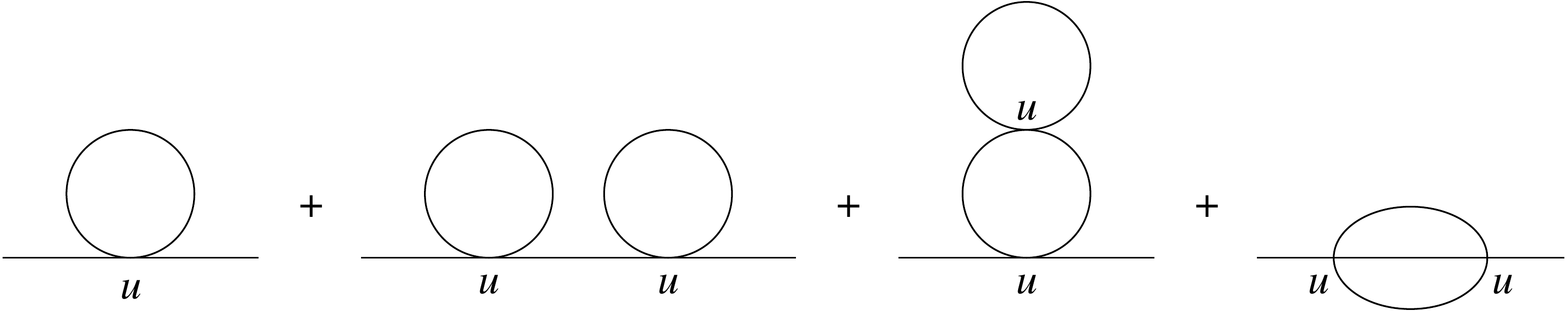}}
\smallskip

\noindent Note that the second graph is a mere repetition of the first; indeed,
upon defining the {\em self-energy} $\Sigma$ as the sum of all 
{\em one-particle irreducible} Feynman graphs that cannot be split into 
lower-order contributions simply by cutting a propagator line, one infers the 
following general structure for the full propagator $C(q)$:

\medskip
\centerline{\includegraphics[width=\columnwidth]{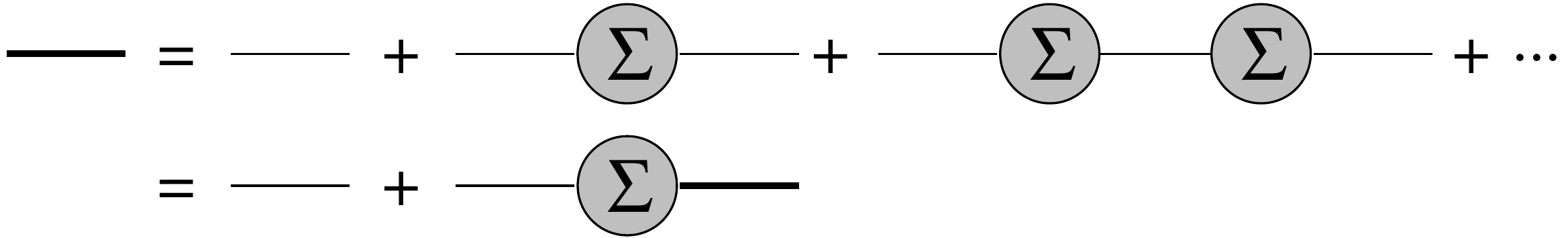}}
\smallskip

\noindent The second line is a graphical depiction of {\em Dyson's equation} 
that reads in Fourier space $C(q) = C_0(q) + C_0(q) \, \Sigma(q) \, C(q)$, 
solved by $C(q)^{-1} = C_0(q)^{-1} - \Sigma(q)$. 

In order to similarly eliminate redundancies for arbitrary $N$-point functions,
one proceeds with a Legendre transformation to construct the generating 
functional for {\em vertex functions}:
\begin{eqnarray} 
  &&\Gamma[\Phi] = - \ln {\cal Z}[h] + \int \! d^dx \sum_\alpha h^\alpha \, 
  \Phi^\alpha \ , \nonumber \\
  &&\Gamma^{(N)}_{\{ \alpha_i \}} = \prod_i^N 
  \frac{\delta \Gamma[\Phi]}{\delta \Phi^{\alpha_i}} \Big\vert_{h=0} \ ,
\label{genfcv}
\end{eqnarray}
where $\Phi^\alpha = \delta \ln {\cal Z}[h] / \delta h^\alpha$.
Through appropriate functional derivatives, these vertex functions can be
related to the corresponding cumulants, for example for the two- and four-point
functions:
\begin{eqnarray}
  &&\Gamma^{(2)}(q) = C(q)^{-1} \ , \nonumber \\
  &&\Big\langle \prod_{i=1}^4 S(q_i) \Big\rangle_c = - \prod_{i=1}^4 C(q_i) \, 
  \Gamma^{(4)}(\{ q_i \}) \ .
\label{cumver}
\end{eqnarray}
By means of these relations one easily confirms that the perturbation series 
for the vertex functions precisely consist of the {\em one-particle 
irreducible} Feynman graphs for the associated cumulants.
Moreover, the perturbative expansion with respect to the non-linear coupling 
$u$ diagrammatically translates to a {\em loop expansion}. 
Explicitly, one obtains for the {\em two-point vertex function} to two-loop 
order (resulting from the first, third, and fourth graph above):
\begin{eqnarray} 
  &&\Gamma^{(2)}(q) = r + q^2 + \frac{n+2}{6} \, u \int_k \frac{1}{r+k^2} 
\label{2lpgm2} \\
  &&\quad - \left( \frac{n+2}{6} \, u \right)^2 \int_k \frac{1}{r+k^2} 
  \int_{k'} \frac{1}{(r+{k'}^2)^2} \nonumber \\ 
  &&\!\!\!\!\!\!\! - \frac{n+2}{18} \, u^2 \! \int_k \! \frac{1}{r+k^2} 
  \int_{k'} \! \frac{1}{r+{k'}^2} \, \frac{1}{r+(q-k-k')^2} \, ; \nonumber
\end{eqnarray} 
for the {\em four-point vertex function} to one-loop order, we just have the
single Feynman diagram

\medskip
\centerline{\includegraphics[width=0.2\columnwidth]{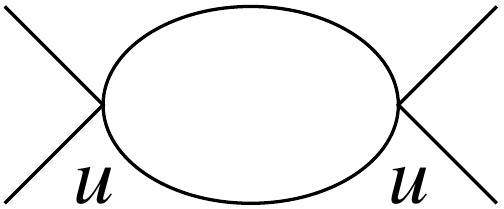}}
\smallskip

\noindent which yields at vanishing external wave vectors
\begin{equation}
  \Gamma^{(4)}(\{ q_i = 0 \}) = u - \, \frac{n+8}{6} \, u^2 \int_k 
  \frac{1}{(r+k^2)^2} \ .
\label{1lpgm4}
\end{equation}

\subsection{UV and IR divergences, renormalization}

Let us now investigate the fluctuation correction (\ref{1lpgm4}) to the 
four-point vertex function.
In dimensions $d < 4$, we can safely set the UV cutoff $\Lambda \to \infty$, 
and obtain after rendering the integral dimensionless:
\begin{eqnarray}
  &&u \int \! \frac{d^dk}{(2 \pi)^d} \, \frac{1}{(r + k^2)^2} = \nonumber \\
  &&\quad \frac{u \, r^{-2 + d/2}}{2^{d-1} \pi^{d/2} \Gamma(d/2)} 
  \int_0^\infty \frac{x^{d-1}}{(1 + x^2)^2} \, dx \ .
\label{intd<4}
\end{eqnarray}
Observe that the true {\em effective} coupling in the perturbation expansion
is not $u$, but $u \, r^{(d-4)/2} \to \infty$ as $r \to 0$, indicating the 
emerging {\em infrared} divergences that render a direct perturbative approach
meaningless in the critical regime.
We conclude once again that for $d < d_c = 4$, fluctuation corrections are 
IR-singular, and consequently expect the critical power laws to be modified as
compared to the mean-field or Gaussian approximations.
In contrast, the integral remains regular in the infrared for $d > 4$, but 
becomes ultraviolet-divergent. 
Indeed, keeping the cutoff finite, we have in dimensions larger or equal to the
{\em upper critical dimension} $d_c = 4$:
\begin{equation}
  \int_0^\Lambda \! \frac{k^{d-1}}{(r + k^2)^2} \, dk \sim \left\{  
  \begin{array}{cc} \ln (\Lambda^2 / r) & \ d = 4 \\ \Lambda^{d-4} & \ d > 4 
  \end{array} \right. \ ,
\label{intd>4}
\end{equation}
which both diverge as $\Lambda \to \infty$.
Notice that at $d_c = 4$, the logarithmic IR and UV divergences are coupled,
signaling the scale invariance of the LGW Hamiltonian in four dimensions.

One may take a convenient shortcut to determine the critical dimension through
simple {\em power counting} in terms of an arbitrary momentum scale $\mu$.
Lengths then scale as $[x] = \mu^{-1}$, wave vectors as $[q] = \mu$, and
fields consequently have the (naive) scaling dimension 
$[S^\alpha(x)] = \mu^\zeta = \mu^{-1+d/2}$.
As noted before, the fields become dimensionless in two dimensions.
For the LGW Hamiltonian (\ref{lgwhmn}) with continuous rotational symmetry in 
order parameter space, that is also just the {\em lower critical dimension}:
For $d \leq d_{lc} = 2$, the system cannot maintain spatially homogeneous
long-range order if $n \geq 3$; and at $d_{lc} = 2$ and for $n = 2$ only 
quasi-long-range order may exist with algebraically decaying correlations and 
temperature-dependent decay exponent (Berezinskii--Kosterlitz--Thouless 
scenario).
For the couplings in (\ref{lgwhmn}) one infers $[r] = \mu^2$, which means that 
the temperature control parameter constitutes a {\em relevant} coupling, and
$[u] = \mu^{4-d}$, which is relevant for $d < 4$, {\em marginal} at $d_c = 4$,
and (dangerously) irrelevant for $d > 4$.
Furthermore, dimensional analysis confirms that fluctuation loops become 
UV-divergent only for the vertex functions $\Gamma^{(2)}(q)$, but only up to
order $q^2$ in a long-wavelength expansion, and $\Gamma^{(4)}(\{ q_i = 0 \})$.
The following table summarizes the mathematical distinctions and their physical
implications in the different dimensional regimes.
\begin{table}[h] \centering
\begin{tabular}{|c|c|c|c|} 
\hline \hline  
  dim. & perturb. & $O(n)$ $\Phi^4$ & critical \\  
  range & series & field theory & behavior \\ 
\hline \hline 
  $d \leq 2$ & IR-sing. & ill-defined & no long-range \\ 
  & UV-conv. & $u$ relevant & order ($n \geq 2$) \\ 
\hline 
  $d < 4$ & IR-sing. & super-ren. & non-classical \\ 
  & UV-conv. & $u$ relevant & exponents \\ 
\hline 
  $d_c = 4$ & log. IR-/ & renorm. & logarithmic \\ 
  & UV-div. & $u$ marginal & corrections \\ 
\hline 
  $d > 4$ & IR-reg. & non-renorm. & mean-field \\ 
  & UV-div. & $u$ irrelevant & exponents \\ 
\hline \hline 
\end{tabular} 
\end{table}

It is useful to perform the loop integrations in {\em dimensional 
regularization}; i.e., to assign the following values to wave vector integrals,
even for non-integer $d$ and $\sigma$:
\begin{eqnarray} 
  &&\int \! \frac{d^dk}{(2 \pi)^d} \, 
  \frac{k^{2 \sigma}}{\left( \tau + k^2 \right)^s} = 
\label{dimreg} \\
  &&\quad \frac{\Gamma(\sigma + d/2) \, 
  \Gamma(s - \sigma - d/2)}{2^d \, \pi^{d/2} \, \Gamma(d/2) \, \Gamma(s)} \ 
  \tau^{\sigma - s + d/2} \ . \nonumber
\end{eqnarray} 
For $d < d_c$, where no UV divergences appear, this result follows directly by 
introducing spherical coordinates in momentum space.
Beyond the upper critical dimension, essentially divergent surface integrals 
are discarded in (\ref{dimreg}).
UV singularities become manifest as {\em dimensional poles} in Euler's $\Gamma$
functions.
 
We may now proceed with the renormalization program.
The goal is to absorb the UV divergences into {\em renormalized} couplings that
through this procedure become scale-dependent.
Beginning with the order parameter susceptibility, we naturally demand that 
$\chi^{-1} = C(q=0)^{-1} = \Gamma^{(2)}(q=0) = \tau = r - r_c$, i.e., $\chi$
diverges at the true critical temperature $T_c$.
From (\ref{2lpgm2}) we thus obtain to first order in $u$,
\begin{eqnarray}   
  &&r_c = - \frac{n+2}{6}\, u \!\int_k \frac{1}{r_c + k^2} + O(u^2) \nonumber\\
  &&\quad = - \frac{n+2}{6} \, u \, S_d\, \frac{\Lambda^{d-2}}{d-2} + O(u^2)\ ,
\label{1lptcs}
\end{eqnarray}
which is to be interpreted as a non-universal fluctuation-induced downward 
shift of the critical temperature, the analog of $r^*$ in Wilson's scheme.
As $\Lambda \to \infty$, $r_c$ becomes quadratically UV-divergent near four
dimensions; this divergence becomes absorbed into the new proper temperature
variable $\tau$ by means of an {\em additive renormalization}.
Inserting (\ref{1lptcs}) into (\ref{2lpgm2}) then yields to order $u$
\begin{equation}
  \chi(q)^{-1} = q^2 + \tau \, \biggl[ 1 - \frac{n+2}{6} \, u \! \int_k \! 
  \frac{1}{k^2 (\tau + k^2)} \biggr] \ .  
\label{1lpsus}
\end{equation} 

At the upper critical dimension, (\ref{1lpsus}) and (\ref{1lpgm4}) are 
logarithmically divergent as $\Lambda \to \infty$, showing up as $1/\epsilon$
poles in the dimensionally regularized integral values.
These UV poles are subsequently absorbed into {\em renormalized} fields 
$S_R^\alpha$ and parameters through the following {\em multiplicative 
renormalization} prescription:
\begin{eqnarray} 
  &&S_R^\alpha = Z_S^{1/2} \, S^\alpha \ \Rightarrow \
  \Gamma_R^{(N)} = Z_S^{-N/2} \, \Gamma^{(N)} \ ; 
\label{fldren} \\
  &&\tau_R = Z_\tau \, \tau \, \mu^{-2} \ , \quad \
  u_R = Z_u \, u \, A_d \, \mu^{d-4} \ , 
\label{cplren}
\end{eqnarray}  
which defines {\em dimensionless renormalized couplings} $\tau_R$ and $u_R$, 
and where $A_d = \Gamma(3-d/2) / 2^{d-1} \, \pi^{d/2}$ is a regular (near 
$d_c = 4$) geometric factor.
To this end, one must carefully avoid the IR-singular regime, i.e., evaluate 
the fluctuation integrals at a safe {\em normalization point}, e.g., 
$\tau_R = 1$ (or $q = \mu$).
In the {\em minimal subtraction} procedure, the renormalization constants 
$Z_S$ in (\ref{fldren}) and $Z_\tau$, $Z_u$ in (\ref{cplren}) are chosen to
contain {\em only} the $1/\epsilon$ poles and their residua.
This leads to the following $Z$ factors 
\begin{eqnarray} 
  &&O(u_R): \ Z_\tau = 1 - \frac{n + 2}{6} \, \frac{u_R}{\epsilon} \ ,
\label{1lpztt} \\ 
  &&\qquad\quad\,\ Z_u = 1 - \frac{n + 8}{6} \, \frac{u_R}{\epsilon} \ ,
\label{1lpztu} \\
  &&O(u_R^2): \ Z_S = 1 + \frac{n + 2}{144} \, \frac{u_R^2}{\epsilon} \ ,
\label{2lpzts}
\end{eqnarray}  
all calculated to first non-trivial order in $u_R$ by means of dimensional 
regularization (\ref{dimreg}) and within the minimal subtraction prescription.
$Z_\tau$ and $Z_u$ follow directly from the one-loop results (\ref{1lpsus}) and
(\ref{1lpgm4}), whereas $Z_S = 1$ to $O(u_R)$ due to the absence of any wave 
vector dependence in the ``Hartree'' loop, whence field renormalization only 
ensues to two-loop order from the rightmost ``sunset'' Feynman diagram in the 
propagator self-energy, and the $1/\epsilon$ pole in the final expression in 
(\ref{2lpgm2}).

\subsection{RG equation and critical exponents}

Through the selection of a normalization point well outside the critical 
regime, the renormalized fields and parameters in (\ref{fldren}), 
(\ref{cplren}) explicitly carry the momentum scale $\mu$.
On the other hand, the {\em unrenormalized} quantities, including the $N$-point
vertex functions, naturally do {\em not} depend on this arbitrary scale $\mu$:
\begin{equation}
  0 = \frac{d}{d \mu} \, \Gamma^{(N)}(\tau,u) = \frac{d}{d \mu} \left[ 
  Z_S^{N/2} \, \Gamma_R^{(N)}(\mu,\tau_R,u_R) \right] .
\label{indgmu}
\end{equation}
Carrying out the derivative with respect to $\mu$ by taking into account the
scale dependence of $Z_s$, $\tau_R$, and $u_R$, (\ref{indgmu}) can be rewritten
as a partial differential equation,
\begin{eqnarray} 
  &&\biggl[ \mu \, \frac{\partial}{\partial \mu} + \frac{N}{2} \, \gamma_S    
  + \gamma_\tau \, \tau_R \, \frac{\partial}{\partial \tau_R} + \beta_u \, 
  \frac{\partial}{\partial u_R} \biggr] \nonumber \\ 
  &&\qquad\qquad \Gamma_R^{(N)}(\mu,\tau_R,u_R) = 0 \ .
\label{rgeqvf}
\end{eqnarray}
This Gell-Mann--Low {\em renormalization group equation} carries crucial
information on the fundamental scale dependence of the {\em renormalized} 
physical system, here rendered explicit for the vertex functions.
In (\ref{rgeqvf}) we have introduced {\em Wilson's flow functions} defined as
\begin{eqnarray}
  &&\gamma_S = \mu \, \frac{\partial}{\partial \mu}\Big\vert_0 \, \ln Z_S
  \nonumber \\
  &&\quad\ = - \frac{n + 2}{72} \, u_R^2 + O(u_R^3) \ , 
\label{gamfld} \\ 
  &&\gamma_\tau = \mu \, \frac{\partial}{\partial \mu}\Big\vert_0 \, \ln
  \frac{\tau_R}{\tau} \nonumber \\
  &&\quad\ = - 2 + \frac{n + 2}{6} \, u_R + O(u_R^2) \ , 
\label{gamtau}
\end{eqnarray}  
where the second lines follow from the lowest-order results (\ref{2lpzts}) and 
(\ref{1lpztt}), and the {\rm RG beta function} for the non-linear coupling $u$,
\begin{eqnarray} 
  &&\beta_u = \mu \, \frac{\partial}{\partial \mu}\Big\vert_0 u_R = u_R 
  \Bigl[ d - 4 + \mu \, \frac{\partial}{\partial \mu} \Big\vert_0 \, \ln Z_u 
  \Bigr] \nonumber \\
  &&\quad\ = u_R \Bigl[ -\epsilon + \frac{n + 8}{6} \, u_R + O(u_R^2) \Bigr]\ ,
\label{betafu}
\end{eqnarray}
where (\ref{1lpztu}) has been inserted.

The first-order linear partial differential equation (\ref{rgeqvf}) can next be
formally solved via the standard {\em method of characteristics}; to this end,
one lets $\mu \to \mu(\ell) = \mu \, \ell$, with a dimensionless scale 
parameter $\ell$; note that in contrast to the convention in Chap.~2.5, the IR 
regime is now reached in the limit $\ell \to 0$.
Inserting this parametrization into the RG equation (\ref{rgeqvf}), one obtains
an equivalent set of coupled first-order ordinary differential equations, 
namely the {\em RG flow equations} for the {\em running couplings}
\begin{equation}
  \ell \, \frac{d{\tilde \tau}(\ell)}{d\ell} = 
  {\tilde \tau}(\ell) \, \gamma_\tau(\ell) \ , \quad
  \ell \, \frac{d{\tilde u}(\ell)}{d\ell} = \beta_u(\ell) \ ,
\label{runcpl}
\end{equation}
with initial values ${\tilde \tau}(1) = \tau_R$, ${\tilde u}(1) = u_R$, and
similar differential equations for the $N$-point vertex functions, which 
involve their overall naive scaling dimensions and the anomalous contributions 
stemming from the field renormalization, as encoded in (\ref{gamfld}).

For example, for the susceptibility $\chi(q) = \Gamma^{(2)}(q)^{-1}$, one has
$\chi_R(\mu,\tau_R,u_R,q)^{-1} = \mu^2\, {\hat \chi}_R(\tau_R,u_R,q/\mu)^{-1}$,
and its RG flow correspondingly integrates to
\begin{equation}
  \chi_R(\ell)^{-1} = \chi_R(1)^{-1} \, \ell^2 \, \exp\biggl[ \int_1^\ell 
  \gamma_S(\ell') \, \frac{d\ell'}{\ell'} \biggr] \ .
\label{susscl}
\end{equation}
Near an {\em infrared-stable} RG fixed point $u^*$, i.e., a zero of the RG beta
function $\beta_u(u^*) = 0$ with $\beta_u'(u^*) > 0$, the flow equation for the
running temperature variable is readily solved:
${\tilde \tau}(\ell) \approx \tau_R \, \ell^{\gamma_\tau^*}$, where 
$\gamma_\tau^* = \gamma_\tau(u^*)$.
Inserting this and the fixed point value $\gamma_S^* = \gamma_S(u^*)$ into
(\ref{susscl}) yields the following general scaling form
\begin{equation} 
  \chi_R(\tau_R,q)^{-1} \approx \mu^2 \, \ell^{2 + \gamma_S^*} \,  
  {\hat \chi}_R(\tau_R \, \ell^{\gamma_\tau^*}, u^*, q / \mu \, \ell)^{-1} \ .
\label{gensus}
\end{equation}
With the {\em matching} condition $\ell = |q| / \mu$, one thus recovers
(\ref{corsch}), with the identifications $\eta = - \gamma_S^*$ and 
$\nu = - 1 / \gamma_\tau^*$ for the critical exponents.

Our perturbative RG analysis of the $O(n)$-symmetric LGW Hamiltonian 
(\ref{lgwhmn}) yielded the one-loop beta function (\ref{betafu}), whose zeros 
are (i) the Gaussian fixed point $u_0^* = 0$, stable for $\epsilon < 0$ or 
$d > d_c = 4$, obviously resulting in the mean-field exponents $\eta = 0$ and 
$\nu = \frac12$; and (ii) the non-trivial {\em Heisenberg fixed point}  
\begin{equation}
  u_H^* = \frac{6 \, \epsilon}{n + 8} + O(\epsilon^2) \ ,
\label{heisfp}
\end{equation}
which exists and becomes IR-stable for $\epsilon > 0$, i.e., in dimensions 
$d < 4$.
This allows us to compute the critical exponents in a systematic 
$\epsilon = 4 - d$ expansion,
\begin{eqnarray} 
  &&\eta = \frac{n + 2}{2 \, (n + 8)^2} \, \epsilon^2 + O(\epsilon^3) \ , 
\label{2lpeta}   \\
  &&\nu^{-1} = 2 - \frac{n + 2}{n + 8} \, \epsilon + O(\epsilon^2) \ .
\label{1lpnur}
\end{eqnarray}  
Aside from the dimensionality, these values only depend on the number of order
parameter components $n$.
Note that (\ref{1lpnur}) reduces to the Ising exponent values (\ref{icrtex}) 
for $n = 1$; in the limit $n \to \infty$ one also finds the correct exponents 
for the exactly solvable {\em spherical model}, namely $\eta = 0$ and 
$\nu = 1 / (2-\epsilon) = 1 / (d-2)$, which diverges at the lower critical
dimension $d_{lc} = 2$.
At the upper critical dimension $d_c = 4$, the solution of the flow equation 
for the running non-linear coupling reads
\begin{equation}
  {\tilde u}(\ell) = \frac{u_R}{1 - \frac{n + 8}{6} \, u_R \, \ln \ell} \ , 
\label{ufld=4}
\end{equation}
whence approximately
\begin{equation}
  {\tilde \tau}(\ell) \sim 
  \frac{\tau_R}{\ell^2 (\ln |\ell|)^{(n + 2) / (n + 8)}} \ ,
\label{tfld=4}
\end{equation}
which in turn implies that the correlation length divergence picks up 
{\em logarithmic} corrections to the mean-field behavior, 
\begin{equation}
  \xi \propto \tau_R^{-1/2} \, (\ln \tau_R)^{(n + 2) / 2 (n+8)} \ . 
\label{xilgcr}
\end{equation}

The field-theoretic formulation of the renormalization group provides us with 
an elegant and powerful tool to extract the proper infrared scaling properties
in low dimensions $d \leq d_c$ from a continuum theory via a careful analysis 
of its ultraviolet singularities that appear in dimensions $d \geq d_c$.
The renormalization group equation carries information on the scale dependence
of physical parameters and correlation functions, and allows to make 
connections between the UV and IR limits provided a stable RG fixed point can 
be identified.
One may then systematically derive scaling laws, and acquire a thorough 
understanding of the origin of universality and its realm of validity for a 
specific physical system.
Moreover, we have seen that a perturbative analysis allows a controlled
computation of critical exponents (and also of scaling functions) in the
framework of a dimensional expansion near the upper critical dimension.
This $\epsilon$ expansion certainly provides useful information on overall 
trends, and can in some instances even be rendered to a precision calculation
if sufficiently high orders in the perturbation series can be evaluated and
subsequently be refined through Borel resummations.
In addition, modern non-perturbative ``exact'' numerical RG methods have 
succeeded in yielding very accurate results.
It should also be stressed that it is of course the very concept of 
universality that also allows us to infer meaningful information from numerical
simulations of simplified lattice or continuum models.

\section{Critical Dynamics}

We now venture to investigate {\em dynamic} critical phenomena near continuous
phase transitions, first in the vicinity of critical point in thermal 
equilibrium, and later at genuine non-equilibrium phase transitions in 
externally driven, non-isolated systems.
The natural time scale separation between the slow kinetics of the order 
parameter (along with any other conserved fields) and fast, non-critical 
degrees of freedom suggests a phenomenological description in terms of 
non-linear  stochastic Langevin-type differential equations, and allows a 
generalization of universal scaling laws to time-dependent phenomena.
Distinct dynamical universality classes ensue dependent on the order parameter
itself representing a conserved quantity or not, and potentially its dynamical
coupling to other conserved hydrodynamic modes \cite{Hohenberg77}.
We begin by writing down scaling laws for dynamical response and correlation
functions, and subsequently introduce effective mesoscopic Langevin equations,
with stochastic noise correlations that near thermal equilibrium are 
constrained by fluctuation-dissipation relations.
It is then demonstrated how such non-linear stochastic partial differential 
equations can be mapped onto a field theory representation via the 
Janssen--De~Dominicis response functional, which in turn may be analyzed by
means of the field-theoretic renormalization group tools developed in the 
previous chapter.
We will specifically construct a dynamic perturbation theory expansion and 
determine the universal scaling behavior along with the dynamic critical 
exponents for the relaxational models A and B with non-conserved and conserved 
order parameter, respectively; for more in-depth treatments, the reader is
referred to Refs.~\cite{Janssen79}--\cite{Tauber07} 
and \cite[chaps. 4,5]{Tauberxx}. 
In addition to purely relaxational kinetics, we shall also address the critical
dynamics of isotropic ferromagnets \cite{Frey94}, as well as generic scale 
invariance and non-equilibrium phase transitions in driven diffusive systems
\cite{Schmittmann95}.

\subsection{Dynamical scaling hypothesis}

Let us first recall the behavior of the static order parameter correlation 
function and susceptibility near a critical point ($h = 0$ and $\tau \to 0$), 
captured by the scaling laws (\ref{corsch}) and (\ref{spcorr}), and induced by 
the divergence of the characteristic {\em correlation length}, 
$\xi(\tau) \sim |\tau|^{- \nu}$.
As spatially correlated regions grow tremendously upon approaching the phase 
transition, one expects the typical relaxation time associated with the order 
parameter kinetics to increase as well, 
$t_c(\tau) \sim \xi(\tau)^z \sim |\tau|^{- z \nu}$.
This phenomenon of {\em critical slowing-down} is governed by the {\em dynamic
critical exponent} $z = \nu_t / \nu$, which can also be visualized as the ratio
between the exponents that describe the divergence of correlations in the 
temporal and spatial ``directions'', respectively. 
We may thus write down a {\em dynamic scaling} ansatz for the corresponding 
wavevector-dependent {\em characteristic frequency} scale,
\begin{equation}
  \omega_c(\tau,q) = |q|^z \, {\hat \omega}_\pm(q \, \xi) \ ,
\label{dynscf}
\end{equation}
with ${\hat \omega}_\pm(y \to \infty) \to {\rm const.}$, whence the 
{\em critical dispersion relation} becomes $\omega_c(0,q) \sim |q|^z$.

We are particularly interested in describing the time dependence for the order 
parameter response and correlation functions:
\begin{eqnarray}
  &&\chi(x-x',t-t') = \frac{\partial \langle S(x,t) \rangle}
  {\partial h(x',t')} \bigg\vert_{h=0} \ , 
\label{dynres} \\
  &&C(x,t) = \langle S(x,t) \, S(0,0) \rangle - \langle S \rangle^2 \ ,
\label{dyncor}
\end{eqnarray}
where we are considering a stationary dynamical regime where spatial and 
temporal time translation invariance holds.
In {\em thermal equilibrium} (only !), the spatio-temporal Fourier transforms 
of these functions are intimately connected through the 
{\em fluctuation-dissipation theorem} (FDT)
\begin{equation}
  C(q,\omega) = 2 k_{\rm B} T \, {\rm Im} \ \frac{\chi(q,\omega)}{\omega} \ .
\label{fldist}
\end{equation}
Generalizing the static scaling laws (\ref{corsch}) and (\ref{spcorr}), we may
then formulate the {\em dynamical scaling hypothesis} for the asymptotic 
critical properties of the time-dependent susceptibility and correlation 
function:
\begin{eqnarray}
  &&\chi(\tau,q,\omega) = |q|^{- 2 + \eta} \, 
  {\hat \chi}_\pm(q \, \xi , \omega \, \xi^z) \ , 
\label{dynscr} \\ 
  &&C(\tau,x,t) = |x|^{-(d - 2 + \eta)} \, {\widetilde C}_\pm(x/\xi, t/t_c) \ .
\label{dynscc}
\end{eqnarray}
As a consequence of the stringent constraints imposed by the FDT 
(\ref{fldist}), the same {\em three} independent critical exponents $\nu$, 
$\eta$, and $z$ characterize the universal scaling regimes in both 
(\ref{dynscr}) and (\ref{dynscc}).
Away from thermal equilibrium, where the FDT restrictions do not apply, the
dynamic response and correlation functions are however in general characterized
by distinct scaling exponents. 
We remark that appropriate variants of the dynamical scaling hypothesis may 
also be postulated for transport coefficients.

\subsection{Langevin dynamics and Gaussian theory}

The critical slowing-down of the order parameter kinetics produces an effective
time-scale separation between the critical degrees of freedom, additional 
conserved hydrodynamic modes that might be present, and all other comparatively
``fast'' variables.
This observation naturally calls for a {\em mesoscopic Langevin description} of
critical dynamics, where the fast degrees of freedom are treated as mere white 
{\em noise} that randomly affects the few slow modes in the system.
In such a {\em coarse-grained} picture, one writes down coupled stochastic
equations of motion for the order parameter and perhaps any other conserved 
fields that reflect their intrinsic microscopic reversible dynamics as well as
irreversible relaxation kinetics, the latter connected in thermal equilibrium 
with the noise strengths through {\em Einstein relations} or FDTs.
Generally the various possible {\em mode couplings} of the order parameter to 
additional conserved, and consequently diffusively slow modes leads to a 
splitting of the static into several {\em dynamic universality classes}
\cite{Hohenberg77,Folk06,Tauberxx}.

Here we shall assume that the order parameter field is decoupled from any other
slow modes, and first focus on its purely relaxational kinetics 
\cite[Chaps.~4,5]{Tauberxx}.
If the order parameter itself is not a conserved quantity, any deviation from
thermal equilibrium will just tend to relax back to the minimizing 
configuration of the free energy, e.g. given by the $O(n$)-symmetric LGW 
Hamiltonian (\ref{lgwhmn}):
\begin{equation} 
  \frac{\partial S^\alpha(x,t)}{\partial t} = 
  - D \, \frac{\delta {\cal H}[{\vec S}]}{\delta S^\alpha(x,t)} 
  + \zeta^\alpha(x,t) \, ,
\label{relmda}
\end{equation}
with Gaussian white noise that is fully characterized by its first two moments,
\begin{eqnarray}
  &&\langle \zeta^\alpha(x,t) \rangle = 0 \ , \nonumber \\
  &&\langle \zeta^\alpha(x,t) \, \zeta^\beta(x',t') \rangle = \\  
\label{stnmda}
  &&\qquad 2 D k_{\rm B} T \delta(x-x') \delta(t-t') \delta^{\alpha \beta} \ .
  \nonumber
\end{eqnarray} 
As can be inferred from the associated Fokker--Planck equation, the 
{\em Einstein relation} that connects the noise correlator strength with the
relaxation constant $D$ and temperature guarantees that the probability
distribution for the field $S^\alpha$ asymptotically approaches the canonical 
stationary distribution ${\cal P}[{\vec S},t] \to {\cal P}_s[{\vec S}] \propto
\exp (-{\cal H}[{\vec S}] / k_{\rm B} T)$ as $t \to \infty$. 

If the order parameter is {\em conserved} under the dynamics, satisfying a 
continuity equation, its spatial fluctuations can only relax {\em diffusively};
as a consequence, one needs to replace the constant relaxation rate $D$ by the 
diffusion operator $- D \, \nabla^2$, both in the Langevin equation 
(\ref{relmda}) and the noise correlation (\ref{stnmda}).
In the following, we shall treat both these situations simultaneously, letting
$D \to D (i \nabla)^a$, where either $a = 0$, corresponding to the purely
relaxational {\em model A} of critical dynamics as appropriate for a 
non-conserved order parameter; or $a = 2$, which describes {\em model B} with a
conserved order parameter field.

Let us again begin with the {\em Gaussian} or {\em mean-field approximation},
where we set the non-linear coupling $u = 0$.
A Fourier transform in space and time according to 
$S^\alpha(x,t) = \int \frac{d^dq}{(2 \pi)^d} \int \frac{d\omega}{2 \pi} \, 
S(q,\omega) \, e^{i q \cdot x - i \omega t}$ of the thus linearized Langevin 
equation (\ref{relmda}) yields
\begin{eqnarray}
  &&\left[ -i \omega + D q^a (r + q^2) \right] S^\alpha(q,\omega) = \nonumber\\
  &&\qquad\qquad D q^a \, h^\alpha(q,\omega) + \zeta^\alpha(q,\omega) \ ,
\label{rellin}
\end{eqnarray}
with $\langle \zeta^\alpha(q,\omega) \rangle = 0$ and
\begin{eqnarray}
  &&\langle \zeta^\alpha(q,\omega) \zeta^\beta(q',\omega') \rangle = 
  2 k_{\rm B} T \, D q^a \nonumber \\
  &&\qquad\quad (2 \pi)^{d+1} \delta(q + q') \delta(\omega + \omega') 
  \delta^{\alpha\beta} \ ,
\label{stnfsp}
\end{eqnarray}
and where the external field term $- \sum_\alpha h^\alpha S^\alpha$ has been 
added to the LGW Hamiltonian (\ref{lgwhmn}).
A derivative of (\ref{rellin}) with respect to the external field then 
immediately gives the {\em dynamic response function} 
\begin{eqnarray} 
  &&\chi_0^{\alpha\beta}(q,\omega)= \frac{\partial \langle S^\alpha(q,\omega)
      \rangle}{\partial h^\beta(q,\omega)} \Big\vert_{h = 0} \nonumber \\
  &&\qquad\qquad = D q^a G_0(q,\omega) \delta^{\alpha\beta} \ ,
\label{relres} \\  
  &&G_0(q,\omega) = \frac{1}{-i \omega + D q^a (r + q^2)} \ . \nonumber 
\end{eqnarray}
Its temporal Fourier backtransform of course satisfies {\em causality}: 
$G_0(q,t)$ vanishes for $t < 0$, and reads
\begin{equation}
  G_0(q,t) = e^{-D q^a (r + q^2) t} \, \Theta(t) \ ,
\label{relret}
\end{equation}
from which we infer the characteristic relaxation rate 
$t_c^{-1} = D q^a (r + q^2)$: 
For model A ($a = 0$), the order parameter relaxes diffusively at the critical
point (i.e., $z = 2$), while for model B ($a = 2$) critical slowing-down 
induces a crossover from $D r q^2$ to $D q^4$ as $r \to 0$ ($z = 4$). 
For $h^\alpha = 0$, the dynamic correlation function is readily obtained from 
(\ref{rellin}) and (\ref{stnfsp}):
\begin{eqnarray}
  &&\langle S^\alpha(q,\omega) \, S^\beta(q',\omega') \rangle_0 = C_0(q,\omega)
  \nonumber \\
  &&\qquad (2 \pi)^{d+1} \delta(q + q') \delta(\omega + \omega') 
  \delta^{\alpha\beta} \ , \nonumber \\
  &&C_0(q,\omega) = \frac{2 k_{\rm B} T \, D q^a}{\omega^2 +  
      [D q^a (r+q^2)]^2} 
\label{relcor} \\
  &&\qquad\quad\;\ = 2 k_{\rm B}T D q^a \, |G_0(q,\omega)|^2 \ , \nonumber \\ 
  &&C_0(q,t) = \frac{k_{\rm B} T}{r + q^2} \, e^{- D q^a \, (r + q^2) |t|} \ . 
\label{relcot}
\end{eqnarray} 
Comparing these results with (\ref{dynscr}) and (\ref{dynscc}), one again 
identifies the static {\em Gaussian critical exponents} $\nu = \frac12$ and 
$\eta = 0$, and the {\em mean-field dynamic exponent} $z = 2 + a$ for the 
purely relaxational models A and B.

\subsection{Field theory representations of Langevin dynamics}

This subsection describes how stochastic Langevin equations of motion can be 
mapped onto continuous field theory representations.
To this end, we consider the following general coupled Langevin equations for 
some mesoscopic stochastic variables $S^\alpha$:  
\begin{eqnarray} 
  &&\frac{\partial S^\alpha(x,t)}{\partial t} = F^\alpha[S](x,t)  
      + \zeta^\alpha(x,t) \ ,
\label{langen} \\ 
  &&\langle \zeta^\alpha(x,t) \zeta^\beta(x',t') \rangle = \nonumber \\ 
  &&\quad 2 L^\alpha[S] \, \delta(x-x') \delta(t-t') \delta^{\alpha \beta} \ .
\label{noigen}
\end{eqnarray}
Naturally we assume $\langle \zeta^\alpha(x,t) \rangle = 0$ here, since a 
non-vanishing mean of the {\em stochastic forces} or {\em noise} could just be 
included in the {\em systematic} forces $F^\alpha[S]$.
Note that the {\em noise correlator} $L^\alpha$ may be an operator, as is the 
case for conserved variables, and could also be a functional of the slow fields
$S^\alpha$. 
The crucial input is that we assume the noise history to represent a 
{\em Gaussian} stochastic process, whose probability distribution if completely
determined by the second moment (\ref{noigen}): 
\begin{equation}
    {\cal W}[\zeta] \propto \exp \, \biggl[ - \frac{1}{4} \! \int \!\! d^dx \!\!
    \int_0^{t_f} \!\! dt \sum_\alpha \zeta^\alpha (L^\alpha)^{-1} \zeta^\alpha 
    \biggr] . \
\label{noidis}
\end{equation}
Switching dynamical variables from the noise $\zeta^\alpha$ to the slow 
stochastic fields $S^\alpha$ yields ${\cal W}[\zeta] \, {\cal D}[\zeta] = 
{\cal P}[S] {\cal D}[S] \propto e^{- {\cal G}[S]} {\cal D}[S]$, with the 
{\em Onsager-Machlup functional} providing the associated exponential weight 
that may be viewed as a field theory action: 
\begin{eqnarray}
  &&{\cal G}[S] = \frac{1}{4} \int \! d^dx \int_0^{t_f} \! dt \sum_\alpha 
  \left( \partial_t S^\alpha - F^\alpha[S] \right) \nonumber \\
  &&\qquad\qquad\quad \left[ (L^\alpha)^{-1} \left( \partial_t S^\alpha 
  - F^\alpha[S] \right) \right] .
\label{onsmch}
\end{eqnarray}
The observant reader will have noticed that the functional determinant stemming
from the variable transformation has been ignored here; however, upon utilizing
a {\em forward} time discretization, i.e., the It\^o interpretation for 
non-linear stochastic processes, this functional determinant turns out to be 
constant, and can simply be absorbed into the functional integral measure.
Notice also that the overall normalization $\int {\cal D}[\zeta] W[\zeta] = 1$
implies the corresponding ``partition function'' to be unity, and hence to 
carry no information, in stark contrast with thermal equilibrium statistical 
mechanics.
While the Onsager--Machlup functional (\ref{onsmch}) provides a desirable
field theory representation of stochastic Langevin dynamics, it is also plagued
by two technical problems:
First, it contains $(L^\alpha)^{-1}$, which for conserved fields entails an
inverse differential operator or Laplacian Green's function; second, it 
includes the square of the systematic force terms $F^\alpha[S]$ and 
consequently highly non-linear contributions.
It is thus beneficial to partially linearize the above action by means of a
Hubbard--Stratonovich transformation, at the expense of introducing an 
{\em additional dynamical field variable}.

In order to completely avoid any possible singularities incorporated in the
inverse operator $(L^\alpha)^{-1}$, we follow another more direct route here.
The goal is to compute averages of observables $A$ that should be functionals
of the slow modes $S^\alpha$ over noise ``histories'': $\langle A[S] 
\rangle_\zeta \propto \int {\cal D}[\zeta] \, A[S(\zeta)] \, W[\zeta]$.
Inserting a rather involved representation of unity in terms of a product of
Dirac delta distributions on each space-time point, and subsequently writing 
these as integrals over {\rm auxiliary fields} ${\widetilde S}^\alpha$ along the
imaginary axis, $1 = \int \! {\cal D}[S] \prod_\alpha \prod_{(x,t)} \delta \, 
\bigl( \partial_t S^\alpha(x,t) - F^\alpha[S](x,t) - \zeta^\alpha(x,t) \bigr) = 
\int \! {\cal D}[i {\widetilde S}] \! \int \! {\cal D}[S] \exp \, \bigl[ 
- \int \! d^dx \! \int \! dt \sum_\alpha {\widetilde S}^\alpha \bigl( 
\partial_t \, S^\alpha - F^\alpha - \zeta^\alpha \bigr) \bigr]$, we arrive at
\begin{eqnarray} 
  &&\langle A[S] \rangle_\zeta \propto \int \! {\cal D}[i{\widetilde S}] 
  \int \! {\cal D}[S] \, A[S] \int \! {\cal D}[\zeta] \nonumber \\
  &&\quad \exp \, \biggl[ - \! \int \! d^dx \! \int \! dt \! \sum_\alpha 
  {\widetilde S}^\alpha \bigl( \partial_t S^\alpha - F^\alpha[S] \bigr) \biggr]
  \nonumber \\ 
  &&\ \exp \, \biggl( - \!\int \!\! d^dx \! \int \!\! dt \sum_\alpha \biggl[ 
  \frac{\zeta^\alpha (L^\alpha)^{-1} \zeta^\alpha}{4} - {\widetilde S}^\alpha 
  \zeta^\alpha \biggr] \biggr) \ . \nonumber
\end{eqnarray} 
Performing the Gaussian integral over the noise $\zeta^\alpha$ finally yields
\begin{eqnarray}
  &&\langle A[S] \rangle_\zeta = \int {\cal D}[S] \, A[S] \, {\cal P}[S] \, , 
  \nonumber \\
  &&{\cal P}[S] \propto \int {\cal D}[i {\widetilde S}] \, 
  e^{- {\cal A}[{\widetilde S},S]} \ .
\label{noisav}
\end{eqnarray}
with the {\em Janssen--De~Dominicis response functional}
\begin{eqnarray}  
  &&{\cal A}[{\widetilde S},S] = \int \! d^dx \int_0^{t_f} \! dt 
\label{janded} \\
  &&\qquad \sum_\alpha \Bigl[ {\widetilde S}^\alpha \bigl( \partial_t \, 
  S^\alpha - F^\alpha[S] \bigr) - {\widetilde S}^\alpha L^\alpha[S] \,
  {\widetilde S}^\alpha \Bigr] \ . \nonumber
\end{eqnarray} 
The stochastic dynamics is now encoded in {\em two} distinct sets of 
mesoscopic fields, namely the original slow variables $S^\alpha$ and the 
associated auxiliary fields ${\widetilde S}^\alpha$.
Once again, the Gaussian noise normalization implies 
$\int \! {\cal D}[i{\widetilde S}] \int \! {\cal D}[S] \, 
e^{- {\cal A}[{\widetilde S},S]} = 1$; furthermore, integrating out the
auxiliary fields ${\widetilde S}^\alpha$ recovers the Onsager--Machlup 
functional (\ref{onsmch}). 

Specifically for the purely relaxational models A and B, the response 
functional reads (we now set $k_{\rm B} T = 1$):
\begin{eqnarray} 
  &&{\cal A}[{\widetilde S},S] = \nonumber \\
  && \int \! d^dx \! \int \! dt \sum_\alpha \, \Bigl( {\widetilde S}^\alpha 
  \left[ \partial_t + D (i \nabla)^a (r - \nabla^2) \right] S^\alpha 
  \nonumber \\
  &&\qquad\qquad - D {\widetilde S}^\alpha (i \nabla)^a {\widetilde S}^\alpha 
  - D {\widetilde S}^\alpha (i \nabla)^a h^\alpha \nonumber \\
  &&\qquad\quad + D \, \frac{u}{6} \sum_\beta {\widetilde S}^\alpha 
  (i \nabla)^a S^\alpha S^\beta S^\beta \Bigr) \, ,
\label{resfab}
\end{eqnarray}
The first two lines here represent the Gaussian action ${\cal A}_0$, and the
term $\sim u$ the non-linear contributions.
By means of (\ref{noisav}) and (\ref{resfab}), the dynamical order parameter 
susceptibility becomes
\begin{eqnarray} 
  &&\chi^{\alpha\beta}(x-x',t-t') = \frac{\delta \langle S^\alpha(x,t) \rangle}
  {\delta h^\beta(x',t')} \Big\vert_{h = 0} \nonumber \\ 
  &&\qquad = D \big\langle S^\alpha(x,t) (i \nabla)^a \, 
  {\widetilde S}^\beta(x',t') \big\rangle \, ;
\end{eqnarray} 
i.e., the response function can be expressed as an expectation value that 
involves both order parameter $S^\alpha$ and auxiliary fields 
${\widetilde S}^\beta$, whence the latter are also referred to as 
{\em ``response'' fields}.
Invoking causality and time inversion symmetry, it is a straightforward 
exercise to derive the {\em fluctuation-dissipation theorem}, equivalent to
(\ref{fldist}):
\begin{eqnarray}
    &&\chi^{\alpha \beta}(x-x',t-t') = \nonumber \\
    &&\quad  \Theta(t-t') \,  
    \frac{\partial}{\partial t'} \, \big\langle S^\alpha(x,t) S^\beta(x',t')  
    \big\rangle \, .
\label{resfdt}
\end{eqnarray}
In analogy with static, equilibrium statistical field theory (\ref{genfcc}), 
one defines the {\em generating functional} for correlation functions and 
cumulants,
\begin{eqnarray}
  &&{\cal Z}[{\tilde j},j] = \Big\langle e^{\int \! d^dx \! \int \! dt  
  \sum_\alpha \left( {\tilde j}^\alpha \, {\widetilde S}^\alpha + j^\alpha \, 
  S^\alpha \right)} \Big\rangle \, ,
\label{resgen} \\
  &&\Big\langle \prod_{ij} S^{\alpha_i} {\widetilde S}^{\alpha_j} 
  \Big\rangle_{(c)} \! = \prod_{ij} \frac{\delta}{\delta j^{\alpha_i}} 
  \frac{\delta (\ln) {\cal Z}[{\tilde j},j]}{\delta {\tilde j}^{\alpha_j}} 
  \Big\vert_{{\tilde j} = j = 0} \, . \nonumber
\end{eqnarray}

\subsection{Dynamic perturbation theory}

We may now proceed and treat the non-linear terms $\sim u$ by means of a 
{\em perturbation expansion},
\begin{eqnarray} 
  &&\Big\langle \prod_{ij} S^{\alpha_i} \, {\widetilde S}^{\alpha_j} 
  \Big\rangle = \frac{\langle \prod_{ij} S^{\alpha_i} \, 
  {\widetilde S}^{\alpha_j} \, e^{-{\cal A}_{\rm int}[{\widetilde S},S]} 
  \rangle_0} {\langle e^{-{\cal A}_{\rm int}[{\widetilde S},S]} \rangle_0}  
  \nonumber \\
  &&= \Big\langle \prod_{ij} S^{\alpha_i} \, {\widetilde S}^{\alpha_j} 
  \sum_{l=0}^\infty \frac{1}{l!} \left( -{\cal A}_{\rm int}[{\widetilde S},S] 
  \right)^l \Big\rangle_0 \, .
\label{dynper}
\end{eqnarray} 
Since the denominator is one owing to noise normalization, there are 
{\em no ``vacuum'' contributions} in this dynamic field theory.
Note furthermore that {\em causality} implies 
$\langle {\widetilde S}^\alpha(q,\omega) \, {\widetilde S}^\beta(q',\omega') 
\rangle_0 = 0$. 
From the Gaussian action ${\cal A}_0$ (with $u = 0$), one immediately recovers 
the expressions (\ref{relres}) and (\ref{relcor}) for $G_0(q,\omega)$ and 
$C_0(q,\omega)$, respectively.
Since the dynamic correlation function can be expressed in terms of the noise
strength and the response function, the graphical representation in terms of
Feynman diagrams can be based entirely on the causal {\em response propagators}
$G_0(q,\omega)$, represented as {\em directed} lines (we use the convention
that time propagates from right to left) that connect ${\widetilde S}^\beta$ 
to $S^\alpha$ fields, and join at either two-point noise or four-point 
non-linear relaxation {\em vertices}, all subject to wave vector and frequency
conservation as a consequence of spatial and temporal time translation 
invariance:

\medskip\smallskip
\centerline{\includegraphics[width=\columnwidth]{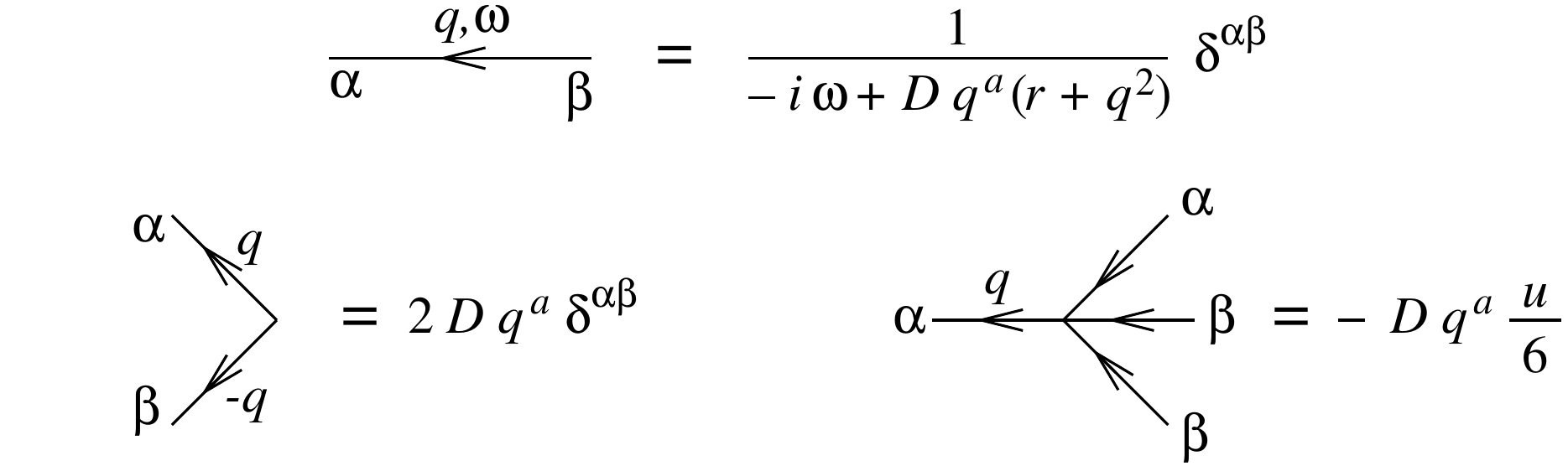}}
\smallskip

Following standard field theory procedures, one next identifies the 
{\em cumulants} as represented by {\em connected} Feynman diagrams, and in 
complete analogy with the static theory establishes {\em Dyson's equation} for 
the full response propagator, 
$G(q,\omega)^{-1} = G_0(q,\omega)^{-1} - \Sigma(q,\omega)$. 
By means of the fields 
${\widetilde \Phi}^\alpha = \delta \ln {\cal Z} / \delta {\tilde j}^\alpha$ and
$\Phi^\alpha = \delta \ln {\cal Z} / \delta j^\alpha$ one constructs the
{\em generating functional} for dynamical {\em vertex functions} via the 
Legendre transform 
\begin{eqnarray} 
  &&\Gamma[{\widetilde \Phi},\Phi] = - \ln {\cal Z}[{\tilde j},j] \nonumber \\
  &&\quad + \int \! d^dx \! \int \! dt \sum_\alpha \left( {\tilde j}^\alpha \, 
  {\widetilde \Phi}^\alpha + j^\alpha \, \Phi^\alpha \right) \ , 
\label{dynvfn} \\
  &&\Gamma^{({\widetilde N},N)}_{\{ \alpha_i \};\{ \alpha_j \}} = 
  \prod_i^{\widetilde N} \! \frac{\delta}{\delta {\widetilde \Phi}^{\alpha_i}}
  \prod_j^N \! \frac{\delta}{\delta \Phi^{\alpha_j}} \, 
  \Gamma[{\widetilde \Phi},\Phi] \Big\vert_{{\tilde j}=0=j} \ . \nonumber
\end{eqnarray} 
Functional derivatives then establish the following connections between 
two-point cumulants and vertex functions,
\begin{eqnarray}
  &&\Gamma^{(1,1)}(q,\omega) = G(-q,-\omega)^{-1} \ , 
\label{dyn2pt} \\ 
  &&\Gamma^{(2,0)}(q,\omega) = - \, \frac{C(q,\omega)}{|G(q,\omega)|^2} 
  \nonumber \\
  &&\qquad = - \frac{2 D q^a}{\omega} \, {\rm Im} \ \Gamma^{(1,1)}(q,\omega)\ ,
\label{corver} 
\end{eqnarray} 
where the last relation follows from the equilibrium FDT (\ref{fldist}).
Moreover, $\Gamma^{(0,2)}(q,\omega) = 0$ as a consequence of causality. 
One thus easily sees that the vertex functions are graphically represented by 
the {\em one-particle (1PI) irreducible} Feynman diagrams.
 
We can now formulate the {\em Feynman rules} for the dynamical perturbation 
expansion for the $l$-th order contribution to the vertex function 
$\Gamma^{({\widetilde N},N)}$, illustrated here for the one-loop graphs for
$\Gamma^{(1,1)}$ and $\Gamma^{(1,3)}$:

\centerline{\includegraphics[width=0.9\columnwidth]{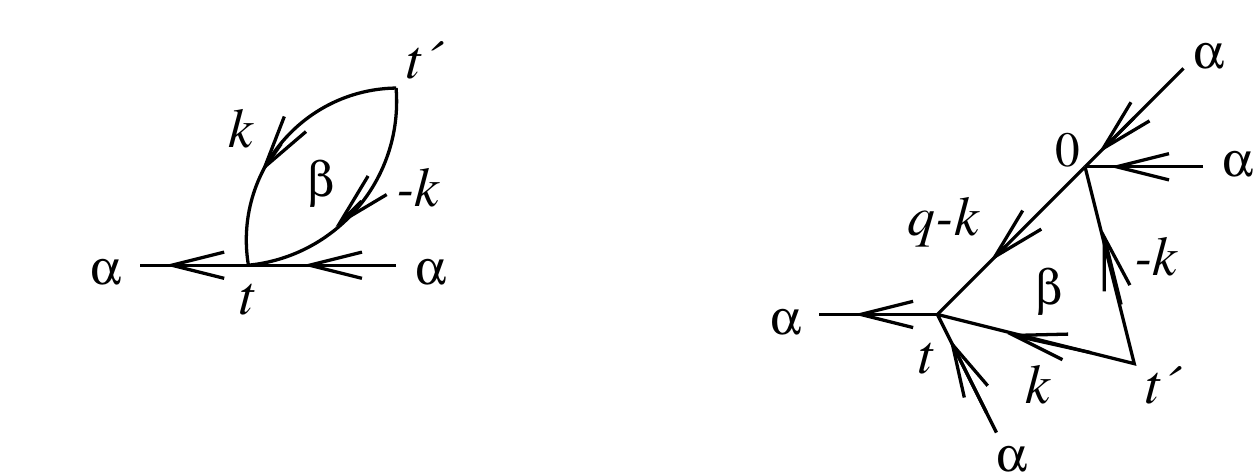}}

\begin{enumerate} 
  \item Draw all topologically different, connected 1PI graphs with 
        ${\widetilde N}$ out- / $N$ incoming lines connecting $l$ 
        relaxation vertices. 
        Do {\em not} allow closed response loops (since $\Theta(0) = 0$ in the
        It\^o calculus). 
  \item Attach wave vectors $q_i$, frequencies $\omega_i$ / times $t_i$, and
        internal indices $\alpha_i$ to all directed lines, obeying 
        {\em ``momentum- energy'' conservation} at each vertex. 
  \item Each {\em directed line} corresponds to a {\em response propagator} 
        $G_0(-q,-\omega)$ or $G_0(q,t_i-t_j)$, the two-point vertex to the 
        {\em noise strength} $2 D q^a$, and the four-point {\em relaxation 
        vertex} to $- D q^a u / 6$. 
        Closed loops imply integrals over the internal wave vectors and 
        frequencies or times, subject to causality constraints, as well as sums
        over the internal vector indices. 
        The residue theorem may be applied to evaluate frequency integrals. 
  \item Multiply with $-1$ and the {\em combinatorial factor} counting all 
        possible ways of connecting the propagators, $l$ relaxation vertices, 
        and $k$ two-point vertices leading to topologically identical graphs, 
        including a factor $1 / l! \, k!$ originating in the expansion of 
        $\exp(-{\cal A}_{\rm int}[{\widetilde S},S])$. 
\end{enumerate} 

The perturbation series then graphically becomes a {\em loop expansion}. 
For example, the propagator self-energy diagrams up to two-loop order are

\smallskip
\centerline{\includegraphics[width=\columnwidth]{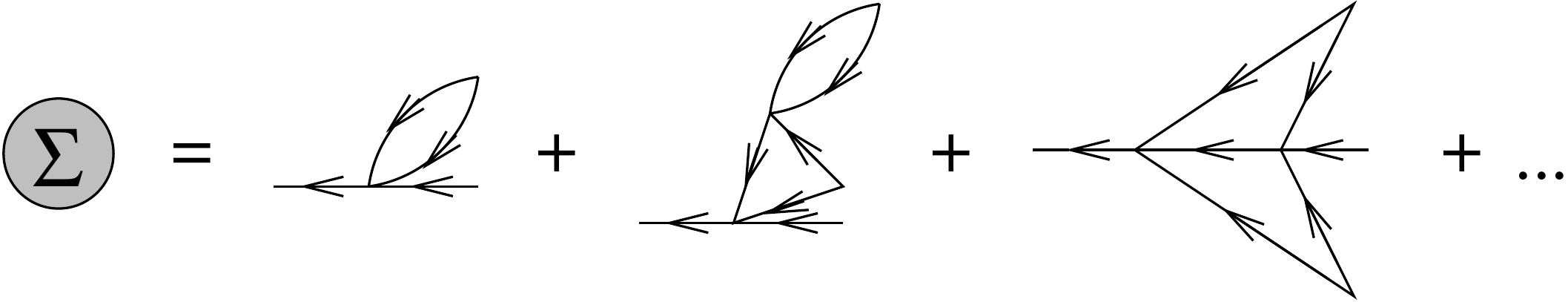}}

With the abbreviation $\Delta(q) = D q^a (r + q^2)$, the corresponding explicit
analytical expressions read: 
\begin{eqnarray} 
  &&\Gamma^{(1,1)}(q,\omega) = i \omega + D q^a \, \biggl[ r + q^2 \nonumber \\
  &&\quad + \frac{n+2}{6} \, u \int_k \frac{1}{r+k^2} - \Bigl( \frac{n+2}{6} 
  \, u \Bigr)^2 \int_k \frac{1}{r+k^2} \nonumber \\ 
  &&\quad \int_{k'} \frac{1}{(r+{k'}^2)^2} - \frac{n+2}{18} \, u^2 \int_k 
  \frac{1}{r+k^2} \nonumber \\
  &&\qquad \int_{k'} \frac{1}{r+{k'}^2} \, \frac{1}{r+(q-k-k')^2} 
\label{gamm11} \\ 
  &&\quad \left( 1 - \frac{i \omega}{i \omega + \Delta(k) + \Delta(k') + 
  \Delta(q-k-k')} \right) \biggr] \ . \nonumber
\end{eqnarray}

The renormalized noise vertex is represented by the vertex function 
$\Gamma^{(2,0)}(q,\omega)$; the first non-vanishing fluctuation correction
appears at two-loop order:

\smallskip
\centerline{\includegraphics[width=0.28\columnwidth]{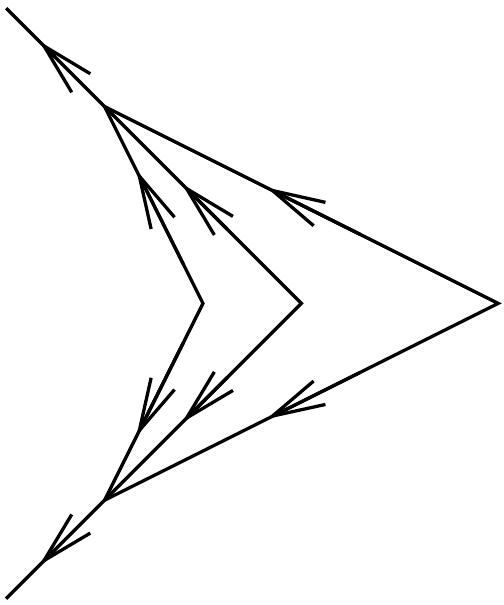}}
\smallskip

which translates to
\begin{eqnarray}
  &&\Gamma^{(2,0)}(q,\omega) = - 2 D q^a \, \biggl[ 1 + D q^a \, \frac{n+2}{18}
  \, u^2 \nonumber \\
  &&\quad \int_k \frac{1}{r+k^2} \int_{k'} \frac{1}{r+{k'}^2} 
  \frac{1}{r+(q-k-k')^2} \nonumber \\
  &&{\rm Re} \, \frac{1}{i \omega + \Delta(k) + \Delta(k') + \Delta(q-k-k')} 
  \biggr] \ .
\label{gamm20}
\end{eqnarray}
Finally, we shall require the renormalized relaxation vertex to one-loop order,
\begin{eqnarray} 
  &&\Gamma^{(1,3)}(-3{\underline k}/2;\{ {\underline k}/2 \}) = 
  D \Bigl( \frac{3q}{2} \Bigr)^a u \nonumber \\
  &&\quad \biggl[ 1 - \frac{n+8}{6} \, u \int_k \frac{1}{r+k^2} \, 
  \frac{1}{r+(q-k)^2} \nonumber \\
  &&\qquad\quad \left( 1 - \frac{i \omega}{i \omega + \Delta(k) + \Delta(q-k)} 
  \right) \biggr] \ ,
\label{gamm13}
\end{eqnarray} 
evaluated at equal incoming external wave vectors and frequencies 
${\underline k} = (q,\omega)$.

\subsection{Critical dynamics of the relaxational models}

We may now proceed with the perturbative renormalization of the purely 
relaxational models A and B, generalizing the methods outlined in Chap.~3 to 
the dynamical action (\ref{resfab}).
The quadratic UV divergence (near the upper critical dimension $d_c = 4$) in 
(\ref{gamm11}) is taken care of by an appropriate {\em additive 
renormalization}; since we are concerned with near-equilibrium kinetics here, 
and $\chi(q,\omega = 0) = \chi(q)$, the result is precisely the $T_c$ shift 
(\ref{1lptcs}) evaluated in the static theory.
In addition to (\ref{fldren}) and (\ref{cplren}), we need two new 
{\em multiplicative renormalization} factors associated with the response
fields and the relaxation rate,
\begin{eqnarray} 
  &&{\widetilde S}_R^\alpha = Z_{\widetilde S}^{1/2} {\widetilde S}^\alpha\ , \
  D_R = Z_D \, D \ ; 
\label{dynren} \\
  &&\Rightarrow \ \Gamma_R^{({\widetilde N},N)} = 
  Z_{\widetilde S}^{-{\widetilde N}/2} Z_S^{-N/2}\, \Gamma^{({\widetilde N},N)}
  \ . \nonumber
\end{eqnarray}
As a consequence of the FDT (\ref{corver}), which must hold for both the
unrenormalized and renormalized vertex functions, these $Z$ factors are 
{\em not} independent in thermal equilibrium, but connected via
$Z_D = \left( Z_S / Z_{\widetilde S} \right)^{1/2}$.

For model A with non-conserved order parameter ($a = 0$), extracting the UV
poles in the minimal subtraction scheme from $\Gamma^{(2,0)}_R(0,0)$ or 
$\frac{\partial}{\partial i \omega} \Gamma^{(1,1)}_R(0,\omega)$ yields   
\begin{equation} 
  Z_D = 1 - \frac{n + 2}{144} \, \Bigl( 6 \ln \frac43 - 1 \Bigr) \, 
  \frac{u_R^2}{\epsilon} \ . 
\label{2lpztd}
\end{equation} 
For model B with conserved order parameter ($a = 2$), on the other hand, the
momentum dependence $\propto q^2$ of the relaxation vertex implies that to 
{\em all orders} in the perturbation expansion
\begin{eqnarray}
  &&\Gamma^{(1,1)}(q=0,\omega) = i \omega \ , \nonumber \\ 
  &&\partial_{q^2} \, \Gamma^{(2,0)}(q,\omega) \big\vert_{q=0} = - 2 D \ ,
\label{modelb}
\end{eqnarray}
whence we arrive at the exact relations $Z_{\widetilde S} \, Z_S = 1$ and 
$Z_D = Z_S$.
The conservation law thus allows us to reduce the dynamical multiplicative to
static renormalizations.

With (\ref{dynren}) taken into account, the {\em renormalization group 
equation} for the renormalized vertex functions 
$\Gamma_R^{({\widetilde N},N)}(\mu,D,\tau_R,u_R)$ becomes in analogy with
(\ref{rgeqvf}):
\begin{eqnarray} 
  &&\biggl[ \mu \frac{\partial}{\partial \mu} + 
  \frac{{\widetilde N} \gamma_{\widetilde S} + N \gamma_S}{2} + 
  \gamma_D \, D_R \, \frac{\partial}{\partial D_R} \nonumber \\
  &&\qquad\ + \gamma_\tau \, \tau_R \, \frac{\partial}{\partial \tau_R} + 
  \beta_u \, \frac{\partial}{\partial u_R} \biggr] \, 
  \Gamma_R^{({\widetilde N},N)} = 0 \, ,
\label{dynrge}
\end{eqnarray}
with the static {\em RG beta function} (\ref{betafu}) and {\em Wilson's flow 
functions} (\ref{gamfld}), (\ref{gamtau}), supplemented with
\begin{eqnarray}
  &&\gamma_{\widetilde S} = \mu \, \frac{\partial}{\partial \mu} \Big\vert_0 \,
  \ln Z_{\widetilde S} \ , 
\label{gamres} \\
  &&\gamma_D = \mu \, \frac{\partial}{\partial \mu}\Big\vert_0 \, \ln
  \frac{D_R}{D} = \frac{\gamma_S - \gamma_{\widetilde S}}{2}
\label{gamrel}
\end{eqnarray}  
owing to the FDT.
Employing characteristics $\mu \to \mu \, \ell $ to solve the Gell-Mann--Low RG
equation (\ref{dynrge}), one has, in addition to (\ref{runcpl}),
\begin{equation}
  \ell \, \frac{d{\widetilde D}(\ell)}{d\ell} = {\widetilde D}(\ell) \, 
  \gamma_D(\ell) \ , 
\label{runrel}
\end{equation}
with ${\widetilde D}(1) = D_R$.

For the {\em dynamic susceptibility} near an infrared-stable RG fixed point, 
the static scaling law (\ref{gensus}) generalizes to
\begin{eqnarray}
  &&\chi_R(\tau_R,q,\omega)^{-1} \approx \mu^2 \ell^{2 + \gamma_S^*} \nonumber
  \\ 
  &&\ {\hat \chi}_R \Bigl( \tau_R \, \ell^{\gamma_\tau^*}, u^*, 
  \frac{q}{\mu \, \ell}, \frac{\omega}{D_R \mu^{2+a} \, 
  \ell^{2 + a + \gamma_D^*}} \Bigr)^{-1} ,
\label{gndsus}
\end{eqnarray} 
which allows us to identify the static {\em critical exponents} as before, 
$\eta = - \gamma_S^*$, and $\nu = - 1 / \gamma_\tau^*$; and in addition 
$z = 2 + a + \gamma_D^*$ for the {\em dynamic critical exponent}.
From the explicit two-loop result (\ref{2lpztd}) one thus obtains for the
$O(n)$-symmetric model A by inserting the Heisenberg fixed point (\ref{heisfp})
to order $\epsilon^2$ in the $4-\epsilon$ expansion
\begin{equation}  
  {\rm model \ A:} \ z = 2 + c \, \eta \ , \ 
  c = 6 \ln \frac43 - 1 + O(\epsilon) \ . \
\label{mdazex}
\end{equation}
Yet if the order parameter is conserved, one has $\gamma_D^* = \gamma_S^*$, 
which implies the {\em exact scaling relation}
\begin{equation}
  {\rm model \ B:} \ z = 4 - \eta \ .
\label{mdbzex}
\end{equation}

\subsection{Critical dynamics of isotropic ferromagnets}

So far we have only considered purely relaxational, dissipative kinetics.
Often, however, the Langevin description of critical dynamics needs to take 
into account {\em reversible} systematic forces contributing to $F[S]$ in 
(\ref{langen}).
The Langevin dynamics of {\em isotropic ferromagnets} provides a prominent 
example \cite{Frey94}, \cite[Chap.~6]{Tauberxx}.
The order parameter here is a three-component vector field, namely the
magnetization density $S^\alpha(x,t)$, which represent the coarse-grained
mesoscopic counterpart of the microscopic local Heisenberg spins, also the
generators of the rotation group $O(3)$.
From the spin operator commutation relation $\bigl[ S^\alpha , S^\beta \bigr] =
i \hbar \sum_{\gamma=1}^3 \epsilon^{\alpha \beta \gamma} S^\gamma$ and 
Heisenberg's equation of motion, or their corresponding classical counterparts
with commutators replaced by Poisson brackets, one readily obtains a spin 
precession term in the dynamics, in addition to the diffusive relaxation of the
conserved magnetization density, and conserved stochastic noise:
\begin{eqnarray}  
  &&\frac{\partial {\vec S}(x,t)}{\partial t} = - g \, {\vec S}(x,t) \times
  \frac{\delta {\cal H}[{\vec S}]}{\delta {\vec S}(x,t)} \nonumber \\
  &&\qquad\qquad + D \nabla^2 \,   
  \frac{\delta {\cal H}[{\vec S}]}{\delta {\vec S}(x,t)} + {\vec \zeta}(x,t)\ ,
\label{mdjlan} \\ 
  &&\big\langle \zeta^\alpha(x,t) \, \zeta^\beta(x',t') \big\rangle = - 2 D 
  \, k_{\rm B} T \nonumber \\ 
  &&\qquad\quad \nabla^2 \delta(x-x') \delta(t-t') \delta^{\alpha \beta} \ .
\label{mdjnoi}
\end{eqnarray}  
The coupled Langevin equations (\ref{mdjlan}) with the three-component LGW
Hamiltonian (\ref{lgwhmn}) and noise correlator (\ref{mdjnoi}) define the 
{\em model J} dynamic universality class.

The associated Janssen--De~Dominicis response functional comprises the model B 
terms, (\ref{resfab}) with $a = 2$, and the additional reversible 
{\em mode-coupling vertex},
\begin{eqnarray}
  &&{\cal A}_J[{\widetilde S},S] = - g \! \int \! d^dx \! \int \! dt 
  \sum_{\alpha,\beta,\gamma} \epsilon^{\alpha \beta \gamma} \,  
  {\widetilde S}^\alpha\nonumber \\
  &&\qquad\qquad\qquad\qquad\ S^\beta \left( \nabla^2 S^\gamma + h^\gamma 
  \right) \ .
\label{resmdj}
\end{eqnarray}
It is diagrammatically represented by a wave vector-dependent three-point 
vertex:

\smallskip
\centerline{\includegraphics[width=0.55\columnwidth]{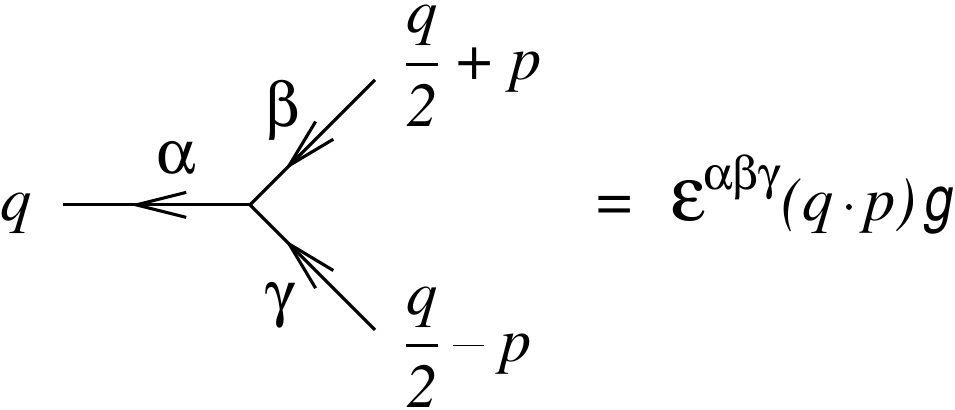}}
\smallskip

Straightforward power counting gives $[g] = \mu^{3-d/2}$ for the mode coupling
strength, which therefore becomes marginal at the {\em dynamical critical 
dimension} $d_c' = 6$, and irrelevant for $d > d_c'$. 
In principle, theories with competing upper critical dimension pose interesting
non-trivial technical problems. 
Recall, however, that we are here considering near-equilibrium dynamics; thus
the static critical properties completely decouple from the system's dynamics.
Indeed, the dynamical critical exponent can be determined exactly from the
underlying rotational symmetry as follows:
We first note that according to (\ref{mdjlan}) an external field $h^\gamma$ 
induces a rotation of the magnetization density vector:   
$\bigl\langle S^\alpha(x,t) \bigr\rangle_h = g \int_0^t \! dt' \sum_\beta 
\epsilon^{\alpha \beta \gamma} \bigl\langle S^\beta(x,t') \bigr\rangle_h \, 
h^\gamma(t)$.
Thus, we obtain an exact identity linking the {\em nonlinear susceptibility} 
$R^{\alpha;\beta\gamma} = \delta^2 \langle S^\alpha \rangle / \delta h^\beta \,
\delta h^\gamma \vert_{h=0}$ with the linear order parameter response function:
\begin{eqnarray} 
  &&\int \! d^dx' \, R^{\alpha;\beta\gamma}(x,t;x-x',t-t') = \nonumber \\
  &&\qquad\ g \, \epsilon^{\alpha \beta \gamma} \, \chi^{\beta\beta}(x,t) \,
  \Theta(t) \, \Theta(t-t') \ .
\label{nlinrs}
\end{eqnarray}

This identity provides crucial information for the renormalization of the UV
singularities.
In addition to (\ref{dynren}), we define the renormalized dimensionless mode
coupling strength
\begin{equation}
  g_R^2 = Z_g \, g^2 \, B_d \, \mu^{d-6} \ ,
\label{mcpren}
\end{equation}
where $B_d = \Gamma(4 - d/2) / 2^d \, d \, \pi^{d/2}$.
Just as for model B, one has to all orders in the perturbation expansion
$\Gamma^{(1,1)}(q=0,\omega) = i \omega$, whence again 
$Z_{\widetilde S} \, Z_S = 1$.
Since (\ref{nlinrs}) must hold for the renormalized susceptibilities as well,
one then infers that $Z_g = Z_S$.
Inspection of diagrams shows that the {\em effective coupling} in the dynamic 
perturbation expansion is $f = g^2 / D^2$.
The associated RG beta function becomes
\begin{equation}
  \beta_f = \mu \, \frac{\partial}{\partial \mu} \Big\vert_0 f_R = 
  f_R \left( d - 6 + \gamma_S - 2 \, \gamma_D \right) \ .
\label{betafj}
\end{equation} 
Consequently, at {\em any non-trivial}, stable RG fixed point 
$0 < f^* < \infty$ the terms in the bracket must cancel each other:
$2 \, \gamma_D^* = d - 6 + \gamma_S^* $ to {\em all orders} in the perturbation
series with respect to $f_R$.
Rotation invariance and the conservation law thus fix the dynamic critical 
exponent in dimensions $d < d_c' = 6$ to be
\begin{equation}
  {\rm model \ J:} \ z = 4 + \gamma_D^* = \frac{d + 2 - \eta}{2} \ .
\label{mdjzex}
\end{equation}
Indeed, an explicit one-loop calculation yields 
$\gamma_D = - f_R + O(u_R^2,f_R^2)$, which leads to the non-trivial model J RG 
fixed point $f_J^* = \frac{\varepsilon}{2} + O(\varepsilon^2)$, where 
$\varepsilon  = 6 - d$.
Note that $\eta = 0$ for $d > d_c = 4$, and $z = 4$ for $d > d_c' = 6$.
Since the mode-coupling vertex does not contribute genuinely new IR 
singularities, dynamic scaling functions for isotropic ferromagnets and related
models can be computed to exquisit precision by means of the {\em mode-coupling
approximation}, which essentially amounts to a self-consistent one-loop theory
for the propagators, ignoring vertex corrections \cite{Frey94}.

\subsection{Driven diffusive systems}

This chapter on the application of field-theoretic RG methods to non-linear 
stochastic Langevin dynamics concludes with two paradigmatic non-equilibrium 
model systems that display {\em generic scale invariance} and a continuous 
phase transition, respectively.
Both are driven lattice gases \cite{Schmittmann95} consisting of particles that
propagate via nearest-neighbor hopping which is biased along a specified `drive'
direction, and is subject to an exclusion constraint, i.e., only at most a
single particle is allowed on each lattice site.
If the system is set up with periodic boundary conditions, the biased diffusion
generates a non-vanishing stationary mean particle current. 
At long times, the kinetics thus reaches a {\em non-equilibrium steady state}
which turns out to be governed by {\em algebraic} rather than exponential 
temporal correlations.
If in addition nearest-neighbor attractive interactions are included, the 
system displays a genuine non-equilibrium continuous phase transition in 
dimensions $d \geq 2$, from a disordered phase to an ordered state 
characterized by phase separation into low- and high-density regions, with the 
phase boundary oriented parallel to the drive and particle current.
As the hopping bias vanishes, the phase transition is naturally described by 
the $d$-dimensional ferromagnetic equilibrium Ising model, since one may map 
the occupation numbers $n_i = 0,1$ to binary spin variables 
$\sigma_i = 2 n_i - 1 = \mp 1$.

We first consider the driven lattice gas with pure exclusion interactions, in
one dimension also called {\em ``asymmetric exclusion process''}.
In order to construct a coarse-grained description for the non-equilibrium 
steady state of this system of particles with conserved density $\rho(x,t)$ 
and hard-core repulsion, driven along the `$\parallel$' spatial direction on a 
$d$-dimensional lattice, one first writes down the {\em continuity equation} 
\begin{equation}
  \frac{\partial}{\partial t} \, S(x,t) + {\vec \nabla} \cdot {\vec J}(x,t) = 0
  \ ,
\label{ddscon}
\end{equation}
where the scalar field $S(x,t) = 2 \rho(x,t) - 1$ represents a local 
magnetization in the spin language, whose mean remains fixed at 
$\langle S(x,t) \rangle = 0$, or $\langle \rho(x,t) \rangle = \frac12$.
Next the current density must be specified; in the $d_\perp$-dimensional 
transverse sector ($d_\perp = d - 1$), one may simply assert a noisy 
diffusion current, whereas along the drive, the bias and exclusion are crucial:
$J_\parallel = - c \, D \, \nabla_\parallel S + 2 D g \rho (1 - \rho) + \zeta$,
where $c$ measures the ratio of diffusivities parallel and transverse to the 
net current.
In the comoving reference frame where $\langle J_\parallel(x,t) \rangle = 0$,
therefore
\begin{eqnarray}
  &&{\vec J}_\perp(x,t) = - D \, {\vec \nabla}_\perp S(x,t) + {\vec \eta}(x,t)
  \ ,
\label{ddscur} \\
  &&J_\parallel(x,t) = - c D \, \nabla_\parallel S(x,t) - \frac12 \, D g \, 
  S(x,t)^2 \nonumber \\
  &&\qquad\qquad\qquad + \zeta(x,t) \ , \nonumber
\end{eqnarray}
with $\langle \eta_i \rangle = 0 = \langle \zeta \rangle$, and the noise 
correlations
\begin{eqnarray} 
   &&\langle \eta_i(x,t) \, \eta_j(x',t') \rangle = \nonumber \\ 
   &&\qquad\qquad 2 D \, \delta(x-x') \delta(t-t') \delta_{ij} \ , 
\label{ddsnoi} \\ 
   &&\langle \zeta(x,t) \, \zeta(x',t') \rangle = 
   2 D {\tilde c} \, \delta(x-x') \delta(t-t') \, . \nonumber
\end{eqnarray} 
It is important to realize that Einstein's relations which connect the noise 
strengths and the relaxation rates need {\em not} be satisfied in the 
non-equilibrium steady state.
Through straightforward rescaling of the field $S(x,t)$, one may however 
formally enforce this connection in the transverse sector, say; the deviation 
from the Einstein relation in the parallel direction is then encoded in 
(\ref{ddscur}) and (\ref{ddsnoi}) through the ratio 
$0 < w = {\tilde c} / c \not= 1$.

The Janssen--De~Dominicis response functional for this driven diffusive system 
becomes
\begin{eqnarray} 
  &&{\cal A}[{\widetilde S},S] = \!\! \int \!\! d^dx \!\! \int \!\! dt \,   
  {\widetilde S} \, \Bigl[ \partial_t S - D \left( \nabla_\perp^2  
  + c \nabla_\parallel^2 \right) S \nonumber \\ 
  &&\qquad + D \left( \nabla_\perp^2  
  + {\tilde c} \, \nabla_\parallel^2 \right) {\widetilde S} 
  - \frac{D \, g}{2} \, \nabla_\parallel S^2 \Bigr] \, ;
\label{ddsact} 
\end{eqnarray}
the action (\ref{ddsact}) represents a ``massless'' field theory, which hence 
displays {\em generic scale invariance} with no specific tuning of any control 
parameters required.
One clearly has to allow for {\em anisotropic scaling} behavior owing to the
very different dynamics parallel to the hopping bias; for example, 
(\ref{dynscr}) for the dynamic response function needs to be generalized to
\begin{equation}
  \chi(q_\perp,q_\parallel,\omega) = |q_\perp|^{-2+\eta} \, {\hat \chi}\left( 
  \frac{q_\parallel}{|q_\perp|^{1 + \Delta}}, \frac{\omega}{|q_\perp|^z} 
  \right) ,
\label{ddssus}
\end{equation}
where $\Delta$ denotes the {\em anisotropy exponent} ($\Delta = 0$ in the 
mean-field approximation).

Following the renormalization procedures in the previous subsections, one first
realizes that the drive generates a three-point vertex 
$\propto g \, i q_\parallel$; consequently no transverse fluctuations are 
affected by this non-linearity, and $Z_{\widetilde S} = Z_S = Z_D = 1$ {\em to 
all orders} in the perturbation expansion in $g$.
The absence of any transverse propagator renormalization immediately implies 
that $\eta = 0$ and $z = 2$ in (\ref{ddssus}), which leaves merely the value of
$\Delta$ to be determined. 
In a similar manner as for model J, one may in fact compute this exponent
{\em exactly}; to this end, one observes that a generalized {\em Galilean 
transformation}
\begin{eqnarray}
  &&S(x_\perp,x_\parallel,t) \to 
\label{galtra} \\
  &&\qquad S'(x_\perp',x_\parallel',t') = S(x_\perp,x_\parallel - D g v \, t,t)
  - v \nonumber
\end{eqnarray}  
leaves the Langevin equation (\ref{ddscon}), (\ref{ddscur}) or equivalent 
action (\ref{ddsact}) invariant.
Thus the (arbitrary) speed $v$ must scale as the order parameter field $S$, 
and since $Z_D = 1 = Z_S$, neither can the coupling $g$ be altered by 
fluctuations, or (\ref{galtra}) would be violated for the renormalized theory. 
Hence $Z_g = 1$ as well, and the only remaining non-trivial renormalizations 
are those for the dimensionless parameters $c_R = Z_c \, c$ and 
${\tilde c}_R = Z_{\tilde c} \, {\tilde c}$. 
  
An explicit one-loop calculation establishes the existence of an IR-stable RG
fixed point for the coupling
\begin{equation} 
  v = g^2 / c^{3/2} \ , \quad v_R = Z_c^{3/2} \, v \, C_d \, \mu^{d-2}\ ,
\label{ddscpl} 
\end{equation}
with $C_d = \Gamma(2-d/2) / 2^{d-1} \pi^{d/2}$, and identifying $d_c = 2$ as 
the upper critical dimension for this problem.
Evaluating the one-loop fluctuation corrections to the longitudinal propagator,
one finds
\begin{eqnarray}
  &&\gamma_c = - \frac{v_R}{16} \left( 3 + w_R \right) \ , 
\label{ddsgam} \\
  &&\gamma_{\tilde c} = - \frac{v_R}{32} \, 
  \left( 3 w_R^{-1} + 2 + 3 w_R \right) \nonumber
\end{eqnarray}
for the anomalous scaling dimensions of $c$ and ${\tilde c}$, or
\begin{eqnarray}
    &&\beta_w = w_R \left( \gamma_{\tilde c} - \gamma_c \right) \nonumber \\ 
    &&\quad\ = - \frac{v_R}{32} (w_R - 1) (w_R - 3) \ , 
\label{ddsbew} \\ 
    &&\beta_v = v_R \left( d - 2 - \frac32 \, \gamma_c \right) 
\label{ddsbev} 
\end{eqnarray} 
for the associated RG beta functions of the ratio $w = {\tilde c} / c$ and the
non-linear coupling $v$. 
At {\em any non-trivial} RG fixed point $0 < v^* < \infty$, (\ref{ddsbew}) 
implies that either $w_N^* = 3$ or $w_E^* = 1$, but the latter is obviously 
{\em stable}; in the asymptotic scale-invariant regime, the Einstein relation 
is evidently restored, and the system effectively equilibrated.
Moreover, in dimensions $d < d_c = 2$, (\ref{ddsbev}) leads to the {\em exact}
scaling exponents
\begin{equation}   
  \Delta = - \frac{\gamma_c^*}{2} = \frac{2 - d}{3} \ , \quad 
  z_\parallel = \frac{z}{1 + \Delta} = \frac{6}{5 - d} \ .
\label{ddsexp}
\end{equation}
At $d = 1$, specifically, one has $z_\parallel = \frac32$, which captures the
dynamic scaling for the asymmetric exclusion process.
In one dimension, the driven lattice gas with exclusion in fact maps onto the
{\em noisy Burgers equation} for equilibrium fluid hydrodynamics, and also to 
the {\em Kardar--Parisi--Zhang equation} for curvature-driven surface or 
interface growth.

We finally briefly summarize the RG analysis for the driven lattice gas with
conserved total density and attractive Ising interactions between the particles
(and ``holes'') at its critical point.
Since the system orders in stripes along the drive direction, only the 
{\em transverse} fluctuations become critical.
Therefore one must amend the response functional (\ref{ddsact}) with a 
higher-order gradient term and non-linearity akin to the scalar model B, see 
(\ref{resfab}) with $a = 2$; yet the noise terms too need only be retained in 
the transverse sector.
For this driven model B, the effective critical action thus becomes
\begin{eqnarray} 
  &&{\cal A}[{\widetilde S},S] \!= \!\! \int \!\! d^dx \!\! \int \!\! dt \,
  {\widetilde S} \, \Bigl[ \partial_t S - D \nabla_\perp^2 \left( r - 
  \nabla_\perp^2 \right) S \nonumber \\ 
  &&\ - D c \nabla_\parallel^2 S + D \left( \nabla_\perp^2 {\widetilde S} 
  - \frac{g}{2} \, \nabla_\parallel S^2 - \frac{u}{6} \, \nabla_\perp^2 S^3 
  \right) \Bigr] \ , \nonumber \\ &&
\label{crddsa}
\end{eqnarray}
and (\ref{ddssus}) is further generalized by adding a relevant temperature 
variable (now $\Delta = 1$ in mean-field theory):
\begin{eqnarray}
  &&\chi(\tau_\perp,q_\perp,q_\parallel,\omega) = |q_\perp|^{-2+\eta} \nonumber
  \\
  &&\qquad {\hat \chi}\left( \frac{\tau}{|q_\perp|^{1/\nu}}, 
  \frac{q_\parallel}{|q_\perp|^{1 + \Delta}}, \frac{\omega}{|q_\perp|^z} 
  \right) \ .
\label{crddss}
\end{eqnarray}

Straightforward power counting for the non-linear couplings yields  
$[g^2] = \mu^{5-d}$ and $[u] = \mu^{3-d}$; therefore the upper critical 
dimension is raised to $d_c = 5$ (compared to both the non-critical driven 
lattice gas and the equilibrium model B), and fluctuation corrections are 
dominated by the drive, while the static coupling $u$ is (dangerously)
{\em irrelevant} near $d_c$.
The three-point vertex $\propto g \, i q_\parallel$ again does not allow any
renormalizations in the transverse sector, whence 
$Z_{\widetilde S} = Z_S = Z_D = 1$; consequently $\eta = 0$, $\nu = \frac12$, and
$z = 4$ in (\ref{crddss}) to all orders in the 
perturbation series, leaving only the anisotropy exponent to be determined.
As before $Z_g = 1$ follows from Galilean invariance, imposing a simple 
structure for the RG beta function for the effective coupling (\ref{ddscpl}):
\begin{equation}
  \beta_v = v_R \left( d - 5 - \frac32 \, \gamma_c \right) \ .
\label{crddsv}
\end{equation}
In dimensions $d < d_c = 5$, at any non-trivial and finite RG fixed point, the
scaling exponents are thus forced to assume the values
\begin{equation}
    \Delta = 1 - \frac{\gamma_c^*}{2} = \frac{8-d}{3} \ , 
    \ z_\parallel = \frac{4}{1 + \Delta} = \frac{12}{11 - d} \ . \
\label{crddse}
\end{equation}
These last examples clearly demonstrate how the powerful field-theoretic RG 
approach can help to exploit the basic symmetries for a given problem, 
allowing to determine certain non-trivial scaling exponents {\em exactly}.

\section{Scale Invariance in Interacting Particle Systems}

This last chapter details how the stochastic kinetics of classical interacting 
(reacting) particle systems, defined through a {\em microscopic} master 
equation, can also be mapped onto a dynamical field theory in the continuum
limit \cite{Cardy97}--\cite{Tauber05}.
For at most binary reactions, one can thus {\em derive} a corresponding 
mescoscopic Langevin representation, typically with multiplicative noise terms.
Furthermore, RG tools may be applied to extract the infrared properties in 
scale-invariant systems, as will be exemplified for diffusion-limited 
annihilation processes \cite{Cardy97,Tauber05,Tauber07}, and for 
non-equilibrium phase transitions from active to inactive, absorbing states, 
where all stochastic fluctuations cease \cite{Tauber05}--\cite{Janssen05}. 
Stochastic models in population dynamics and ecology are naturally formulated 
in a chemical reaction language, and hence amenable to these field-theoretic
tools \cite{Mobilia07,Tauber11}.

\subsection{Chemical reactions and population dynamics}

Let us thus consider (classical) particles of various species $A,B,\ldots$ on a
$d$-dimensional lattice that propagate by hops to nearest-neighbor sites, and 
either spontaneously decay or produce offspring, and/or upon encounter with 
other particles, undergo certain ``chemical'' reactions with prescribed rates. 
Our goal is to systematically construct a continuum description of such
stochastic particle systems that however faithfully encodes the associated
{\em intrinsic reaction noise}, and consequently allows us to properly address
the effects of statistical fluctuations and spatio-temporal correlations 
\cite{Cardy97,Tauber05,Tauber07}.

To set the stage, we introduce three characteristic examples that we shall 
explore in more detail below.
First, we address the general single-species {\em irreversible annihilation} 
reaction $k \, A \to m \, A$ with integers $m < k$, and rate $\lambda_k$.
Assuming the system to be well-mixed, one may neglect spatial variations
and focus on the mean particle density $a(t) = \langle a(x,t) \rangle$.
Ignoring in addition any non-trivial correlations, one can write down the
{\em rate equation} for this stochastic process, which in essence thus 
constitutes the simplest mean-field approximation:
\begin{equation}   
  \frac{\partial a(t)}{\partial t} = - (k - m) \, \lambda_k \, a(t)^k \ .
\label{annreq} 
\end{equation}
For $k = 1$ (and $m = 0$), this just describes spontaneous exponential decay, 
$a(t) = a(0) \, e^{- \lambda_1 \, t}$; for $k \geq 2$, (\ref{annreq}) is easily
integrated with the result
\begin{equation}
  a(t) = \bigl[ a(0)^{1-k} + (k-m) (k-1) \, \lambda_k \, t \bigr]^{-1/(k-1)} .
\label{mfasol}
\end{equation}
For the $k$th order annihilation reaction, the particle density decays 
algebraically $a(t) \sim (\lambda_k \, t)^{-1/(k-1)}$ at long times 
$t \gg \lambda_k^{-1}$, with an amplitude that does not even depend on the 
initial density $a(0)$ anymore.
The replacement of an exponential decay by a power law signals scale invariance
and indicates the potential importance of fluctuations and correlations.
Indeed, the annihilation kinetics generates particle {\em anti-correlations}, 
whence the long-time kinetics is dominated by the ensuing {\em depletion zones}
in low dimensions $d \leq d_c(k)$ that need to be traversed by any potentially 
reacting particles.
As a consequence, one obtains {\em slower} decay power laws than predicted by
the mean-field rate equation (\ref{mfasol}). 
   
Next, we allow {\em competing reactions}, namely decay $A \to \emptyset$ (the 
empty state) with rate $\kappa$, and the reversible process 
$A \rightleftharpoons A + A$ with forward / backward rates $\sigma$ and 
$\lambda$, respectively.
Again, we begin with an analysis of the rate equation for this set of 
reactions,
\begin{equation}
  \frac{\partial a(t)}{\partial t} = (\sigma - \kappa) \, a(t) 
  - \lambda \, a(t)^2 \ ,
\label{dprreq}
\end{equation} 
which obviously predicts a {\em continuous non-equilibrium phase transition} at
$\sigma_c = \kappa$: 
For $\sigma > \kappa$, the mean particle density approaches a finite value, 
$a(t \to \infty) \to a_\infty = (\sigma - \kappa) / \lambda$.
One refers to this state as an {\em active} phase; ongoing reactions cause the 
particle number to fluctuate about its average.
On the other hand, for $\sigma < \kappa$, the density can only decrease, whence
ultimately $a(t) \to 0$; in this {\em inactive} phase, all reaction processes
terminate since they all require the presence of a particle.
Such a state is therefore called {\em absorbing}: once reached, the stochastic 
dynamics cannot escape from it anymore.
Right at the critical point $\sigma = \kappa$, one recovers the long-time
algebraic decay of the pair annihilation process, 
$a(t) \sim (\lambda \, t)^{-1}$; this suggests the interpretation of 
(\ref{mfasol}) as a {\em critical} power law induced by the precise 
cancellation of the contributions from first-order reactions that enter 
linearly proportional to the particle concentration.
The obvious issues to be addressed by a more refined theoretical treatment are:
How can internal reaction noise and correlations be systematically 
incorporated? 
What is the upper critical dimension $d_c$ below which fluctuations crucially 
alter the mean-field power laws? 
Can certain {\em universality classes} be identified, and the associated 
critical exponents be computed, at least perturbatively in a dimensional
expansion near $d_c$? 

Finally, let us address a prominent textbook example from population dynamics,
namely the classic {\em Lotka--Volterra predator-prey competition}.
Invoking the stochastic chemical reaction framework, this model is defined via 
spontaneous death $A \to \emptyset$ (rate $\kappa$) and birth $B \to B + B$
(rate $\sigma$) processes for the ``predators'' $A$ and ``prey'' $B$; absent 
any interactions between these two species, the predators must go extinct, 
while the prey population explodes exponentially.
Interesting species competition and potentially coexistence is created by the 
binary {\em predation} reaction $A + B \to A + A$ (with rate $\lambda$).
The associated coupled rate equations for the presumed homogeneous population 
densities read
\begin{eqnarray}
  &&\frac{\partial a(t)}{\partial t} = \lambda \, a(t) b(t)  - \kappa \, a(t) 
  \ , \nonumber \\   
  &&\frac{\partial b(t)}{\partial t} = \sigma \, b(t) - \lambda \, a(t) b(t)\ .
\label{lvmreq} 
\end{eqnarray}
In this mean-field approximation, one easily confirms the existence of a 
conserved first integral for the ordinary differential equations 
(\ref{lvmreq}): 
The quantity 
$K(t) = \lambda [a(t) + b(t)] - \ln [a(t)^\sigma b(t)^\kappa] = K(0)$ remains 
unchanged under the temporal evolution.
Consequently, the mean-field trajectories are closed orbits in the phase space
spanned by the population densities, and the dynamics is characterized by
regular {\em population oscillations}, determined by the {\em initial} state.
This is clearly not a biologically realistic feature, and indeed represents an 
artifact of the implicit mean-field factorization for the non-linear predation 
process.
Upon including the internal reaction noise and spatial degrees of freedom with
diffusively spreading particles, as, e.g., in individual-based Monte Carlo
simulations, one in fact observes striking {\em ``pursuit and evasion'' waves} 
in the species coexistence phase that generate complex dynamical patterns, 
locally discernible as {\em erratic} population oscillations which ultimately
become overdamped in finite systems.
Moreover, if the local ``carrying capacity'' is finite, i.e., only a certain
maximum number of particles may occupy each lattice site, there emerges a 
predator {\em extinction threshold} which indicates a continuous phase 
transition to an absorbing state, namely the lattice filled with prey.
Stochastic fluctuations as well as reaction-induced noise and correlations are 
thus crucial ingredients to properly describe the large-scale features of 
spatially extended Lotka--Volterra systems even and especially far away from 
the extinction threshold (for recent overviews, see 
Refs.~\cite{Mobilia07,Tauber11}).

\subsection{Coherent-state path integral for master equations}

In the following, reacting particle systems on a $d$-dimensional lattice shall
be {\em defined} through the associated chemical {\em master equation} 
governing a Markovian stochastic process with prescribed, time-independent 
transition rates.
Any possible configuration at time $t$ of the stochastic dynamics is then
uniquely characterized by a list of the  integer occupation numbers 
$n_i = 0,1,2,\ldots$ for each particle species at sites $i$.
The master equation governs the temporal evolution of the corresponding 
probability distribution $P(\{ n_i \};t)$ through a {\em balance} of gain and 
loss terms induced by the reaction processes; for example, for the binary 
reactions $A + A \to \emptyset$ and $A + A \to A$ with rates $\lambda$ and 
$\lambda'$: 
\begin{eqnarray} 
  &&\!\!\!\!\!\!\!\!\!\!\frac{\partial}{\partial t} \, P(n_i;t) = 
  \lambda \, (n_i + 2) \, (n_i + 1) \, P(\ldots,n_i+2,\ldots;t) \nonumber \\ 
  &&\quad + \lambda' \, (n_i + 1) \, n_i \, P(\ldots,n_i+1,\ldots;t) 
\label{masteq} \\ 
  &&\quad - (\lambda + \lambda') \, n_i \, (n_i-1) \, P(\ldots,n_i,\ldots;t)
  \ , \nonumber
\end{eqnarray} 
with, say, an uncorrelated {\em initial Poisson distribution} $P(\{ n_i \},0) = 
\prod_i \left( {\bar n}_0^{n_i} \, e^{-\bar n_0} / n_i ! \right)$. 

Since the dynamics merely consists of increasing or decreasing the particle
occupation numbers on each site, it calls for a representation through 
{\em second-quantized bosonic ladder operators}, at least if arbitrary many 
particles are permitted per site, with standard commutation relations 
$[a_i, a_j] = 0$, $[a_i,a_j^\dagger] = \delta_{ij}$, and an empty vacuum state 
$| 0 \rangle$ that is annihilated by all operators $a_i$, 
$a_i | 0 \rangle = 0$.
The Fock space of states $| \{ n_i \} \rangle$ with $n_i$ particles on sites 
$i$ is then constructed through multiple creation operators acting on the 
vacuum,
$| \{ n_i \} \rangle = \prod_i \bigl( a_i^\dagger \bigr)^{n_i} \, | 0 \rangle$;
note that a different normalization from standard many-particle quantum
mechanics has been implemented here.
Thus, $a_i \, |n_i \rangle = n_i \, |n_i-1 \rangle$ and
$a_i^\dagger \, |n_i \rangle = |n_i + 1 \rangle$, whence the states
$| \{ n_i \} \rangle$ are eigenstates of ${\hat n}_i = a_i^\dagger a_i$ with
eigenvalues $n_i$.
Next one defines the formal {\em state vector} 
$| \Phi(t) \rangle = \sum_{\{ n_i \}} P(\{ n_i \};t) \, | \{ n_i \} \rangle$,
whose temporal evolution is determined by the master equation (\ref{masteq}),
and may be written in terms of a time-independent quasi-Hamiltonian or 
Liouvillian $H$ that can be decomposed into a sum of local operators:
\begin{equation}
  \frac{\partial}{\partial t} \, | \Phi(t) \rangle = - H \, | \Phi(t) \rangle 
  \ , \ H = \sum_i H_i(a_i^\dagger,a_i) \ .
\label{masham}
\end{equation}
Note that (\ref{masham}) constitutes a {\em non-Hermitian imaginary-time 
Schr\"odinger equation}, with the formal solution 
$| \Phi(t) \rangle = \exp (- H t) \, | \Phi(0) \rangle$.

For example, the quasi-Hamiltonian in this {\em Doi--Peliti bosonic operator 
formulation} reads for diffusion-limited annihilation and coagulation reactions
\begin{eqnarray}
  &&H = D \sum_{<ij>} \bigl( a_i^\dagger - a_j^\dagger \bigr) \,
  \bigl( a_i - a_j \bigr) 
\label{annham} \\
  &&\quad - \sum_i \Bigl[ \lambda \, \bigl( 1 - {a_i^\dagger}^2 \bigr) \, a_i^2
  + \lambda' \bigl( 1 - a_i^\dagger \bigr) \, a_i^\dagger a_i^2 \Bigr] \, ,
  \nonumber
\end{eqnarray}
where the first line represents nearest-neighbor hopping, and the second
encodes the processes in (\ref{masteq}).
For each stochastic reaction, $H$ contains two contributions: the first one
directly reflects the physical {\em process} under considerations, i.e., 
annihilation and production of particles, whereas the second term carries 
information on the {\em order} of the reaction (which powers of the particle
concentrations enter the rate equations). 
In order to access the desired {\em statistical averages} with the 
time-dependent probability distribution $P(\{ n_i \};t)$, one needs the
{\em projection} state $\langle {\cal P} | = \langle 0 | \prod_i e^{a_i}$, with
$\langle {\cal P} | 0 \rangle = 1$; the mean value for any observable $F$, 
necessarily a function of all occupation numbers $n_i$, at time $t$ then
follows as
\begin{eqnarray}
  &&\langle F(t) \rangle = \sum_{\{ n_i \}} F(\{ n_i \}) \, P(\{ n_i \};t) 
  \nonumber \\
  &&\qquad\;\ = \langle {\cal P} | \, F(\{ a_i^\dagger \, a_i \}) \, | \Phi(t) 
  \rangle \ . 
\label{mastav}
\end{eqnarray}
{\em Probability conservation} implies that 
$1 = \langle {\cal P} | \Phi(t) \rangle = \langle {\cal P} | e^{- H \, t} |
\Phi(0) \rangle$, and therefore $\langle {\cal P} | H = 0$.
By means of $\bigl[ e^a , a^\dagger \bigr] = e^a$, one may commute the
product $e^{\sum_i a_i}$ through the quasi-Hamiltonian $H$, which effectively
results in the operator shifts $a_i^\dagger \to 1 + a_i^\dagger$.
$H$ must consequently vanish if all creation operators are replaced with $1$, 
$H_i(a_i^\dagger \to 1,a_i) = 0$.
In averages, one may thus also replace $a_i^\dagger a_i \to a_i$; e.g., for the
particle density one obtains simply $a(t) = \langle a_i \rangle$, while the 
two-point occupation number operator product becomes 
$a_i^\dagger a_i \, a_j^\dagger a_j \to a_i \delta_{ij} + a_i a_j$. 

Starting with the Hamiltonian (\ref{masham}), and based on the expectation
values (\ref{mastav}), one may invoke standard procedures from quantum 
many-particle theory to construct a {\em path integral representation} based on
{\em coherent states}, defined as eigenstates of the annihilation operators 
$a_i$ with arbitrary complex eigenvalues $\phi_i$:
$a_i \, |\phi_i \rangle = \phi_i \, |\phi_i \rangle$.
It is straightforward to confirm
\begin{eqnarray} 
  &&|\phi_i \rangle = \exp \, \Bigl( - \frac12 \, |\phi_i|^2 + \phi_i \, 
  a_i^\dagger \Bigr) \, | 0 \rangle \ , 
\label{cohstt} \\ 
  &&1 = \int \prod_i \frac{d^2 \phi_i}{\pi} \, |\{ \phi_i \} \rangle \, 
  \langle \{ \phi_i \}| \, .
\label{cohovc} 
\end{eqnarray}
The {\em closure relation} (\ref{cohovc}) demonstrates that the coherent states
for each site $i$ form an {\em overcomplete} basis of Fock space.
Upon splitting the time evolution into infinitesimal steps, and inserting 
(\ref{cohovc}) into (\ref{mastav}) with the formal solution of (\ref{masham}), 
one eventually arrives at
\begin{eqnarray} 
  &&\langle F(t) \rangle \propto \int \! \prod_i {\cal D}[\phi_i] \, 
  {\cal D}[\phi_i^*] \, F(\{ \phi_i \}) \, e^{- {\cal A}[\phi_i^*,\phi_i]} , 
  \nonumber \\ 
  &&{\cal A}[\phi_i^*,\phi_i] = \sum_i \Bigl[ - \phi_i(t_f) 
\label{masdpa} \\
  &&\qquad + \int_0^{t_f} \! dt \bigl[ \phi_i^* \, \partial_t \phi_i 
  + H(\phi_i^*,\phi_i) \bigr] \! - {\bar n}_0 \phi^*_i(0) \Bigr] . \nonumber
\end{eqnarray} 

In the end we take the {\em continuum limit} $\phi_i(t) \to a_0^d$ $\psi(x,t)$
(with lattice constant $a_0$), and $\phi_i^*(t) \to {\hat \psi}(x,t)$; for
diffusively propagating particles, the ensuing ``bulk'' action becomes
\begin{eqnarray} 
  &&{\cal A}[{\hat \psi},\psi] = \int \! d^dx \int_0^{t_f} \! dt \, \Bigl[  
  {\hat \psi} \left( \partial_t - D \, \nabla^2 \right) \psi \nonumber \\
  &&\qquad\qquad\qquad\qquad\qquad + {\cal H}_r \bigl( {\hat \psi}, \psi \bigr)
  \Bigr] \, .  
\label{masact}
\end{eqnarray}  
Here, ${\cal H}_r$ denotes the contributions stemming from the stochastic
reaction kinetics; e.g., for {\em pair annihilation and coagulation},
\begin{equation}
  {\cal H}_r \bigl( {\hat \psi},\psi \bigr) = - \lambda \bigl( 1 - 
  {\hat \psi}^2 \bigr) \psi^2 - \lambda' \bigl( 1 - {\hat \psi} \bigr) 
  {\hat \psi} \, \psi^2 \ .
\label{annact}
\end{equation}
Appropriate factors of $a_0$ were absorbed into the continuum diffusion
constant $D$ and reaction rates $\lambda$,$\lambda'$.
It is worthwhile emphasizing that the actions (\ref{masdpa}) based on a master 
equation should be viewed as {\em microscopic} stochastic field theories, which
may well require additional coarse-graining steps.
Yet the internal stochastic dynamics of the master equation is faithfully and 
consistently accounted for, since aside from the continuum limit no 
approximations have been invoked; specifically,  no assumptions on the form or 
strength of any noise terms have been made.
As exemplified next for the action (\ref{annact}), the Doi--Peliti 
coherent-state path integral representation of stochastic master equations may 
serve as a convenient starting point for systematic analytical approaches such 
as field-theoretic RG studies.

\subsection{Diffusion-limited annihilation processes}

Pair annihilation $A + A \to \emptyset$ or coagulation $A + A \to A$ represent
the perhaps simplest but non-trivial diffusion-limited reactions.
In order to reach beyond the mean-field rate equation approximation 
(\ref{mfasol}), we explore the corresponding Doi--Peliti field theory 
(\ref{masact}) with the specific reaction Hamiltonian (\ref{annact}).
First we note that the associated {\em classical field equations}   
$\delta {\cal A} / \delta \psi = 0 = \delta {\cal A} / \delta {\hat \psi}$ are
solved by ${\hat \psi} = 1$ (which just reflects probability conservation) and 
\begin{equation}  
  \frac{\partial \psi(x,t)}{\partial t} = D \, \nabla^2 \, \psi(x,t) - 
  (2 \lambda + \lambda') \, \psi(x,t)^2 \ ,
\label{andifr}
\end{equation}
i.e., the rate equation for the local density field $\psi(x,t)$ augmented by
diffusive spreading.
It is convenient to shift the conjugate field about the mean-field solution, 
${\hat \psi}(x,t) = 1 + {\widetilde \psi}(x,t)$, which turns the reactive
action into
\begin{equation} 
  {\cal H}_r \bigl( {\widetilde \psi},\psi \bigr) = (2 \lambda + \lambda') \,
  {\widetilde \psi} \, \psi^2 + (\lambda + \lambda') \, {\widetilde \psi}^2 \, 
  \psi^2 \ .
\label{anshac}
\end{equation}
Since the annihilation and coagulation processes $\propto \lambda, \lambda'$ 
generate the very same vertices, we conclude that aside from non-universal 
amplitudes, both diffusion-limited reactions should follow {\em identical} 
scaling behavior.
One may also formally interpret the ensuing field theory as a 
Janssen--De~Dominicis response functional (\ref{janded}) originating from a
``Langevin equation'' (\ref{andifr}) with added white noise $\zeta(x,t)$, whose
second moment (\ref{noigen}) is given by the functional 
$L[\psi] = - (\lambda + \lambda') \, \psi^2 < 0$.
This negative variance, which reflects the emerging {\em anti-}correlations for
surviving particles that are induced by the annihilation reactions, of course 
implies that a Langevin representation is not truly feasible for this 
stochastic process.
One must also keep in mind that the fields $\psi$ and ${\hat \psi}$ are 
complex-valued; indeed, the reaction noise can be recast as {\em ``imaginary'' 
multiplicative noise} $\propto i \psi(x,t) \, \zeta(x,t)$ in the associated 
stochastic differential equation.
  
As coagulation thus falls into the same universality class as annihilation, let
us more generally study the Doi--Peliti action for {\em $k$-particle 
annihilation} $k \, A \to \emptyset$,
\begin{eqnarray}
  &&{\cal A}[{\hat \psi},\psi] = \int \! d^dx \! \int \! dt \, \Bigl[ 
  {\hat \psi} \, \bigl( \partial_t - D \nabla^2 \bigr) \psi \nonumber \\
  &&\qquad\qquad\qquad\qquad\ - \lambda_k \, \bigl( 1 - {\hat \psi}^k \bigr) \,
  \psi^k \Bigr] \, .
\label{kannac}
\end{eqnarray} 
The corresponding mean-field rate equation (\ref{annreq}) predicts algebraic 
decay $a(t) \sim (\lambda_k \, t)^{-1/(k-1)}$ at long times. 
Since the field ${\hat \psi}$ appears to higher than quadratic power for 
$k \geq 3$, {\em no} (obvious) equivalent Langevin description is possible for 
triplet and higher-order annihilation reactions.  
With $[{\hat \psi}(x,t)] = 1$ and $[\psi(x,t)] = \mu^d$, as 
$\langle \psi(x,t) \rangle = a(t)$ is just the particle density, power counting
gives $[\lambda_k] = \mu^{2 - (k-1) d}$; the upper critical dimension thus is
$d_c(k) = 2/(k-1)$ for $k$th order annihilation, and one expects the mean-field
power laws (\ref{mfasol}) to be accurate for $k > 3$ in all physical dimensions
$d \geq 1$; for triplet reactions, one should encounter merely logarithmic 
corrections in one dimension, where genuine non-trivial exponents ensue only 
for pair annihilation.  
  
Even for $k = 2$, one cannot construct any loop graphs from the vertices in 
(\ref{kannac}) that would modify the massless diffusion propagator, implying 
that $\eta = 0$ and $z = 2$.
This leaves the task to determine the reaction {\em vertex renormalization}, 
which can also be achieved to {\em all orders} by summing the diagrammatic
geometric series (essentially a {\em Bethe--Salpeter equation}; here for 
$k = 3$):

\medskip
\centerline{\includegraphics[width=\columnwidth]{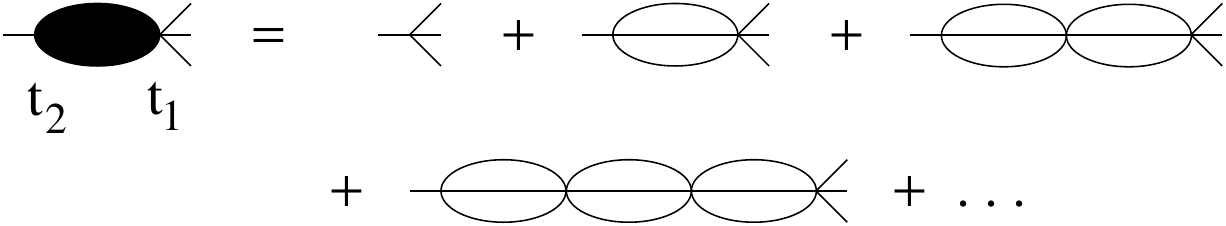}} 
\smallskip    

\noindent With the factor 
$B_{kd} = k! \, \Gamma(2 - d/d_c) \, d_c / k^{d/2} \, (4 \pi)^{d/d_c}$, one 
finds the renormalized reaction rate
\begin{eqnarray}
  &&g_R = Z_g \, \frac{\lambda}{D} \, B_{kd} \, \mu^{- 2 (1 - d/d_c)} \ , 
  \nonumber \\
  &&Z_g^{-1} = 1 + 
  \frac{\lambda \, B_{kd} \, \mu^{- 2 (1 - d/d_c)}}{D \, (d_c - d)} \ ,
\label{renanr}
\end{eqnarray}  
and the exact RG beta function and stable fixed point
\begin{equation}
  \beta_g = - \frac{2 g_R}{d_c} \left( d - d_c + g_R \right) \ , \quad  
  g^* = d_c - d \ .
\label{annrgf}
\end{equation}

Next we write down the Gell-Mann--Low RG equation for the particle density 
$a(t)$, applying the matching condition $(\mu \ell)^2 = 1 / D t$:
\begin{eqnarray}
  &&\left[ d + 2 D t \, \frac{\partial}{\partial (D \, t)} - d \, n_0 \,  
  \frac{\partial}{\partial n_0} + \beta_g \, \frac{\partial}{\partial g_R}
  \right] \nonumber \\
  &&\qquad\quad a(\mu,D,n_0,g_R,t) = 0 \ ,
\label{andrge}
\end{eqnarray}
with the solution
\begin{eqnarray}
  &&a(\mu,D,n_0,g_R,t) = \nonumber \\
  &&\qquad (D \mu^2 \, t)^{- d/2} \, {\hat a}\bigl( n_0 \, 
  (D \mu^2 \, t)^{d/2} , {\widetilde g}(t) \bigr) \ .
\label{andrgs}
\end{eqnarray}
The particle density at time $t$ naturally depends on its initial value $n_0$, 
clearly a {\em relevant} parameter in the RG sense. 
One therefore needs to establish through explicit calculation that the (tree 
level) scaling function ${\hat a}$ remains finite to {\em all orders} as 
$n_0 \to \infty$.
In the end, (\ref{andrgs}) yields for pair annihilation,
\begin{eqnarray} 
    k = 2: &&d < 2: \ a(t) \sim (D \, t)^{-d/2} \ , \nonumber \\ 
    &&d = 2: \ a(t) \sim (D \, t)^{-1} \ln (D \, t) \ , 
\label{k=2ann} \\ 
    &&d > 2: \ a(t) \sim (\lambda \, t)^{-1} \ ; \nonumber
\end{eqnarray}
while for the triplet reaction  
\begin{eqnarray} 
    k = 3: &&d = 1 \, : \ 
    a(t) \sim \left[ (D \, t)^{-1} \ln (D \, t) \right]^{1/2} \ , \nonumber \\
    &&d > 1: \ a(t) \sim (\lambda \, t)^{-1/2} \ . 
\label{k=3ann}
\end{eqnarray} 
At low dimensions $d \leq d_c(k) = 2/(k-1)$, the density decay is slowed down
as compared to the mean-field power laws by the emergence of depletion zones 
around the surviving particles.
Further annihilations require that the reactants traverse the diffusion length 
$L(t) \sim (D \, t)^{1/2}$, which sets the typical separation scale; the
corresponding density must then scale as $L(t)^{-d}$.
Beyond the upper critical dimension $d_c(k)$, the system becomes effectively
well-mixed, diffusion plays no limiting role, and the time scale is set by the
reaction rate.

\subsection{Phase transitions from active to absorbing states}

Turning to our second example in the introductory remarks, we now investigate 
diffusing particles subject to the {\em competing reactions} $A \to \emptyset$ 
and $A \rightleftharpoons A + A$; adding the diffusion term to (\ref{dprreq}), 
we arrive at the rate equation for the local particle density,
\begin{equation}
  \frac{\partial a(x,t)}{\partial t} = - D \left( r - \nabla^2 \right) a(x,t)  
  - \lambda \, a(x,t)^2 \ , 
\label{fiskol}
\end{equation}
where $r = (\kappa - \sigma) / D$; in mathematical biology and ecology, the
partial differential equation (\ref{fiskol}) is known as the 
{\em Fisher--Kolmogorov equation}, and for example has been used to study 
population invasion fronts into empty regions.  
We shall instead focus on the critical region where the control parameter 
$r \to 0$, and a {\em continuous non-equilibrium phase transition} from an
active to an inactive and absorbing state occurs.

The Doi--Peliti field theory action (\ref{masact}) capturing the above 
reactions reads
\begin{eqnarray} 
  &&{\cal A}[{\hat \psi},\psi] = \int \! d^dx \! \int \! dt \, \Bigl[ 
  {\hat \psi} \left( \partial_t - D \, \nabla^2 \right) \psi \quad
\label{fikoac} \\
  &&\ - \kappa \bigl( 1 - {\hat \psi} \bigr) \psi + \sigma \bigl( 1 - 
  {\hat \psi} \bigr) {\hat \psi} \, \psi - \lambda \big( 1 - {\hat \psi} \bigr)
  {\hat \psi} \, \psi^2 \Bigr] \, . \nonumber
\end{eqnarray} 
Upon shifting and rescaling the fields according to 
${\hat \psi}(x,t) = 1 + \sqrt{\lambda / \sigma} \, {\widetilde S}(x,t)$ and 
$\psi(x,t) = \sqrt{\sigma / \lambda} \, S(x,t)$, one arrives at
\begin{eqnarray}  
  &&{\cal A}[{\widetilde S},S] = \int \! d^dx \! \int \! dt \, \Bigl(
  {\widetilde S} \left[ \partial_t + D \left( r - \nabla^2 \right) \right] S 
  \nonumber \\
  &&\qquad\qquad - u \, \bigl( {\widetilde S} - S \bigr) \, {\widetilde S} \, S
  + \lambda \, {\widetilde S}^2 \, S^2 \Bigr) \, ,
\label{dpcrac} 
\end{eqnarray}
where the three-point vertices have been symmetrized and now are proportional
to the coupling $u = \sqrt{\sigma \, \lambda}$ with scaling dimension 
$[u] = \mu^{2-d/2}$.
The associated {\em upper critical dimension} is therefore $d_c = 4$, and the
annihilation four-point vertex $[\lambda] = \mu^{2-d}$ consequently is 
{\em irrelevant} in the RG sense near $d_c$.
In the effective critical action, one may set $\lambda \to 0$, whereupon 
(\ref{dpcrac}) reduces to the familiar {\em Reggeon field theory} action, which
is invariant under {\em rapidity inversion} 
$S(x,t) \leftrightarrow - {\widetilde S}(x,-t)$.
Viewing (\ref{dpcrac}) as a Janssen--De~Dominicis functional (\ref{janded}), it
is equivalent to the Langevin equation that amends the Fisher--Kolmogorov 
equation (\ref{fiskol}) with a noise term,
\begin{equation} 
  \!\!\! \frac{\partial S(x,t)}{\partial t} = D \bigl( \nabla^2 - r \bigr) 
  S(x,t) - u S(x,t)^2 \nonumber + \zeta(x,t) \, , \,\
\label{dplang}
\end{equation}
with $\langle \zeta(x,t) \rangle = 0$ and the multiplicative ``square-root''
noise correlator
\begin{equation}
  \langle \zeta(x,t) \zeta(x',t') \rangle = 2 u \, S(x,t) \, \delta(x-x') 
  \delta(t-t') \, . \
\label{dpnois}
\end{equation} 

Drawing a space-time plot (time running from right to left) for the branching
$A \to A + A$, death $A \to \emptyset$, and coagulation $A + A \to A$ 
processes, starting from a single occupied site, as depicted below, one 
realizes that they generate a {\em directed percolation} (DP) cluster, with 
``time'' playing the role of a specified ``growth'' direction.
The field theory (\ref{dpcrac}) (with $\lambda = 0$) thus also describes the 
universal scaling properties of critical DP.

\centerline{\includegraphics[width = 0.7\columnwidth,angle=180]{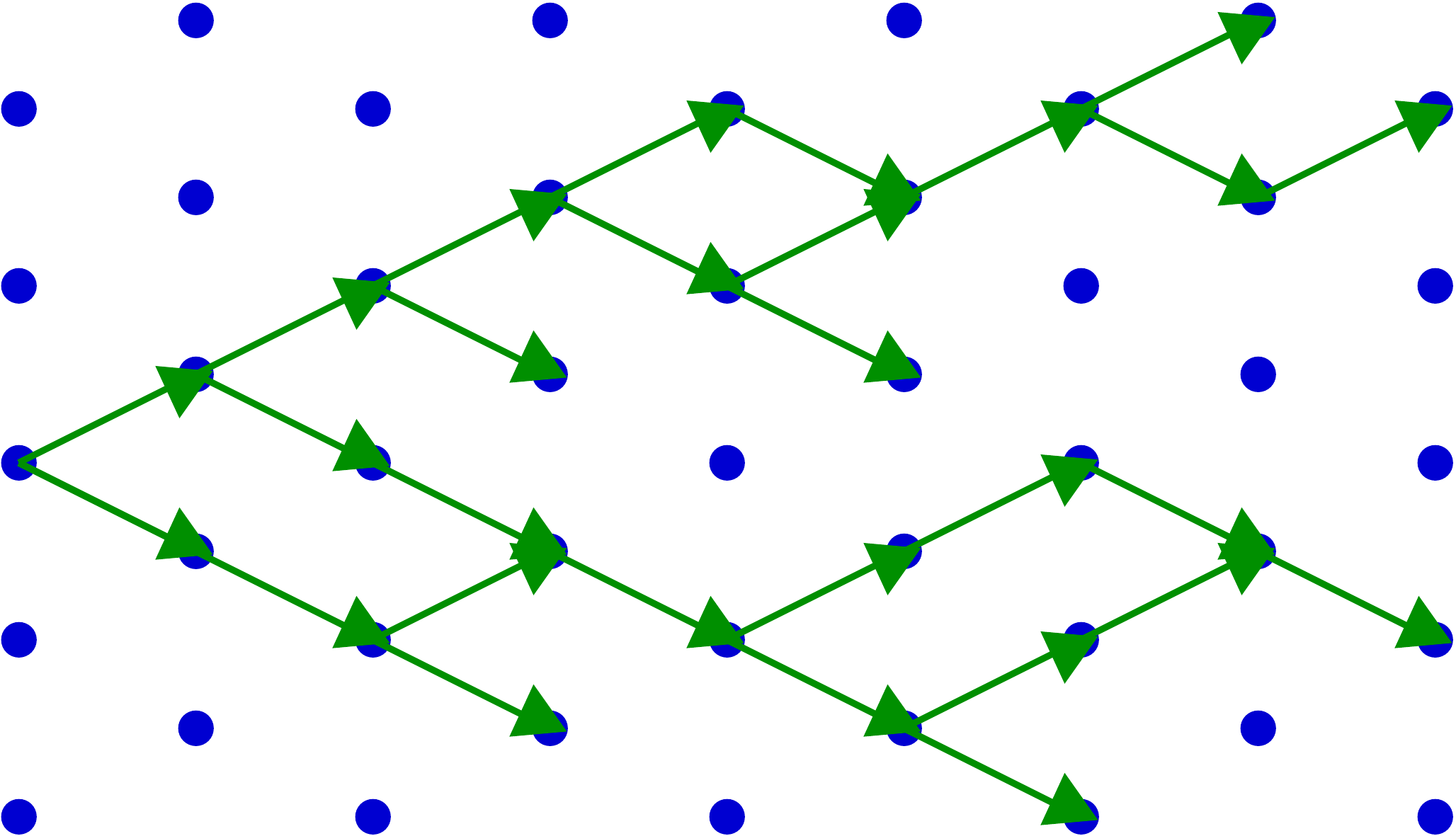}}
\bigskip 

Indeed, one expects that active to absorbing state phase transitions should 
{\em generically} be captured by this DP universality class,
namely in the absence of coupling to other slow conserved modes, disorder, or
special additional symmetries.
The origin for this remarkable DP conjecture becomes evident in a complementary
coarse-grained phenomenological approach that will be framed in the language of
epidemic spreading \cite{Janssen05}.
Consider the following {\em simple epidemic process}:
\begin{enumerate} 
  \item A {\em ``susceptible'' medium} is locally {\em ``infected''}, depending
        on the density of ``sick'' neighbors. 
        Infected regions may later recover.
  \item The ``disease'' {\em extinction} state is {\em absorbing}.
  \item The disease {\em spreads diffusively} via infection, see 1.
  \item Other fast microscopic degrees of freedom are incorporated as random
        {\em noise}.
        Yet according to condition 2, noise alone cannot regenerate the 
        disease. 
\end{enumerate} 

These decisive features can be encoded in a mesoscopic Langevin stochastic 
differential equation for the local density $n(x,t)$ of ``active'' (infected) 
individuals,
\begin{equation}
  \frac{\partial n(x,t)}{\partial t} = D \left( \nabla^2 - R[n(x,t)] \right) 
  n(x,t) + \zeta(x,t) \, , \,\
\label{seplan}
\end{equation}
with the reactive functional $R[n]$, $\langle \zeta(x,t) \rangle = 0$, and 
noise correlator $L[n] = n \, N[n]$.
In the spirit of Landau theory, near extinction one may expand these 
functionals in a Taylor series for small densities,
\begin{equation}
  r \approx 0: \, R[n] = r + u \, n \, + \ldots , \ N[n] = v + \ldots , \quad
\label{seplex}
\end{equation}
where higher-order terms are {\em irrelevant} in the RG sense.
After rescaling, the corresponding Janssen--De~Dominicis response functional 
(\ref{janded}) becomes identical to the Reggeon field theory action.
 
We now proceed to analyze the dynamic perturbation theory and renormalization 
for the DP action (\ref{dpcrac}) to one-loop order.
The only singular vertex functions are the propagator self-energy 
$\Gamma^{(1,1)}(q,\omega)$ and the three-point functions 
$\Gamma^{(1,2)} = - \Gamma^{(2,1)}$, owing to rapidity inversion symmetry, with
the lowest-order Feynman graphs:

\medskip
\centerline{\includegraphics[width=0.7\columnwidth]{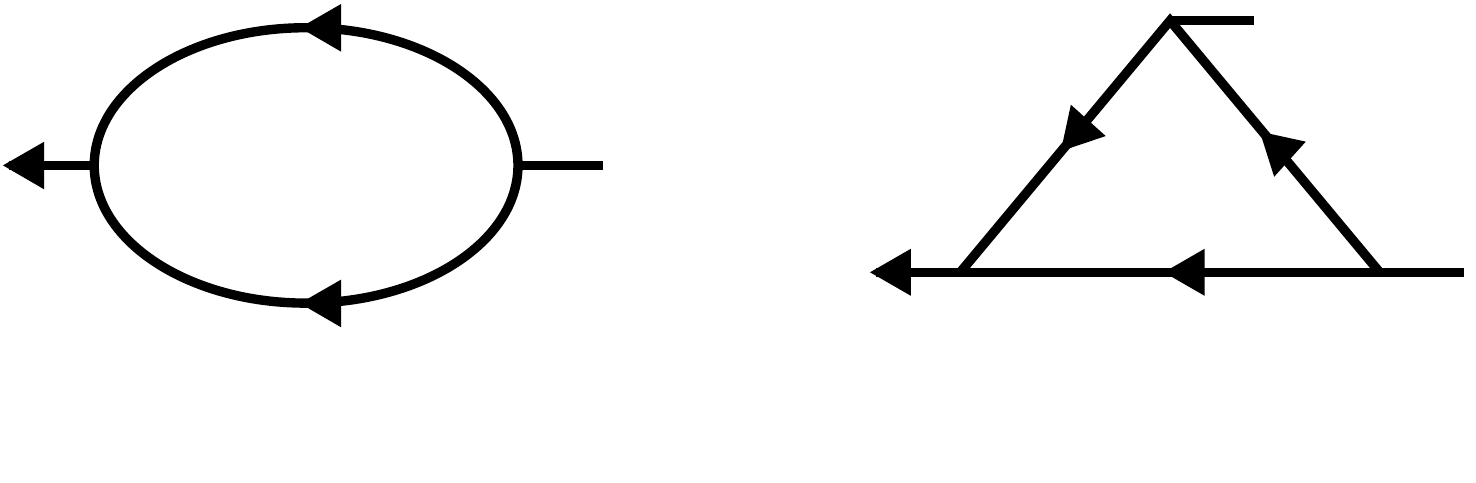}} 
\vskip -0.4truecm

\noindent Explicit evaluation of the self-energy yields
\begin{eqnarray}
  &&\Gamma^{(1,1)}(q,\omega) = i \omega + D (r + q^2) \nonumber \\
  &&\qquad + \frac{u^2}{D} \int_k \frac{1}{i \omega / 2 D + r + q^2/4 + k^2}\ .
\label{dpgm11}
\end{eqnarray}
As in critical statics, one needs to first ensure the {\em criticality 
condition}, namely that $\Gamma^{(1,1)}(0,0) = 0$ at the true percolation 
threshold $r = r_c$.
To one-loop order, (\ref{dpgm11}) results in the shift (additive 
renormalization) 
\begin{equation}
  r_c = - \frac{u^2}{D^2} \int_k \frac{1}{r_c + k^2} + O(u^4) \ ,
\label{dpthsh}
\end{equation}
and inserting $\tau = r - r_c$ in (\ref{dpgm11}) subsequently leads to
\begin{eqnarray}
  &&\Gamma^{(1,1)}(q,\omega) = i \omega + D \left( \tau + q^2 \right)
\label{dpsg11} \\
  &&\quad - \, \frac{u^2}{D} \int_k \frac{i \omega / 2 D + \tau + q^2/4}
  {k^2 \left( i \omega / 2 D + \tau + q^2/4 + k^2 \right)} \ . \nonumber
\end{eqnarray}
The three-point vertex function at vanishing external wave vectors and 
frequencies finally becomes
\begin{equation}
  \Gamma^{(1,2)}(\{ 0 \},(\{ 0 \}) = - 2 u \left[ 1 - \frac{2 u^2}{D^2} \!
  \int_k \frac{1}{\left( \tau + k^2 \right)^2} \right] . \nonumber 
\label{dpgm12}
\end{equation}

For the {\em multiplicative renormalizations}, we follow (\ref{fldren}) and 
(\ref{cplren}), but with the slight modification 
$u_R = Z_u \, u \, A_d^{1/2} \mu^{(d-4)/2}$, as well as (\ref{dynren}) with
$Z_{\widetilde S} = Z_S$ owing to rapidity inversion invariance. 
Thus one obtains Wilson's RG flow functions to one-loop order,
\begin{eqnarray}
  &&\gamma_S = \frac{v_R}{2} + O(v_R^2) \ , \quad 
  \gamma_D = - \frac{v_R}{4} + O(v_R^2) \ , \nonumber \\
  &&\gamma_\tau  = - 2 + \frac{3 v_R}{4} + O(v_R^2) \ ,
\label{dpwilf}
\end{eqnarray}
with the effective coupling and associated beta function
\begin{eqnarray}
  &&v_R = \frac{Z_u^2}{Z_D^2} \, \frac{u^2}{D^2}\, A_d\, \mu^{d-4} \ ,
\label{dpecpl} \\
  &&\beta_v = v_R \Bigl[ - \epsilon + 3 v_R + O(v_R^2) \Bigr] \ .
\label{dpbeta}
\end{eqnarray} 
Below the upper critical dimension $d_c = 4$, the {\em IR-stable RG fixed 
point} 
\begin{equation} 
  v_{\rm DP}^* = \frac{\epsilon}{3} + O(\epsilon^2) 
\label{dprgfp}
\end{equation}
appears, and solving the RG equation for the two-point {\em correlation 
function} in its vicinity yields
\begin{eqnarray}
  &&C_R(\tau_R,q,\omega)^{-1} \approx q^2 \, \ell^{\gamma_S^*} \nonumber \\ 
  &&\qquad {\hat C}_R\Bigl( \tau_R \, \ell^{\gamma_\tau^*},v^*, 
  \frac{q}{\mu \ell}, \frac{\omega}{D_R \, \mu^2 \ell^{2 + \gamma_D^*}} 
  \Bigr)^{-1} .
\label{dpcorf}
\end{eqnarray}
This allows us to identify the three independent {\em critical exponents} for 
directed percolation to order $\epsilon = 4-d$:
\begin{eqnarray}  
  &&\eta = - \gamma_S^* = - \frac{\epsilon}{6} + O(\epsilon^2) \ , \nonumber \\
  &&\nu^{-1} = - \gamma_\tau^* = 2 - \frac{\epsilon}{4} + O(\epsilon^2) \ ,
\label{dpcrex} \\  
  &&z = 2 + \gamma_D^* = 2 - \frac{\epsilon}{12} + O(\epsilon^2) \ . \nonumber
\end{eqnarray} 

The DP universality class also applies to many active to absorbing state phase
transitions with more than just one particle species.
As an example, consider the predator extinction threshold in the spatially
extended stochastic two-species Lotka--Volterra model with finite carrying 
capacity discussed in the chapter introduction.
The associated Doi--Peliti field theory action reads
\begin{eqnarray}
  &&S[{\hat a},a;{\hat b},b] = \int \!\! d^dx \! \int \!\! dt \, \Bigl[ 
  {\hat a} \, \bigl( \partial_t - D_A \nabla^2 \bigr) \, a \nonumber \\
  &&\qquad + \kappa \, \bigl( {\hat a} - 1 \bigr) \, a 
  + {\hat b} \, \bigl( \partial_t - D_B \nabla^2 \bigr) \, b 
\label{lvdpft} \\
  &&\quad + \sigma \, \bigl( 1 - {\hat b} \bigr) \, {\hat b} \, b \, 
  e^{-\rho^{-1} \, {\hat b} b} + \lambda \, \bigl( {\hat b} - {\hat a} \bigr) 
  \, {\hat a} \, a \, b \Bigr] \, , \nonumber
\end{eqnarray}
where diffusive spreading has been assumed, and the exponential term in the
prey production term takes into account the local restriction to a maximum 
particle density $\rho$.
As usual, one applies the field shifts ${\hat a} = 1 + {\tilde a}$, 
${\hat b} = 1 + {\tilde b}$; realizing that $\rho^{-1}$ constitutes an
irrelevant perturbation (since the density scales as $[\rho] = \mu^d$), we
furthermore expand to lowest order in $\rho^{-1}$, which yields 
\begin{eqnarray}
  &&S[{\tilde a},a;{\tilde b},b] = \int \!\! d^dx \! \int \!\! dt \, 
  \Bigl[ {\tilde a} \, \bigl( \partial_t - D_A \nabla^2 + \kappa \bigr) \, a 
  \nonumber \\
  &&\qquad + {\tilde b} \, \bigl( \partial_t - D_B \nabla^2 - \sigma \bigr) \,
  b - \sigma \, {\tilde b}^2 \, b 
\label{dpshft} \\
  &&\ + \sigma \, \rho^{-1} \, \bigl( 1 + {\tilde b} \bigr)^2 \, {\tilde b}
  \, b^2 - \lambda \, \bigl( 1 + {\tilde a} \bigr) \, 
  \bigl( {\tilde a} - {\tilde b} \bigr) \, a \, b \Bigr] . \nonumber
\end{eqnarray}
Near the predator extinction threshold, the prey almost fill the entire system.
We therefore define the properly {\em fluctuating fields} $c = b_s - b$ with 
$b_s \approx \rho$ and $\langle c \rangle = 0$, and 
${\tilde c} = - {\tilde b}$.
Rescaling to $\phi = \sqrt{\sigma} \, c$ and
${\tilde \phi} = \sqrt{\sigma} \, {\tilde c}$, and noting that asymptotically
$\sigma \to \infty$ under the RG flow since $[\sigma] = \mu^2$, the ensuing
action simplifies drastically.
At last, we add the {\em growth-limiting reaction} $A + A \to A$ (with rate 
$\tau$); the fields $\phi$ and ${\tilde \phi}$ can then be integrated out,
leaving Reggeon field theory (\ref{dpcrac}) as the resulting effective action, 
with the non-linear coupling $u = \sqrt{\tau \, \lambda \, b_s}$.

\section{Concluding Remarks}

These lecture notes can of course only provide a very sketchy and vastly
incomplete introduction to the use of field theory tools and applications of 
the renormalization group in statistical physics.
I have merely focused on continuous phase transitions in equilibrium, dynamic 
critical phenomena in simple relaxational models, and a few examples for 
universal scaling behavior in non-equilibrium dynamical systems.
Among the many topics not covered or even mentioned here are critical phenomena
in finite and disordered systems; universality classes of critical dynamics 
with reversible couplings to other conserved modes; universal short-time and 
non-equilibrium relaxation scaling properties in the aging regime; depinning 
transitions and driven interfaces in disordered media; spin glasses and 
structural glasses; and of course phase transitions and generic scale 
invariance in quantum systems.
Nor have I addressed powerful representations through supersymmetric or 
conformally invariant quantum field theories, Monte Carlo algorithms, or 
numerical non-perturbative RG approaches, since the latter will be covered 
elsewhere in this volume; for their applications to non-equilibrium systems, 
see, e.g., Ref.~\cite{Canet11}.

Nevertheless, I hope to have conveyed the message that methods from field 
theory are ubiquitous in statistical physics, and the renormalization group has
served as a remarkably powerful mathematical tool to address at least the
universal scaling aspects of cooperative behavior governed by strong 
correlations and fluctuations.
Thus, the RG has become a cornerstone of our understanding of complex 
interacting many-particle systems, and its language and basic philosophy now
pervade the entire field, with applications that increasingly reach out beyond
fundamental physics to material science, chemistry, biology, ecology, and even
sociology.

Finally, I would like to express my gratitude to the organizers (and their 
funding agencies) for the opportunity to attend and lecture at the 49th 
Schladming Theoretical Physics Winter School.
I thoroughly enjoyed the stimulating and informal atmosphere in the Styrian 
Alps, and profited from many discussions with colleagues and students.
I can only hope that all attendants learned as much about new and exciting 
developments in theoretical physics from my fellow lecturers as I did.





\begin{thebibliography}{00}

\bibitem{Wilson74} 
	K.G.~Wilson and J.~Kogut,
        {\em The renormalization group and the $\epsilon$ expansion},
        Phys. Rep. {\bf 12 C}, 75--200 (1974).

\bibitem{Fisher74} 
	M.E.~Fisher,
        {\em The renormalization group in the theory of critical behavior},
        Rev. Mod. Phys. {\bf 46}, 597--616 (1974).

\bibitem{Ma76} 
	S.-k.~Ma,
        {\em Modern theory of critical phenomena},
        Benjamin--Cummings (Reading, 1976).

\bibitem{Patashinskii79} 
	A.Z.~Patashinskii and V.L.~Pokrovskii,
        {\em Fluctuation theory of phase transitions},
        Pergamon Press (New York, 1979).

\bibitem{Goldenfeld92}
	N.~Goldenfeld,
        {\em Lectures on phase transitions and the renormalization group},
        Addison--Wesley (Reading, 1992).

\bibitem{Binney93} 
	J.J.~Binney, N.J.~Dowrick, A.J.~Fisher, and M.E.J.~Newman,
        {\em The theory of critical phenomena}, 
        Oxford University Press (Oxford, 1993).

\bibitem{Cardy96} 
	J.~Cardy,
        {\em Scaling and renormalization in statistical physics},
        Cambridge University Press (Cambridge, 1996).

\bibitem{Mazenko03} 
	G.F.~Mazenko,
        {\em Fluctuations, order, and defects},
        Wiley--Inter\-science (Hoboken, 2003).

\bibitem{Tauberxx} 
	U.C.~T\"auber,
        {\em Critical Dynamics: A Field Theory Approach to Equilibrium and 
             Non-Equilibrium Scaling Behavior}, 
        under contract with Cambridge University Press (Cambridge); 
        see {\tt http://www.phys.vt.edu/\~{}tauber/utaeuber.html}.

\bibitem{Ramond81} 
	P.~Ramond,
        {\em Field theory --- A modern primer},
        Benjamin--Cummings (Reading, 1981). 

\bibitem{Amit84} 
	D.J.~Amit,
	{\em Field theory, the renormalization group, and critical phenomena},
        World Scientific (Singapore, 1984).

\bibitem{Parisi88} 
	G.~Parisi, 
        {\em Statistical field theory},
        Addison--Wesley (Redwood City, 1988).

\bibitem{Itzykson89} 
	C.~Itzykson and J.M.~Drouffe,
        {\em Statistical field theory}, Vol.~I, 
        Cambridge University Press (Cambridge, 1989).

\bibitem{Bellac91} 
	M.~Le~Bellac,
        {\em Quantum and statistical field theory},
        Oxford University Press (Oxford, 1991).

\bibitem{Zinn93} 
	J.~Zinn-Justin,
        {\em Quantum field theory and critical phenomena},
        Clarendon Press (Oxford, 1993).

\bibitem{Hohenberg77} 
	P.C.~Hohenberg and B.I.~Halperin, 
        {\em Theory of dynamic critical phenomena}, 
        Rev. Mod. Phys. {\bf 49}, 435--479 (1977). 

\bibitem{Janssen79} 
	H.K.~Janssen,
	{\em Field-theoretic methods applied to critical dynamics},
        in: {\em Dynamical critical phenomena and related topics}, ed. 
        C.P.~Enz, Lecture Notes in Physics, Vol.~{\bf 104}, 
        Springer (Heidelberg), 26--47 (1979).

\bibitem{Folk06} 
	R.~Folk and G.~Moser,
  	{\em Critical dynamics: a field theoretical approach},
  	J. Phys. A: Math. Gen. {\bf 39}, R207--R313 (2006).

\bibitem{Tauber07} 
	U.C.~T\"auber,
        {\em Field theory approaches to nonequilibrium dynamics}, 
        in: {\em Ageing and the Glass Transition}, eds. M.~Henkel, 
        M.~Pleimling, and R.~Sanctuary, Lecture Notes in Physics, 
        Vol.~{\bf 716}, Springer (Berlin), 295--348 (2007).

\bibitem{Frey94} 
	E.~Frey and F.~Schwabl,
        {\em Critical dynamics of magnets},  
        Adv. Phys. {\bf 43}, 577--683 (1994).

\bibitem{Schmittmann95} 
	B.~Schmittmann and R.K.P.~Zia,
        {\em Statistical mechanics of driven diffusive systems},
        in: {\em Phase Transitions and Critical Phenomena}, ed. C.~Domb and 
        J.L.~Lebowitz, Vol.~{\bf 17}, Academic Press (London, 1995).

\bibitem{Cardy97} 
	J.L.~Cardy,
        {\em Renormalisation group approach to reaction-diffu\-sion problems}, 
        in: {\em Proceedings of Mathematical Beauty of Physics}, ed. 
        J.-B.~Zuber, Adv. Ser. in Math. Phys. {\bf 24}, 113 (1997).
 
\bibitem{Mattis98} 
	D.C.~Mattis and M.L.~Glasser, 
        {\em The uses of quantum field theory in diffusion-limited reactions},
        Rev. Mod. Phys. {\bf 70}, 979--1002 (1998).
 
\bibitem{Tauber05} 
	U.C.~T\"auber, M.J.~Howard, and B.P.~Vollmayr-Lee,
        {\em Applications of field-theoretic renormalization group methods to 
        reaction-diffusion problems},  
        J. Phys. A: Math. Gen. {\bf 38}, R79--R131 (2005).

\bibitem{Hinrichsen01} 
	H.~Hinrichsen,
        {\em Nonequilibrium critical phenomena and phase transitions into 
             absorbing states} 
        Adv. Phys. {\bf 49}, 815--958 (2001).

\bibitem{Odor04} 
	G.~\'Odor,
        {\em Phase transition universality classes of classical, 
             nonequilibrium systems}, 
        Rev. Mod. Phys. {\bf 76}, 663--724 (2004).

\bibitem{Janssen05}
	H.K.~Janssen and U.C.~T\"auber, 
        {\em The field theory approach to percolation processes},  
        Ann. Phys. (NY) {\bf 315}, 147--192 (2005).

\bibitem{Mobilia07}
	M. Mobilia, I.T. Georgiev, and U.C.~T\"auber,
        {\em Phase transitions and spatio-temporal fluctuations in stochastic
             lattice Lotka--Volterra models},
        J. Stat. Phys. {\bf 128}, 447--483 (2007).

\bibitem{Tauber11} 
	U.C.~T\"auber,
        {\em Stochastic population oscillations in spatial preda\-tor-prey 
             models},
        J. Phys.: Conf. Ser. {\bf 319}, 012019 - 1--14 (2011).

\bibitem{Canet11}
	L. Canet, H. Chat\'e, and B. Delamotte,
	{\em General framework of the non-perturbative renormalization group
             for non-equilibrium steady states},
        J. Phys. A: Math. Gen. {\bf 44}, 495001 - 1--26 (2011).

\end{thebibliography}
\end{document}